\documentclass [10pt,twocolumn]{IEEEtran}
\usepackage{algorithm,algorithmic,amsbsy,amsmath,amssymb,epsfig,bbm,mathrsfs,multirow,amsthm}
\usepackage[T1]{fontenc}
\usepackage[latin9]{inputenc}
\usepackage{amsmath}
\usepackage{fixltx2e}
\usepackage{algorithm}
\usepackage{array,multirow,graphicx}
\usepackage{color}
\usepackage{cite}
\usepackage{float}
\usepackage{makecell}
\usepackage[x11names,dvipsnames,table]{xcolor} 
\usepackage{colortbl}
\usepackage{multirow}
\definecolor{mygray}{gray}{0.6}
\definecolor{myblue}{rgb}{0.8,0.85,1} 

\usepackage{array, tabularx, boldline}
\usepackage{graphicx}
\usepackage{cellspace}
\begin{document}
\title{Resource Management in Cloud Networking Using Economic Analysis and Pricing Models: A Survey}

\author{Nguyen Cong Luong, Ping Wang, \textit{Senior Member, IEEE}, Dusit Niyato, \textit{Fellow, IEEE}, Wen Yonggang, \textit{Senior Member, IEEE}, and Zhu Han, \textit{Fellow, IEEE}
\thanks{N.~C.~Luong, P.~Wang, D.~Niyato, and Y. Wen, are with School of Computer Science and Engineering, Nanyang Technological University, Singapore. E-mails: clnguyen@ntu.edu.sg, wangping@ntu.edu.sg, dniyato@ntu.edu.sg, and ygwen@ntu.edu.sg.}
\thanks{Z.~Han is with Electrical and Computer Engineering and Computer Science, University of Houston, Houston, TX, USA. E-mail: hanzhu22@gmail.com.} 
}

\maketitle
\begin{abstract}
This paper presents a comprehensive literature review on applications of economic and pricing models for resource management in cloud networking. To achieve sustainable profit advantage, cost reduction, and flexibility in provisioning of cloud resources, resource management in cloud networking requires adaptive and robust designs to address many issues, e.g., resource allocation, bandwidth reservation, request allocation, and workload allocation. Economic and pricing models have received a lot of attention as they can lead to desirable performance in terms of social welfare, fairness, truthfulness, profit, user satisfaction, and resource utilization. This paper reviews applications of the economic and pricing models to develop adaptive algorithms and protocols for resource management in cloud networking. Besides, we survey a variety of incentive mechanisms using the pricing strategies in sharing resources in edge computing. In addition, we consider using pricing models in cloud-based Software Defined Wireless Networking (cloud-based SDWN). Finally, we highlight important challenges, open issues and future research directions of applying economic and pricing models to cloud networking.

\end{abstract}

\section{Introduction}
\label{sec:intro}
Cloud computing is becoming the platform of choice for a number of applications due to the advantages of high computing power, low service cost, high scalability, accessibility, and availability. Cloud computing is used as an integral part of society in various domains and disciplines such as education \cite{sultan2010cloud}, commerce \cite{rahimi2014mobile}, health care services \cite{kuo2011opportunities}, transportation \cite{li2011cloud}, and social networks \cite{chard2010social}. Cloud computing is expected to bring huge new revenue opportunities. Recent reports have showed that the global revenue generated from cloud services is more than \$200 billion in 2016, and there will be approximately 3.6 billion Internet users accessing cloud services by 2018 (http://www.statista.com). On-demand services provided by cloud computing include, for example, Software-as-a-Service (SaaS), Platform-as-a-Service (PaaS), and Infrastructure-as-a-Service (IaaS). Google Docs \cite{GoogleDoc}, Google App Engine \cite{GoogleApp}, and Amazon's Elastic Compute Cloud (Amazon's EC2) \cite{AmazonEC2} are among the popular commercial services available in cloud computing.


The cloud computing infrastructure is typically hosted in data centers. Therefore, the current cloud services are based on parallel implementations in distributed data centers connected with each other through high speed networks. For example, the EU-funded Scalable and Adaptive Internet Solution (Eu-funded SAIL) project \cite{SAIL_project} investigates a combination of cloud computing infrastructure and networking capabilities, called \textit{cloud networking}. Cloud networking actually considers the network beyond the data centers with the aim of providing both on-demand computing and network resources. With cloud networking, the resources and services can be provisioned from interconnecting distributed data centers owned by one or multiple providers, called \textit{cloud data center networking}. The cloud resources and services can also be integrated with mobile networks, i.e., \textit{mobile cloud networking}. Moreover, \textit{edge computing} models are deployed in cloud networking to bring the cloud resources and services close to users, and thus minimize overall costs, jitter, latencies, and network load. The aforementioned models of cloud networking along with the integration of the Software-Defined Networking (SDN) technology are expected to support and satisfy a large number of users and applications in terms of flexibility, cost and availability of the services.
 
However, managing network and cloud resources together in cloud networking has many challenges. It is crucial to have an integrated view of the existing physical and virtual topologies and characteristics of the resources, as well as the status of all network entities. Besides, the provisioning and placement of virtual resources must be done in the best way possible, taking into account the available resources of both the cloud and networks. Moreover, the reconfigurations must often be performed to resize or release the existing virtual resources due to, e.g., the dynamic network environments (node or link failure), and the variability/elasticity of resource demand. Inefficient resource management negatively affects performance and cost as well as impairing system functionality.

To address the aforementioned challenges, it is vital to develop resource management approaches which guarantee the scalability, efficiency, manageability, adaptability and reliability for cloud networking. Traditional approaches, e.g., the system optimization, merely focus on the system performance metrics given system parameters and constraints rather than economic factors, e.g., the profit, cost, and revenue. Therefore, economic and pricing approaches have been recently explored, developed, and adopted for resource management in cloud networking. Compared with the system optimization approaches, the economic and pricing approaches provide the following advantages:
\begin{itemize}
\item In cloud networking, the profits of cloud providers have to be maximized while meeting the user demands. Thus, the profit guarantee for all cloud providers is a primary goal. Pricing models based on, e.g., the profit maximization or cost minimization, have been efficiently used to achieve the goal.
\item There are various actors/stakeholders in cloud networking which belong to different entities, e.g., end-users, infrastructure providers, service providers, brokers, and network operators. They have different objectives, e.g., the profit, revenue, cost and utility, as well as different constraints, e.g., the budget and technology. Their objectives often conflict with each other, and this makes economic and pricing models become effective tools in cloud networking. More specifically, through the use of negotiation mechanisms, economic and pricing approaches can determine optimal solutions for selfish entities given their constraints.
\item The demand for cloud computing and network resources depends on many users' attributes, e.g., the willingness to pay and performance requirements. Pricing strategies which rely on the demand elasticity such as price discrimination have been recently used as ideal solutions to optimize the provisioning of resources and profits of providers.
\item Video on Demand (VoD) undoubtedly is among the most important services in cloud networking. Several commercial video delivery services have been introduced and become popular, e.g., YouTube and Netflix. However, the bandwidth cost of the service is typically very significant. Pricing mechanisms, e.g., smart data pricing, have been applied to regulate the user demands and maximize the bandwidth utilization.
\item Besides the high bandwidth utilization for providers, guaranteeing quality of service (QoS), e.g., a small delay, for users is very important. Pricing approaches provide very efficient solutions for the joint optimization of both providers and users.
\item To reduce service delay for users, cloud networking has developed edge computing models which employ devices at network edges to provide closer cloud resources and services to users. Pricing and payment strategies stimulate users to use the edge resources rather than distant data centers while still guaranteeing profits for cloud providers.
\end{itemize}

Although there are several surveys related to cloud networking, they do not focus on economic and pricing approaches, which are emerging as a promising tool. For example, a survey of applications of network virtualization for cloud computing was given in \cite{jain2013network}. The survey of technologies of the Network-as-a-Service (NaaS) paradigm for supporting network-cloud convergence was presented in \cite{duan2012survey}. There are also surveys related to the architecture of SDNs, e.g., \cite{xia2015survey}, \cite{kreutz2015software}, \cite{farhady2015software}, \cite{yang2015software}, and applications of edge computing \cite{yi2015survey}. There are surveys related to the pricing approaches, e.g., \cite{he2012internet}, \cite{dasilva2000pricing}, \cite{gizelis2011survey}. However, they addressed the issues in Internet or wireless networks only. To the best of our knowledge, there is no survey specifically discussing the use of economic and pricing models to deal with resource management in cloud networking. This motivates us to develop the survey with the comprehensive literature review on the economic and pricing models in cloud networking.


For convenience, the related works in this survey are classified based on various models of cloud networking and then their issues as shown in Table~\ref{table_classification}. The models of cloud networking considered in this survey are cloud data center networking, mobile cloud networking, edge computing, and cloud-based Video-on-Demand (VoD) systems. Furthermore, some pricing approaches for the resource management in cloud-based Software Defined Wireless Networking (cloud-based SDWN) are discussed. Advantages and disadvantages of each approach are highlighted.

The rest of this paper is organized as follows. Section~\ref{sec:cloud_general} describes a general architecture of cloud networking. Section~\ref{sec:Intro_Price} introduces the fundamentals of economic and pricing models. Section~\ref{sec:cloud_data_center} discusses how to apply economic and pricing models for resource management in cloud data center networking such as bandwidth, request, and workload allocation. Applications of economic and pricing models for resource allocation in mobile cloud networking are given in Section \ref{sec:mobile_cloud_networking}. Section \ref{sec:social_edge_computing} reviews economic and pricing models to address issues concerning the bandwidth allocation, task allocation, and storage sharing in edge computing. Section \ref{sec:VoD_system} considers economic and pricing approaches for bandwidth allocation and Peer-to-Peer (P2P) caching in cloud-based VoD system. In addition, applications of economic and pricing models for bandwidth allocation and mobile data offloading in cloud-based SDWN are given in Section \ref{sec:SDN_NFV}. We outline important challenges, open issues, and future research directions in Section~\ref{sec:Open_issues}. Finally, we conclude the paper in Section~\ref{sec:Conclusion}. The list of abbreviations appeared in this paper are given in Table~\ref{tab:table_abb}. 

\begin{table}[h!]
  \caption{Major abbreviations}
  \label{tab:table_abb}\centering
  \begin{tabularx}{8.7cm}{|Sl|X|}
    \hline
  \cellcolor{mygray} \textbf{Abbreviation} &   \cellcolor{mygray} \textbf{Description} \\   
    \hline
    BBU & BaseBand processing Units \\
    \hline
    Cloud-RAN& Cloud-Radio Access Network \\
     \hline
    CAPEX   & CAPital EXpenditure\\
   \hline
    CWMSN& Cloud-based Wireless Multimedia Social Network \\ 
    \hline
    IaaS& Infrastructure-as-a-Service \\
    \hline
    IoT  & Internet of Things\\
      \hline
    MCN&Mobile Cloud Networking \\
        \hline
    MNO&Mobile Network Operator \\
      \hline
    NFV   & Network Function Virtualization \\
       \hline
    NUM  & Network Utility Maximization\\
        \hline
    OPEX & OPerational EXpenditure\\
        \hline
    P2P&Peer-to-Peer \\
         \hline
    PaaS&Platform-as-a-Service \\
        \hline
    RRH  & Remote Radio Head\\ 
        \hline
    SaaS& Software-as-a-Service \\
        \hline  
    SDN & Software-Defined Networking \\
        \hline
    SDWN & Software-Defined Wireless Networking \\
     \hline
    SLA & Service-Level Agreement\\
     \hline
     SP & Service Provider\\
      \hline
    VCG & Vickrey-Clarke-Groves\\
      \hline
    VM& Virtual Machine \\
       \hline
    VoD& Video on Demand \\
      \hline
  \end{tabularx}
\end{table}

\begin{center}
\scriptsize
\begin{table*}[ht]
\caption{A taxonomy of the applications of economic and pricing models for resource management in cloud networking}
{
\hfill{}
\begin{tabular}{|l|c|c|c|c|c|c|}\hline
\scriptsize
\diaghead{\theadfont Diag ColumnmnHead II}%
{\textbf{Design issues}}{\textbf{System models}}&
\thead{\color{blue}Cloud data \\ \color{blue} center networking \\ \textbf{(Section IV)}} &\thead{\color{blue} Mobile cloud \\\color{blue} networking\\ \textbf{(Section V)}}&\thead{\color{blue} Edge \\ \color{blue} computing\\  \textbf{(Section VI)}} &\thead{\color{blue} Cloud-based\\ \color{blue}VoD system\\ \textbf{(Section VII)}}& \thead{\color{blue} Cloud-based \\ \color{blue}SDWN\\ \textbf{(Section VIII)}}\\ \hline
\thead{Bandwidth allocation}&\checkmark &\checkmark&\checkmark &\checkmark &\checkmark\\ \hline
\thead{Resource allocation}& &\checkmark&\checkmark && \\ \hline
\thead{Task allocation}&&&\checkmark & &\\ \hline
\thead{Request allocation}&\checkmark &&&& \\ \hline
\thead{Workload allocation}&\checkmark &&&&\\ \hline
\thead{Storage sharing}&&&\checkmark&& \\ \hline
\thead{P2P caching}&&&&\checkmark&\\ \hline
\thead{Mobile data offloading}&&&&&\checkmark\\ \hline
\end{tabular}}
\label{table_classification}
\hfill{}
\end{table*}
\end{center}

\section{General architecture of cloud networking}
\label{sec:cloud_general}

\subsection{Definition of cloud networking}
\label{sec:cloud_networking_definition}
 The term cloud networking is understood in a multi-administrative domain scenario in which network and data center domains interact with each other through predefined interfaces \cite{murray2012cloud}, \cite{mouftah2013communication}. Specifically, cloud networking extends network virtualization beyond the data centers to provide cloud and network resources to clients/users. Network resources can be virtual routers, bandwidth, virtual firewalls, or any network management software.

The definition also shows a key difference between the cloud networking and traditional computer networks, that is the \textit{network virtualization}. By using network virtualization, the cloud networking reduces the cost for both providers and clients through real-time, on-demand resource and service provisioning. The resources are assigned and used by the client's needs, and the client only pays for what is used \cite{chowdhury2010survey}. On the contrary, resource allocation in traditional computer networks is static, and a client needs to pay for every cost regardless of whether the resource has been used or not.

\subsection{Architecture of cloud networking}
\label{sec:cloud_networking_architecture}
The goal of a cloud networking architecture is to enable an efficient composition of cloud and network resources in a cloud environment. To achieve the goal, several architectures were proposed for cloud networking. They can be based on intra-data center networking and inter-data center networking which are commonly called the cloud data center networking \cite{duan2012survey}, \cite{murray2012cloud}, \cite{xiang2014greening}, \cite{bitar2013technologies}, \cite{levin2015enabling}. They can be based on mobile cloud networking \cite{jamakovic2013mobile}, \cite{karagiannis2014mobile} or edge computing models \cite{lewis2014tactical} \cite{stojmenovic2014fog}, \cite{ahlgren2011content}, \cite{beck2014mobile}.

\begin{figure*}[ht]
\centering
\includegraphics[width=16.3cm,height = 13.9cm]{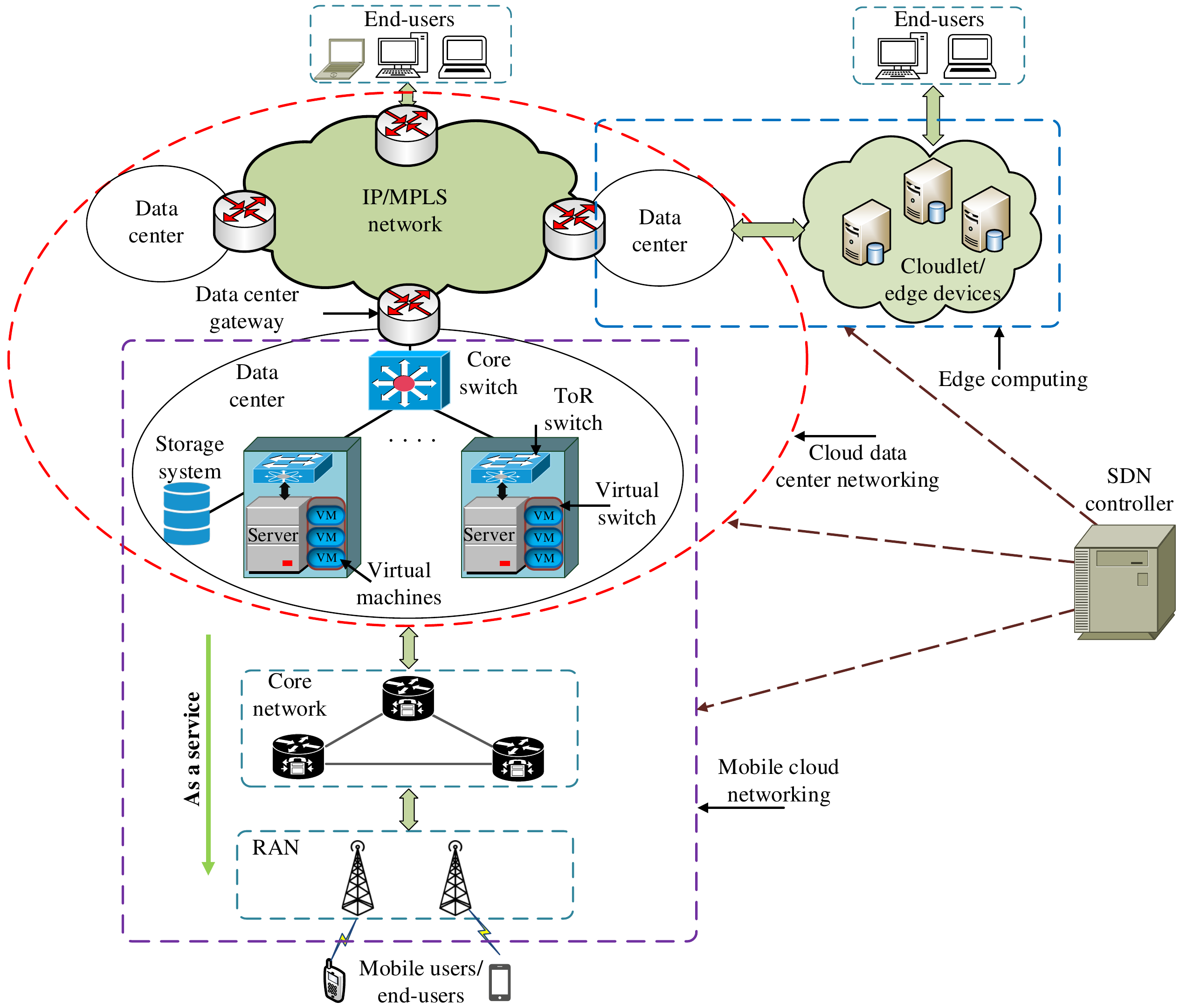}
 \caption{A general architecture of cloud networking.}
 \label{cloud_networking_architecture}
\end{figure*}

Based on these architectures, we provide a general, unified architecture for cloud networking as shown in Fig.~\ref{cloud_networking_architecture}. The architecture has three major parts: (i) cloud data center networking, (ii) mobile cloud networking, and (iii) edge computing. Their descriptions are given in what follows. Note that these parts can be independent from each other. 
Stakeholders or actors commonly participating in cloud networking are as follows.

\begin{itemize}
\item \textit{Cloud provider:} A cloud provider, e.g, an IaaS cloud provider, owns and manages data centers and system software.
\item \textit{Network provider:} A network provider provides network connectivities among data centers of cloud providers or between end-users and data centers. In cloud networking, the network providers aim at cooperating with cloud providers to allocate cloud network resources and services to end-users or cloud users.
\item \textit{Cloud tenant/cloud user:} A cloud tenant can be a service provider, an organization or an enterprise, which uses cloud resources to host applications offered to its end-users. Netflix (https://www.netflix.com/) is an example of a service provider of video on demand.
\item \textit{Cloud service broker:} A cloud service broker (or broker for the sake of shortness) acts as an intermediary between cloud users/end-users and cloud providers.
\item \textit{End-users:} The users generate resource and service requests or workloads that need to be processed using cloud resources.
\end{itemize}

\subsubsection{Cloud data center networking}
\label{sec:cloud_data_center_definition}
 A data center is a large group of networked computer servers which are capable of providing the remote storage, processing, or distribution of large amounts of data. We provide brief descriptions of the components and resources in both intra- and inter-data center networking to which they will be referred in this survey.
\begin{itemize}
\item \textit{Intra-data center networking:}
Intra-data center networking refers to the interconnection between
servers and storage resources through a networking system within a
data center. The networking system includes virtual switches,
Top-of-Rack (ToR) switches, core switches, and non-broking switch.
\begin{itemize}
\item \textit{Virtual Machine (VM):} VM is a software program or operating system which is able to perform tasks such as running applications and programs as a separate computer \cite{smith2005architecture}. Multiple VMs can exist within a physical server or machine through virtualization techniques. In cloud networking, a VM can be migrated among servers within a data center or between data centers owned by different providers.
\item \textit{Virtual switch:} A virtual switch is generally a software-based Ethernet switch function running inside a server. It can support Ethernet and/or IP services and provide switching and routing context separation among tenants/users sharing the same server. 
\item \textit{Network slicing:} Network slicing allows compartmentalizing VMs of the same application into the same virtual networks \cite{yuan2012game} and guarantees virtual resource isolation and virtual network performance.
\item \textit{ToR switch:} A ToR switch supports Ethernet virtual LAN (VLAN) services or simple IP routing for the data center. The ToR switch aggregates Ethernet links from the servers. ToR switches are connected to one or two core switches in a data center.
\item \textit{Non-blocking switch:} A switch is called non-blocking if it is able to connect all ports such that any routing request to any free output port can be established successfully without interfering other traffics.
\item \textit{Core switch:} A core switch hosts multiple ToR switches and large-scale virtual LAN services or simple IP routing for the data center.
\end{itemize}
\item \textit{Inter-data center networking:}
Data centers can be interconnected across the Wide Area Network (WAN) using inter-data center networking. Some commonly referred entities in the inter-data center networking are as follows.
\begin{itemize}
\item \textit{Data center gateway:} A data center gateway provides connectivity among data centers and to Internet and VPN customers. The data center gateway can provide virtual routing and switching capabilities.
\item \textit{IP/MPLS network:} An Internet Protocol/Multi-Protocol Label Switching (IP/MPLS) network is a packet-switched network that employs the Internet Protocol (TCP/IP) enhanced with the MPLS standard.
\item \textit{Resource pool:} A resource pool is a collective set of resources in data centers.
\item \textit{Federated cloud networking:} Federated or federation cloud networking refers to the cooperation among cloud providers to establish the federated cloud resource. For the federated cloud networking, a cloud provider can ``borrow'' cloud resources from other providers if its own resources are overloaded. This is called \textit{outsourcing}. Also, a cloud provider can ``rent out'' its resources to other cloud providers if its resources are free. This is called \textit{insourcing}.
\end{itemize}
\end{itemize}

\subsubsection{Mobile cloud networking}
\label{sec:cloud_mobile_networking_definition}
Mobile Cloud Networking (MCN) is the EU FP7 Large-scale Integrating Project (IP) (cordis.europa.eu/fp7/ict/future-networks). It focuses on integrating the cloud computing and network function virtualization technologies to mobile networks \cite{carella2015mobile}. MCN is able to provision services involving mobile network, decentralized computing, and storage as one on-demand unified service. The main characteristics of MCN are as follows \cite{jamakovic2013mobile}:
\begin{itemize}
\item MCN improves the real-time performance of mobile network functions, e.g., the baseband unit processing, mobility management, and QoS control, based on the high-performance cloud computing infrastructure. Thus, MCN enables adapting to the elasticity of the load.
\item MCN provides an entirely new mobile cloud application platform as well as novel revenue streams for Telco by orchestrating infrastructure and services across different domains including wireless, mobile core networks, and data centers.
\item MCN has the 3GPP LTE compliant architecture to exploit and support cloud computing.
\item MCN introduces a new business actor, i.e., the MCN provider, in addition to typical stakeholders, e.g., the cloud computing provider, application provider, and users.
\end{itemize}

A wide range of services is offered by MCN: (i) typical cloud computing atomic services, e.g., the computing, storage, and networking, (ii) support services, e.g., Monitoring as a Service (MaaS), (iii) virtualized network infrastructure services, e.g., Radio Access Network-as-a-Service (RANaaS) and Evolved Packet Core-as-a-Service (EPCaaS), (iv) new virtualized applications and services, e.g., Content Delivery Networks-as-a-Service (CDNaaS), and (v) End-to-End (E2E) services. In particular, RANaaS allows to partially move functionalities of RAN, i.e., digital processing functions, to a data center depending on the actual needs and network characteristics \cite{rost2014cloud}. When all RAN functionalities are shifted towards the data center, and only RF functions are performed at Remote Radio Head (RRH), we have the concept of \textit{Cloud-RAN} or \textit{Centralized-RAN} \cite{peng2015fronthaul}. The RANaaS implementation has the following major characteristics \cite{sabella2013ran}:
\begin{itemize}
\item \textit{On-demand provisioning:} Mobile network resources and services are provisioned according to the demand elasticity of mobile users.
\item \textit{Virtualization of RAN resources and functions:} They aim at optimizing usage, management, and scalability of the mobile network.
\item \textit{Resource pooling:} This allows virtual operators to share more dedicated resources and services, and thus enabling more business opportunities.
\item \textit{Elasticity:} This characteristic enables scaling network resources at the data centers or controlling the number of active RRHs.
\item \textit{Service metering:} Operators provision and charge RAN operation services, e.g., the usage of RRHs, on a measurable and controllable basis.
\item \textit{Multi-tenancy:} This feature ensures the security in the mobile network by enabling isolation mechanisms and charging of different users.
\end{itemize}

\subsubsection{Edge computing}
\label{sec:cloud_edge_computing_definition}
Edge computing is a paradigm which pushes the frontier of computing applications, data, and services away from central nodes, e.g., the data centers, to the periphery or edges of the network \cite{garcia2015edge}. Edge computing covers a wide range of technologies including cloudlet, remote/micro/community clouds, nano data centers, volunteer computing system, local cloud/fog computing, client-assisted cloud system, sensing networks, e.g., the wireless sensor network and crowdsensing network, and distributed Peer-to-Peer (P2P). Edge computing has the following major advantages \cite{ahmed2016survey}:
\begin{itemize}
\item It significantly reduces the data traffic, cost, and latency and improves QoS since cloud resources and services are located close to users.
\item It alleviates the major bottleneck and the risk of a potential point of failure since it does not rely on centralized computing.
\item It enhances security since data is encrypted as the data is moved towards the network edge.
\item It provides high levels of scalability, reliability, and automation.
\end{itemize}

\subsubsection{Software-Defined Networking (SDN)}
\label{sec:cloud_SDN_definition}
A traditional network architecture is composed of three planes of functionality, i.e., data, control and management planes. The control plane makes forwarding/routing decisions on the user traffic based on forwarding/routing tables. The data plane is responsible for forwarding the user traffic using the decisions from the control plane. The control and data planes are always coupled and embedded in the same networking devices, e.g., switches and routers, to guarantee network resilience. However, such architecture is rigid and complex to manage and control \cite{raghavan2012software}, \cite{pan2011survey}. The Software-Defined Networking (SDN) \cite{kreutz2015software}, \cite{zerrik2014towards}, \cite{costanzo2015software} is an emerging networking paradigm towards simple and flexible network management for network operators. SDN is defined as \textit{``a network architecture where the forwarding state in the data plane is managed by a remotely controlled plane decoupled from the former''} \cite{kreutz2015software}. In other words, SDN decouples the control plane from the network devices to become an external entity, the so-called \textit{SDN controller}. The SDN architecture has four major features below:
\begin{itemize}
\item The control and data planes are decoupled, and network devices just act as forwarding elements.
\item Forwarding decisions are flow-based instead of destination-based, meaning that all packets in the same flow receive identical service policies at the forwarding devices. This allows to unify different types of network devices, e.g., routers, switches, firewalls, load-balancers, and traffic shapers.
\item The control logic is moved to an external entity, i.e., the SDN controller or the Network Operating System (NOS).
\item The network is programmable through software applications running on the SDN controller which interacts with
the underlying data plane devices.
\end{itemize}

These features of SDN make the networks more programmable and easily partitionable and virtualizable. In practice, SDN has been used to address many issues in a wide range of network environments \cite{nunes2014survey}. For example, it was used to address the security and resource allocation in enterprise networks \cite{nayak2009resonance}, \cite{handigol2009plug}, flow control, virtual data center embedding, and resource utilization maximization in cloud networking \cite{bitar2013technologies}, \cite{jain2013b4}, \cite{rabbani2013tackling}, mobility management and load balancing in wireless access networks \cite{suresh2012towards}, wavelength path control and QoS-aware unified control in optimal networks \cite{liu2011openflow}, and network management in home and small business networks \cite{calvert2011instrumenting}. In particular for the cloud networking, using SDN makes a number of network devices become simple forwarding elements which are cheap and easy to deploy. This reduces both capital and operational expenditures for cloud and service providers. It is also expected to significantly improve benefits for all stakeholders, especially when SDN can be combined with the economic and pricing models which will be discussed in the next section.

\section{Overview and fundamentals of economic and pricing theory applied in cloud networking}
\label{sec:Intro_Price}
\begin{figure*}[ht]
 \centering
\includegraphics[width=\linewidth]{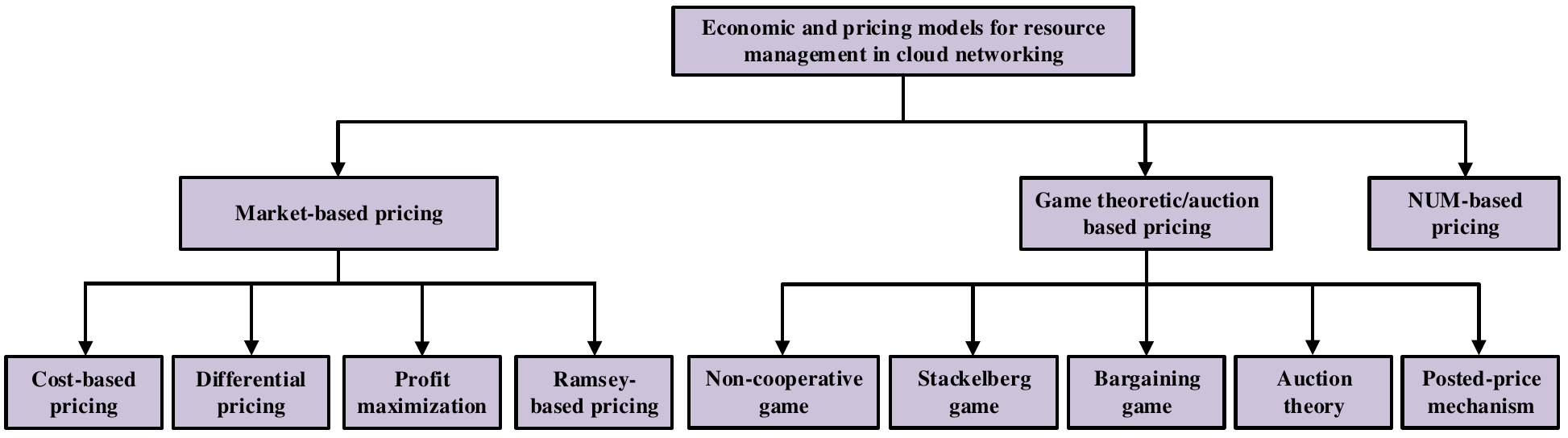}
 \caption{A taxonomy of economic and pricing models in cloud networking.}
 \label{taxonomy_pricing_model}
\end{figure*}

Economic and pricing approaches have been applied to address many issues in cloud networking due to the aforementioned benefits. In this section, we classify the economic and pricing approaches commonly used for resource management in cloud networking as shown in Fig.~\ref{taxonomy_pricing_model}. The classification is based on how the prices are set, i.e., market-based pricing, game theoretic and auction based pricing, and Network Utility Maximization (NUM) based pricing.

\subsection{Market-Based Pricing}
\label{subsec:Economic_concept}

In the following, we present the pricing models based on economic and financial concepts which have been applied in the cloud networking. We first present a simple pricing model, i.e., cost-based pricing, and then describe more complex pricing models including differential pricing, profit maximization pricing, and Ramsey pricing.

\subsubsection{Cost-based pricing}
\label{subsec:Cost_pricing}
Cost-based pricing is a common pricing strategy to determine the
price of a service based on calculating the total cost of the
service and adding a percentage of the cost as a desired profit. The
objective of using the cost-based pricing is to ensure that the
price makes the service provider profitable, or at least the price
covers the total cost of the service provider. The total cost
generally consists of a fixed cost and a variable cost. The fixed
cost is the cost that does not change when the number of sales of
the services changes. For example, hardware costs, e.g., servers and
network devices. On the contrary, the variable cost varies according
to the number of sales of the services produced. For example, the
resource costs, e.g., energy and bandwidth costs, the cost of data
transfer between different data centers, and the cost of cloud
server usage, are the variable cost for generating cloud
services. The advantage of the cost-based pricing is the ease of setting the price since the price is a function of the internal cost, i.e., the cost required to generate the service
\cite{Costas2003costeltit}. However, this pricing strategy does not
consider external market factors, e.g., the pricing strategies of
other providers and the perceive value and willingness to pay of
buyers.

In cloud networking, the cost-based pricing has been used by cloud
providers for evaluating the service cost in geo-diverse data
center networks \cite{greenberg2008cost}, \cite{agarwal2010volley}.
It has been also employed to analyze the cost saving when SDN and the
Network Function Virtualization (NFV) in the cloud are enabled \cite{zhang2015cost},
\cite{knoll2015life}. However, the internal cost information in
cloud networking may not be easy to obtain due to the variable cost
diversity. For example, the variable cost could depend on the geography of data centers. Further
details on the cost-based pricing model can be found in
\cite{pindyck2005microeconomics}, \cite{nagle2008strategy}.


\subsubsection{Differential pricing}
\label{sec:Differential_pricing}
The cost-based pricing ignores the requirements and preferences of
cloud users or tenants. To maximize the profit of providers, differential pricing, also called price discrimination, can be used. Consider a cloud resource market consisting of a cloud provider with
its cloud resources, i.e., the computing resource and network bandwidth. Using the differential pricing, the cloud provider may charge different prices to different cloud users based on their demand and willingness to pay. By setting higher prices for one type of user, the use of the differential pricing actually transfers the user surplus to the provider. Here, the user surplus is the difference between the total money that users are willing to pay and the total money that they actually pay. Thus, although this pricing guarantees a high revenue for the provider, it can be unfair to cause one type of user to pay a greater price than another type of user. In cloud networking, the differential pricing has been applied for bandwidth allocation among groups of users having different elasticities on cloud resources or among cloud users having different flexibility in resource usage as proposed in \cite{divakaran2014bandwidth}. In current cloud service markets, the differential pricing is used to set prices of the cloud services based on the requirements of the users. For example, the Alibaba group (https://intl.aliyun.com/) offers lower prices to users which require the cloud services for long term, e.g., 1 year.

\subsubsection{Profit maximization}
\label{subsec:Profit_maximization}
Profit maximization is the process of determining the output quantity and the corresponding price which yield the highest profit for a provider \cite{korrapati2014validated}. We present briefly how to find the optimal quantity and price based on the profit maximization in the following. Assume that a cloud provider needs to determine the number of cloud resource (i.e., the computing and network bandwidth) units, denoted by $Q$ and the corresponding price $P$ for their cloud users. The profit of the cloud provider is $\pi=R(P,Q)-C(Q)$, where $R(\cdot,\cdot)$ is the total revenue and $C(\cdot)$ is the total cost. The total cost may involve a fixed cost and a variable cost. The revenue is the amount of money that the cloud provider receives from selling $Q$ resource units to its users. The optimal quantity of cloud resource units, i.e., $Q^*$, is determined such that the profit is maximized, i.e., $Q^* = \max \limits_{Q} \pi$. The optimal quantity allows to find the optimal price based on the demand curve. The demand curve is typically a linear curve to show the relationship between the price of a resource unit and the quantity of resource units that users are willing to buy. A general demand curve can be expressed as $P=a-bQ$, where $a$ and $b$ are proper parameters. Thus, at $Q=Q^*$, the optimal price is $P^*=a-bQ^*$.

\begin{figure}[ht]
 \centering
\includegraphics[width=\linewidth]{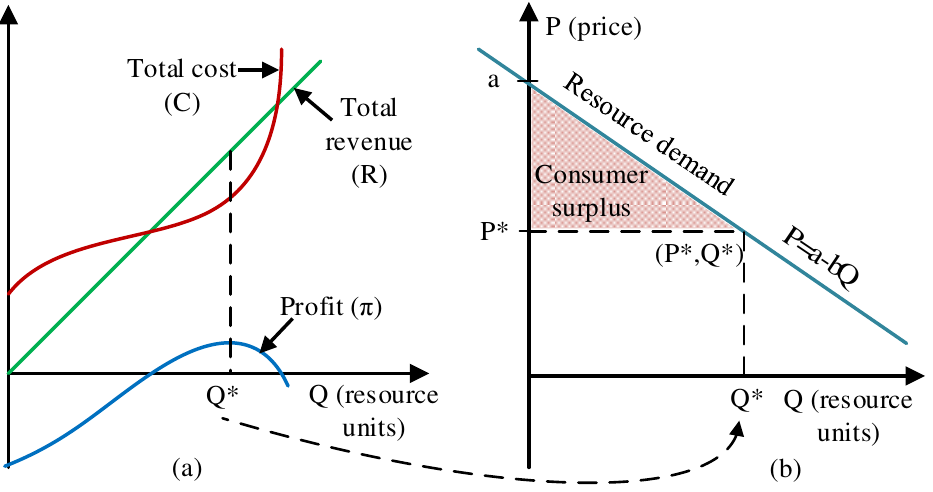}
 \caption{Pricing based on profit maximization.}
 \label{profit_maximization_theory}
\end{figure}

The optimal quantity and price can be determined using the graph as
shown in Fig.~\ref{profit_maximization_theory}(a) and
Fig.~\ref{profit_maximization_theory}(b).
Fig.~\ref{profit_maximization_theory}(a) shows the curves of the
total cost, total revenue, and profit. The optimal quantity $Q^*$ is
determined at the positive peak value of the profit curve. Then, the
optimal price $P^*$ is obtained from the demand curve in
Fig.~\ref{profit_maximization_theory}(b). However, in real markets,
it is not easy to determine the demand curve. The user demand can be
random or assumed to follow some distributions, e.g., the Gaussian
\cite{tsai2016virtualized}. Moreover, the profit maximization does
not consider the market competition in determining the quantity and
price. In cloud networking, the profit maximization has been adopted
to allocate computing and network resources to users
\cite{tsakalozos2011flexible} or to assign resource requests from
users to cloud providers \cite{rebaiimproving}.
\subsubsection{Ramsey pricing}
\label{sec:Ramsey_pricing}
In Ramsey pricing, different prices of the same commodity are applied to different markets depending on the demand elasticity of the commodity \cite{shepherd1992ramsey}. Ramsey pricing is
similar to the differential pricing. However, unlike the
differential pricing that maximizes the profit of the provider,
Ramsey pricing aims to maximize the social welfare of users subject to a
predefined threshold on the provider's profit. Specifically, assume
that a cloud provider provides cloud resource units in two
independent markets. The cloud provider determines different prices
$(p_1,p_2)$ in the two markets. In the independent market setting,
the demands of the resource corresponding to two prices are
$(q_1(p_1),q_2(p_2))$. The marginal cost of offering one cloud
resource unit in both markets is $c$, and the cloud provider has a
fixed cost. The objective of the cloud provider is to determine
$(p_1,p_2)$ to maximize social welfare subject to the constraint that the profit of
the cloud provider is not less than a threshold. Here, the social
welfare is the consumer surplus which is the area on the left of the
demand curve and above the price as shown in
Fig.~\ref{profit_maximization_theory}(b). The threshold is a fixed
profit value $\pi^*$ which is predefined by the cloud provider.
Therefore, the optimization problem of the cloud provider can be
formulated as follows:
\begin{align}
\label{Ramsey_pricing}
 \max \limits_{(p_1,p_2)} \sum_{i=1}^2 \Bigg{(} \int_{p_i}^{\infty} q_i(p) dp \Bigg{)} \\
\text{ s.t.} \sum_{i=1}^2 (p_i-c)q_i(p_i) \geq \pi^*.\notag \\
\notag
\end{align}

The problem in (\ref{Ramsey_pricing}) can be solved using the Lagrange multiplier method. The relationship between the optimal price and the demand elasticity in market $i$, $i \in \{1,2\}$, is expressed as follows:
\begin{equation}
\label{Ramsey_solution}
 \frac{p^*_i-c}{p^*_i}=\frac{1-\lambda^*}{\lambda^*}\frac{1}{\epsilon_i},
\end{equation}
where $\lambda$ is the Lagrange multiplier, and $\epsilon_i= \frac{dq_i(p_i^*)}{dp_i} \frac{p_i^*}{q_i}$ is the coefficient of price elasticity of demand for the resource in market $i$. The price elasticity of demand gives the percentage change in quantity demanded according to a fall or a rise in its price. We get the following relationship from (\ref{Ramsey_solution}):
\begin{equation}
\label{Ramsey_division2}
 \frac{p^*_1-c}{p^*_1}/ \frac{p^*_2-c}{p^*_2}=\frac{\epsilon_2}{\epsilon_1}.
\end{equation}

The expression in (\ref{Ramsey_division2}) shows that the price of a resource in a market should be relatively low when the demand elasticity in the market is high. On the contrary, in a market with inelastic or less elastic demand, the provider can set a high price because the demand for the resource does not change significantly according to a fall or a rise in the price. Clearly, the provider desires to have a market with inelastic demand since it can increase the price to earn more revenue. However, determining demand elasticity in markets is challenging. Moreover, the cloud provider cannot apply this pricing in the long run since users charged with a higher price will seek alternatives \cite{united1992paying}. One application of the Ramsey pricing in cloud networking is to regulate traffic flows of users among data centers \cite{wanisefficient}, which will be discussed in Section \ref{sec:cloud_data_center_bandwidth_other}.

\subsection{Game Theory and Auction Based Pricing}
\label{subsec:Game_theory}

Game theory and auctions are the study of multiparticipant decision making problems in which a choice of a participant, i.e., a player, potentially affects the interests of other participants \cite{gibbons1992primer}. In the context of cloud networking, participants can be cloud providers, service providers, cloud tenants, and users. In the following, we briefly present game theoretic models and auction mechanisms which have been widely used to determine resource prices in the cloud networking. First, some important terminologies are defined below~\cite{shi2012game}.

\begin{itemize}
\item{\textit{Player}:} A player is a participant which makes a decision in the game.
\item{\textit{Payoff}:} A payoff, i.e., a utility, a profit, or an interest, reflects the desired outcome of the player.
\item{\textit{Strategy}:} Player's strategy is a set of actions/instructions that the player can follow to achieve a desired outcome. The payoff depends on not only the player's own
action, but also the actions of others.
\item{\textit{Rationality}:} A player is rational if its strategy always aims at maximizing its own payoff.
\end{itemize}

\subsubsection{Non-cooperative game}
\label{subsec:Noncooperative_game}
In the non-cooperative game, each player maximizes only its own payoff neither being concerned about the payoff of the other players nor about the social welfare of the network \cite{alskaif2015game}. In this game, the players are selfish, and they do not form coalitions or make agreements with each other.

Consider a cloud resource market in which cloud providers as the sellers compete for selling resources to users. The sellers are typically selfish, and therefore the market can be modeled as a non-cooperative game among the sellers along with their pricing strategies. Assume that there are $N$ players, and $P_i$ is a set of pricing strategies of player $i$, where $P=P_1 \times \dots \times P_N$ is the Cartesian product of the individual strategy sets. Let $p_i \in P_i$ be the pricing strategy of player $i$. A vector of strategies of $N$ players is $\mathbf{p}=(p_1,\dots,p_N)$, and a vector of corresponding payoffs is $ \boldsymbol{\pi}=(\pi_1(\mathbf{p}),\dots,\pi_N(\mathbf{p})) \in R^N$, where $\pi_i(\mathbf{p})$ is the payoff of player $i$ given the player's chosen strategy and strategies of the others. Each player chooses its best strategy $p_i^*$ which maximizes its payoff. A set of strategies $\mathbf{p^*}=(p_1^*,\dots,p_N^*) \in P$ is the Nash equilibrium if no player can gain higher payoff by changing its own strategy when the strategies of the others remain the same \cite{friedman1971non}, i.e.,
\begin{equation}
\label{Nash_equilibrium}
\forall i, p_i \in P_i : \pi_i(p_i^*,\mathbf{ \overline{p}_i^*}) \geq \pi_i(p_i, \mathbf{\overline{p}_i^*}),
\end{equation}
where $\mathbf{\overline{p}_i}=(p_1,\dots,p_{i-1}, p_{i+1},\dots, p_N)$ is a vector of strategy choices of all players except player $i$.

The inequality in (\ref{Nash_equilibrium}) shows the stable state of the game in which the players have no incentive to change their own strategies since the payoffs will be worse off. However, in some cases, there is no Nash equilibrium at all, or there may exist multiple Nash equilibria which can make players not be clear about which one to choose. Therefore, checking the existence and uniqueness of the Nash equilibrium is important when setting prices based on the non-cooperative game.

The non-cooperative game theory has been widely used for the resource management in cloud networking. For example, it has been used to model the bandwidth sharing among peers in cloud-assisted P2P streaming systems \cite{chakareski2015cost} or among brokers in cloudlet systems \cite{guanvalue}. It has been adopted to maximize the profits of cloud providers as presented in \cite{pal2013economic}.

\subsubsection{Stackelberg game}
\label{subsec:Stackelberg_game}
The non-cooperative game discussed above assumes that players announce their pricing strategies simultaneously, and the players know each other's strategies at the same time. However, this may not always hold in real markets. Therefore, sequential games can be used in which players can announce their strategies following a certain predefined order. This is the Stackelberg game \cite{amir1999stackelberg}. In the Stackelberg game, the player decides its own strategic choice after observing the strategies of other players \cite{kim2014game}. It was proved that even if the players have to choose their strategies first, their payoffs are not less than those at the Nash equilibrium \cite{han2012game}, i.e., due to the first-mover advantage. The following provides the definition and properties of the Stackelberg game.

Assume that there are two cloud resource sellers 1 and 2 in the market. $P_1$ and $P_2$ are the sets of pricing strategies of sellers 1 and 2, respectively.
Seller 1 chooses its pricing strategy $p_1$ from set $P_1$ to maximize its payoff or profit function $\pi_1(p_1,p_2)$, and seller 2 chooses its pricing strategy $p_2$ from set $P_2$ to maximize its payoff function $\pi_2(p_1,p_2)$. Without loss of generality, assume that seller 2 selects its strategy before seller 1 decides its selection. Seller 2 is namely the leader, and seller 1 is called the follower. We have the following definition \cite{leitmann2013multicriteria}:

\textbf{Definition 1.} If there exists a mapping $F: P_2 \rightarrow P_1 $ such that, for any fixed $p_2 \in P_2 $, $\pi_1(Fp_2,p_2) \geq \pi_1(p_1,p_2) $, $\forall p_1 \in P_1 $, and if there exists $p_{2s2} \in P_2 $ such that $\pi_2(Fp_{2s2},p_{2s2}) \geq \pi_2(Fp_2,p_2) $, then the pair $(p_{1s2}, p_{2s2}) \in P_1 \times P_2$, where $p_{1s2}= Fp_{2s2} $, is called a Stackelberg strategy pair.

Definition 1 means that the Stackelberg strategy is optimal for the leader when the follower responds to the leader with the follower's optimal strategy. Let $D_1= \{(p_1,p_2) \in P_1 \times P_2: p_1 = Fp_2\}$ denote \textit{the rational reaction set} of seller 1 when seller 2 chooses strategy $ p_2 \in P_2$. Seller 1 is referred to as a rational player. Similarly, when seller 1 is the leader, let $D_2$ denote the rational reaction set of seller 2. The sets $D_1$ and $D_2$ have significant importance which is indicated in the following two propositions.

\textbf{Proposition 1.} A strategy pair $(p_{1s2}, p_{2s2})$ is the Stackelberg strategy with seller 2 as the leader iff $(p_{1s2}, p_{2s2}) \in D_1 $ and
\begin{equation}
\label{Stackelberg_Nash}
 \pi_2(p_{1s2}, p_{2s2}) \geq \pi_2(p_1,p_2) , \forall (p_1,p_2)\in D_1.
\end{equation}

\textbf{Proposition 2.} A strategy pair $(p_{1N}, p_{2N})$ is the Nash strategy pair iff $(p_{1N}, p_{2N}) \in D_1 \cap D_2 $.

The expression in (\ref{Stackelberg_Nash}) and Proposition 2 show that $\pi_2(p_{1s2}, p_{2s2}) \geq \pi_2(p_{1N},p_{2N})$. In other words, for the leader, the Stackelberg strategy guarantees to achieve the payoff at least as good as the corresponding Nash equilibrium. This is because when choosing the Stackelberg strategy, the leader actually imposes a solution which will be favorable to itself.

In cloud networking, the Stackelberg game has been applied for allocating the cloud provider's bandwidth to virtual networks \cite{yuan2012game} and for reducing the access of users to servers in the cloud \cite{lin2015autotune}. The Stackelberg game has been also applied in cloud computing. For example, it was used to maximize revenue of the cloud provider while maximizing server clients' utilities \cite{al2009brief} or to maximize revenue of the cloud provider while guaranteeing QoS for its end-users \cite{di2013optimal}. Besides, the Stackelberg game has been used in Internet of Things (IoT). For example, it was adopted to maximize the profits of different participants of IoT industry value chain \cite{lv2012competition} and to improve the QoS and the network's robustness in sensing networks \cite{danak2006inner}.

\subsubsection{Bargaining game}
\label{subsec:Bargaining_game}
In the bargaining game or Nash bargaining game, two or more players must reach an agreement regarding how to distribute a monetary amount. Consider trading bandwidth in cloud networking between a cloud provider, i.e., a seller, and a cloud tenant, i.e., a buyer. A successful bargain is reached if and only if the bandwidth is allocated at a mutually acceptable price. Let $p^0_s$ be the smallest price that the seller can accept for selling the bandwidth and $p^0_b$ be the buyer's greatest price that the buyer is willing to pay for the bandwidth. The pair $(p^0_s, p^0_b)$ is called the disagreement point or threat point that the seller and the buyer expect to receive if their negotiations fail to reach a settlement \cite{muthoo1999bargaining}.

The strategy of the seller is to offer the selling price $p_s^*$ to maximize its expected profit $\pi_s(p_s,p^0_s)$, i.e., $\pi_s (p_s^*,p^0_s) \geq \pi_s (p_s,p^0_s) $, $\forall p_s$. Similarly, the strategy of the buyer is to offer the buying price $p_b^*$ to maximize its profit $\pi_b (p_b,p^0_b)$, i.e., $\pi_b (p_b^*,p^0_b) \geq \pi_b (p_b,p^0_b) $, $\forall p_b$. If $p_b^* \geq p_s^*$, a bargain is enacted and the transaction price for trading the bandwidth can be set by \cite{chatterjee1983bargaining}, $p^* = kp_b^* + (1-k) p_s^*$, with $0 \leq k \leq 1$. When $k=1/2$, the transaction price is determined by splitting the difference between the buyer's and seller's offers. A pair of the best response offer strategies $(p_s^*, p_b^*)$ constitutes the Nash bargaining solution. At this agreement point, the seller earns $(p^*-p^0_s)$, and the buyer earns $(p^0_b-p^*)$.

Some other scenarios in cloud networking where the bargaining game has been applied are allocating requests of users to data centers \cite{xu2012general} and sharing cloud resources among service providers \cite{ding2015service}. In cloud computing, the bargaining game has been used for negotiating the price among the cloud resource brokers and grid service providers as proposed in \cite{samimi2011review} and for pricing and allocating virtual resource instances for independent tasks and workflow tasks as presented in \cite{iyer2011resource}.

\subsubsection{Auction}
\label{subsec:Auction_theory}
An auction is the economic mechanism the goals of which are to allocate commodities and establish corresponding prices via a process known as bidding \cite{mcafee1987auctions}. There are some common terminologies used in the auction as follows:

\begin{itemize}
\item{\textit{Bidder}:} A bidder is a buyer which wants to purchase resources. In cloud networking, bidders can be end-users or cloud tenants.
\item{\textit{Seller}:} A seller, e.g., a cloud provider, offers its resources and services for sale.
\item{\textit{Auctioneer}:} An auctioneer acts as an intermediate agent to conduct an auction, determine, and announce the winner. In many cases, an auctioneer is a seller itself.
\item{\textit{Price}:} A price in an auction may be a bidding price or an asking price. The bidding price is the price that the bidder is willing to pay for a requested resource, and the asking price is the price of a resource that the seller is willing to offer.
\end{itemize}

There exist several studies on auctions as well as their applications. There are a survey of the auction theory \cite{klemperer2004auctions}, a survey of auction on Internet \cite{chui1999auction}, or a survey of auction approaches for resource management in wireless networks \cite{zhang2013auction}. In what follows, we discuss typical types of auctions which have been commonly applied to resource management in cloud networking.
\begin{figure*}[ht]
 \centering
\includegraphics[width=\linewidth]{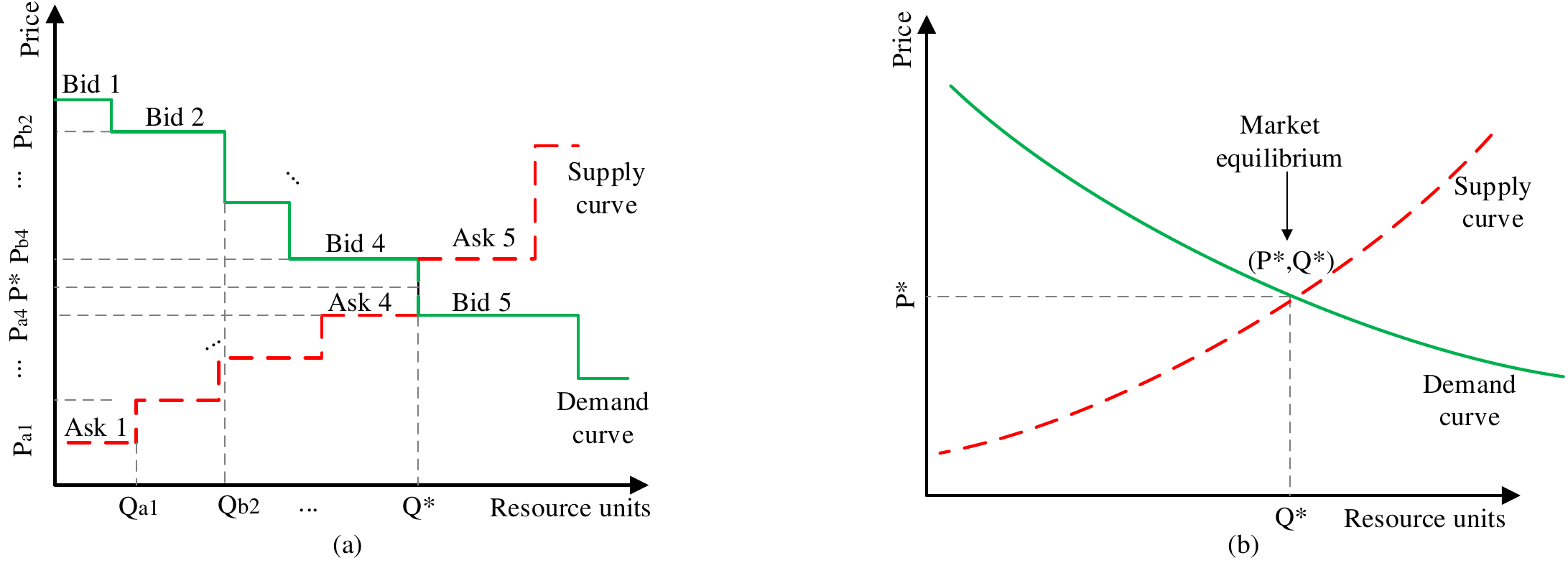}
 \caption{(a) Discrete supply and demand curves of double auction, and (b) continuous supply and demand curves from economics.}
 \label{Double_auction_resourc}
\end{figure*}

\textit{(a) Conventional auctions:}
A conventional auction is known as the \textit{open-outcry} auction. In the open-outcry auction, bids of buyers are disclosed to each other during the auction. There are two types of the conventional auction~\cite{vijay2002auction}.
\begin{itemize}
\item{\textit{English auction:}} The English auction is an ascending-bid auction, meaning that the bidding price submitted by buyers increases monotonically. Specifically, buyers submit their bidding prices for the resource sequentially or simultaneously to the auctioneer. The auction will terminate if there is no new higher price submitted. The buyer with the highest price wins the resource and pays the price $p^*$, i.e., a hammer price, which satisfies $p_s^0 \leq p^* \leq \max \limits_{i} B_i$, where $p_s^0$ is the lowest price that the seller can accept to sell, and $B_i$ is buyer $i$'s budget. Generally, $p^*$ changes depending on the number of buyers in the auction and may not equal $p_s^0$.
\item{\textit{Dutch auction:}} Contrary to the English auction, the Dutch auction is a descending-bid auction in which the auctioneer or seller initially sets a high asking price for the resource and then decreases the price until one of the buyers accepts the price. The winning buyer pays the final price and receives the resource. This simple allocation enables the Dutch auction to spend less time than the English auction \cite{rodriguez2007simple}. 
\end{itemize}
\textit{(b) Vickrey and Vickrey-Clarke-Groves (VCG) auctions:}
Vickrey and VCG auctions are the sealed-bid auctions in which buyers submit simultaneously their sealed bids to the auctioneer. Different from the open-outcry auctions, in the sealed-bid auction, buyers do not know bidding strategies of each other and cannot change their own bids during the auction.
\begin{itemize}
\item{\textit{Vickrey auction:}} A Vickrey auction, also known as the second-price sealed-bid auction, is one of the two most important \textit{k-th}-price sealed-bid auctions. In the Vickrey auction, the winning buyer pays the second-highest price rather than the price that it submitted \cite{lucking2000vickrey}, i.e., $p^*= \max \limits_{p\in P \backslash\{p_i\}}p$, where $p_i$ is the highest price of the winner. In other words, the winner pays a price less than its expected price \cite{sandholm1996limitations}. Therefore, the Vickrey auction motivates buyers to bid truthfully, and such an auction achieves strategy-proofness, or incentive compatibility, or truthfulness. The truthfulness is an important property because an auction which does not hold this property may be vulnerable to market manipulation and produce very poor outcomes \cite{klemperer2002really}.

\item{\textit{Vickrey-Clarke-Groves (VCG) auction:}} The VCG auction is a generalization of the Vickrey auction with multiple commodities \cite{ausubel2006lovely}. The VCG auction allocates commodities in a socially optimal manner and charges the winner with the loss of the social value due to its getting the commodity.
\begin{itemize}
\item Assume that there is a set $T$ of $M$ commodities for sale $T= \{t_1, t_2,\dots, t_M\} $, where $t_i$ is the $i$th commodity, and a set of $N$ buyers, i.e., bidders, $B= \{1, 2,\dots, N\} $.
\item Let $b_i(t_j)$ denote a bid of bidder $i$ for commodity $t_j$ and $V_N^M$ denote the social welfare value, i.e., the social cost, created by $M$ commodities.
\end{itemize}

Similar to the Vickrey auction, if $b_i(t_j)$ is the highest, bidder $i$ wins to obtain commodity $t_j$. According to the VCG auction rule, bidder $i$ pays the price which is equal to
\begin{equation}
V_{N\backslash\{i\}}^M - V_{N\backslash\{i\}}^{M\backslash\{t_j\}},
\label{VCG}
\end{equation}
where $V_{N\backslash\{i\}}^M$ represents the social welfare value if bidder $i$ does not participate in the auction, and $V_{N\backslash\{i\}}^{M\backslash\{t_j\}} $ is the attainable social welfare value after bidder $i$ wins commodity $t_j$. The expression in (\ref{VCG}) indicates that winner $i$ needs to pay the price which is the loss in attainable welfare suffered by the remaining bidders since it got the commodity $t_j$.

\end{itemize}

\textit{(c) Forward, reverse, and double auctions:}
\label{subsubsec:Auction_reverse}
Considering the sides of sellers and buyers, we can classify auctions as follows:
\begin{itemize}
\item{\textit{Forward and reverse auctions:}} In the forward auction, there are many potential buyers and one seller. On the contrary, in the reserve auction, there are one buyer and many potential sellers.
\item{\textit{Double auction:}} In the double auction, buyers and sellers simultaneously submit their bids and asks to an auctioneer, respectively \cite{friedman1993double}. The basic idea of the double auction is to match asks from sellers and bids from buyers by assigning commodities from the sellers to the buyers and payments from the buyers to the sellers accordingly. To further understand the double auction process, we consider a cloud resource market in which the sellers are cloud providers and the buyers are cloud tenants/users. Upon receiving bids and asks, the auctioneer sorts the buyers' bids in a non-ascending order and the sellers' asks in a non-descending order. The auctioneer finds the largest index $k$ at which the asking price is less than the bidding price, i.e., $p^a_k \leq p^b_k$. The transaction price $p^*$, i.e., a hammer price or clearing price, can be determined as $p^*=(p^a_k + p^b_k)/2$. The buyer receives resources, and the seller gets payment $p^*$. The process is repeated to match the remaining buyers and sellers as well as to determine corresponding clearing prices.


In fact, the double auction has a very similar concept to the supply and demand model \cite{mullen1996market}. As shown in Fig.~\ref{Double_auction_resourc}(a), the asks from sellers and the bids from buyers form the supply and demand curves, respectively. The x-axis represents the supplied resource units, and y-axis represents the asking or the bidding prices. For example, seller $1$ sells $Q_{a1}$ units of resources at price $P_{a1}$ (i.e., ``Ask 1'' in Fig.~\ref{Double_auction_resourc}(a)), and buyer 2 bids to buy $Q_{b2}$ units of resources at price $P_{b2}$ (i.e., ``Bid 2'' in Fig.~\ref{Double_auction_resourc}(a)), and so on. Fig.~\ref{Double_auction_resourc}(a) is actually a discretized form of the standard supply and demand model which is shown in Fig.~\ref{Double_auction_resourc}(b). The supply and demand curves intersect at a point which is called the supply-demand equilibrium~\cite{krishna2002auction}, i.e., the market equilibrium point $(P^*,Q^*)$. The clearing price $P^*$ at the equilibrium can be determined by $(P_{a4}+ P_{b4})/2$ (as shown in Fig.~\ref{Double_auction_resourc}(a))).

The double auction can hold important properties: individual rationality (i.e., no participant loses when joining the auction), balanced budget (i.e., the auctioneer gains money), truthfulness (i.e., buyers and sellers submit truthfully their bids and asks), and economic efficiency (i.e., the social welfare is the best possible).

\end{itemize}

\textit{(d) Combinatorial auction:}
 In the combinatorial auction, each bid of a buyer indicates a combination or package of discrete commodities rather than an individual commodity \cite{cramton2006combinatorial}. Given a set of bids, the auctioneer finds an optimal allocation of the commodities to the buyer, i.e., the winner. The combinatorial auction has more advantages compared with standard auctions, e.g., the sealed-bid auction. These advantages include little global information requirement, economic efficiency, utility maximization for buyers, and revenue maximization for sellers. However, a big challenge in the combinatorial auction is the winner determination problem. This problem is NP-hard, and there does not exist a polynomial-time algorithm to find the optimal allocation. However, many algorithms have been proposed to find approximate solutions for the problem, e.g., the Lagrangian relaxation approach \cite{hsieh2010combinatorial}.

\textit{(e) Shapley value:}
The aforementioned auction schemes only consider how to determine the winner and payment. In practice, the winner can be a group of users, which is called a \textit{virtual winner} \cite{pan2009fair}. How to fairly allocate resources and share the payment among users in the virtual winner is still an important problem. The Shapley value method can be combined with an auction to solve the problem. This section provides the definition, properties, and applications of the Shapley value.

The Shapley value is a concept in the cooperative game theory which provides a unique and fair allocation/distribution of the total surplus/profit generated by the coalition among players \cite{roth1988shapley}. Specifically, in a coalition, players may contribute differently to obtain an overall surplus/profit. The Shapley value provides a method to measure the importance of each player in the cooperation and to assign the distribution of generated surplus among the players. Formally, assume that there is a coalitional game $(v,\mathbb{N})$, where $\mathbb{N}$ denotes the coalition of $|\mathbb{N}|$ players, and $v$ or $v(\mathbb{N})$ describes the total expected surplus obtained from the cooperation of the $|\mathbb{N}|$ players. The amount of surplus that player $i$ in the coalitional game receives is defined by
\begin{equation}
\label{Shapley_value}
\phi_i(v)=\sum_{\mathbb{S}\subseteq \mathbb{N}\backslash {i}} \frac{|\mathbb{S}|!(|\mathbb{N}|-|\mathbb{S}|-1)}{|\mathbb{N}|!} (v(\mathbb{S}\cup{i})-v(\mathbb{S})),
\end{equation}
where $v(\mathbb{S})$ is the total surplus obtained from the cooperation of $|\mathbb{S}|$ players in a coalition $\mathbb{S}$. The expression in (\ref{Shapley_value}) means that to calculate the Shapley value of player $i$ in the coalition game $(\mathbb{N},v)$, we need to determine the marginal contribution of the player in a potential coalition $(\mathbb{S},v)$ as $v(\mathbb{S}\cup{i})-v(\mathbb{S})$ that is the difference between the total surplus of the game with player $i$ and the game without player $i$. We then take the average of this contribution over all possible different permutations in which the coalition can be formed. The Shapley value satisfies the following axioms \cite{kaewpuang2013framework}, \cite{shapley2016value}:
\begin{itemize}
\item \textit{Joint efficiency:} The total surplus of all players equals that of the coalition, i.e., $\sum_{i \in \mathbb{N}} \phi_i(v)=v(\mathbb{N})$.
\item \textit{Zero payoff to the dummy:} A dummy player is the one which does not contribute anything to the value of any coalition that it joins, i.e., $v(\mathbb{S}\cup{i})=v(\mathbb{S})$, for all $\mathbb{S}$. The payoff to the dummy player is zero.
\item \textit{Symmetry:} If two players have the same contribution, i.e., $v(\mathbb{S}\cup{i})=v(\mathbb{S}\cup{j})$, they receive the same payoff.
\item \textit{Additivity:} The payoff of any player is equal to the sum of all the payoffs that the player will receive as a member of all possible coalitions.
\end{itemize}

It was shown in \cite{winter2002shapley}, \cite{van2013reconciling} that the Shapley value given in (\ref{Shapley_value}) is unique satisfying the four above axioms. This is desirable from the perspective of cooperative providers. In cloud networking, the Shapley value has been used for sharing profit among cloud providers from their resource pooling. Moreover, it has been also adopted for sharing the cost among cloud users from using bandwidth \cite{niyato2011cooperative} and for the fair profit sharing among Internet service providers \cite{ma2010internet}.

The summary of the above auctions along with their applications in cloud networking is given in Table \ref{table_auction_sum}. As seen, auction mechanisms have been used for resource management in cloud networking. Moreover, the Vickrey auction, i.e., the second-price sealed-bid auction, has been more frequently used compared with the other auction mechanisms due to its privacy and truthfulness guarantee \cite{suzuki2001efficient}.

\begin{table*}[h]
\caption{A summary of key features and suitable scenarios of auctions used in cloud networking.}
\label{table_auction_sum}
\scriptsize

\begin{centering}
\begin{tabular}{|>{\centering\arraybackslash}m{2cm}|>{\centering\arraybackslash}m{2.5cm}|>{\centering\arraybackslash}m{5cm}|>{\centering\arraybackslash}m{5cm}|>{\centering\arraybackslash}m{1.5cm}|}
\hline
\cellcolor{mygray} &\cellcolor{mygray} &\cellcolor{mygray} &\cellcolor{mygray} &\cellcolor{mygray} \tabularnewline
\cellcolor{mygray} \multirow{-2 }{*}{\textbf{Auction type}} &\cellcolor{mygray} \multirow{-2}{*} {\textbf{Market structure}} &\cellcolor{mygray} \multirow{-2}{*} {\textbf{Key descriptions}} &\cellcolor{mygray} \multirow{-2}{*} {\textbf{Suitable scenarios}} &\cellcolor{mygray} \multirow{-2}{*}{\textbf{Solution}} \tabularnewline
\hline
\hline
 English auction \cite{vijay2002auction} & A seller, multiple buyers, and an auctioneer & Open-outcry ascending-price auction, and winning buyer pays the second highest price &\begin{itemize} \item {Economics: seller's revenue maximization} \item{Cloud networking: bandwidth allocation} \end{itemize} &Nash equilibrium\tabularnewline \cline{2-5}
\hline
Dutch auction \cite{vijay2002auction} & A seller, multiple buyers, and an auctioneer & Open-outcry descending price auction, and winning buyer pays the final price &\begin{itemize} \item {Economics: the best price guarantee for buyer} \item{Cloud networking: bandwidth allocation} \end{itemize} &Nash equilibrium\tabularnewline \cline{2-5}
\hline
 Vickrey auction/second-price sealed-bid auction \cite{lucking2000vickrey}& A seller, multiple buyers, and an auctioneer & Sealed-bid auction, and winning buyer pays the second highest price &\begin{itemize} \item {Economics: buyer's expected utility maximization} \item{Cloud networking: resource reservation, task allocation, and storage sharing} \end{itemize} &Nash equilibrium\tabularnewline \cline{2-5}
\hline
VCG \cite{ausubel2006lovely}& A seller, multiple buyers, and an auctioneer & A generation of the Vickrey auction for multiple commodities, winning buyer pays a price equal to the loss of the social value due to its getting commodities &\begin{itemize} \item {Economics: social welfare maximization} \item{Cloud networking: bandwidth allocation} \end{itemize} &Bayesian Nash
equilibrium\tabularnewline \cline{2-5}
\hline
 Double auction\cite{friedman1993double}& Multiple sellers, multiple buyers, and an auctioneer & Buyers and sellers submit respectively their bids and asks, and the auctioneer matches asks and bids &\begin{itemize} \item {Economics: ordinary markets with multiple sellers and buyers which need to be cleared} \item{Cloud networking: bandwidth reservation, resource sharing, and task allocation} \end{itemize} &Market
equilibrium\tabularnewline \cline{2-5}
\hline
Combinatorial auction \cite{cramton2006combinatorial}& A seller, multiple buyers, and an auctioneer &Buyers bid on combinations/packages of commodities, and winner determination problem is to find the optimal allocation of commodities. &\begin{itemize} \item {Economics: markets where buyers compete on many different but related commodities} \item{Cloud networking: bandwidth allocation} \end{itemize} &Optimal solution\tabularnewline \cline{2-5}
\hline
\end{tabular}
\par\end{centering}
\end{table*}

\subsubsection{Posted-price mechanism}
\label{subsec:Posted_price}
The posted-price mechanism is typically used in online procurement markets, e.g., a digital market, in which sellers arrive in a sequential order. The posted-price mechanism assigns a specific price to each seller when the seller arrives. The seller can ``take or leave'' this price. Typically, the seller accepts the price if its actual cost is less than the price offered by the mechanism. The seller then gives its buyers ``take-it-or-leave-it'' offer price, meaning that a buyer can accept or reject the offer \cite{hartline2001dynamic}. Based on the buyer's responses, the mechanism will set prices for subsequent sellers' commodities. Since the posted-price mechanism may set different prices for the same commodity, it is similar to the differential pricing mechanism, i.e., the price discrimination, as presented in Section \ref{sec:Differential_pricing}. Thus, the term ``posted-price discrimination'' sometimes appears in the literature \cite{wu2012consumer}. Besides, making a comparison with the auction mechanisms, the posted-price mechanism is less complex since it decides immediately an offer price without soliciting an ask from each seller upon the seller's arrival \cite{badanidiyuru2012learning}. In the context of cloud networking, the posted-price mechanism has been used when resource sellers, e.g., users in the social cloud, arrive in a sequential order to offer their storage services. In commercial clouds, e.g., Amazon's \textit{elastic cloud compute (EC2)}, the cloud provider uses this mechanism to post a certain price on the ``take-it-or-leave-it'' basis \cite{nallur2013decentralized}. Extended to IoT, this mechanism has been also used for the data aggregation in crowdsensing networks in which the data sellers are phone users \cite{sun2014collection}, \cite{sun2013behavior}.

\subsection{Network Utility Maximization (NUM)-based pricing}
\label{subsec:Utility_maximization}
In this section, we explain the Network Utility Maximization (NUM), known as a dual-based distributed algorithm for the resource allocation. NUM is essentially the problem of maximizing the total utility of users in a network, given the capacity constraint of the network \cite{mas1995microeconomic}. The original NUM problem only considers utility functions of users. In the context of cloud networking, when users utilize resources from a cloud provider, they incur a total cost to the cloud provider. Therefore, the modified NUM problem which takes into account the total cost should be investigated. Consider the cloud resource reservation scenario in which $N$ cloud tenants, e.g., video content providers, reserve network bandwidth from a cloud provider to deliver their videos to end-users. The goal of the cloud provider is to maximize the social welfare. Thus, the resource allocation problem can be expressed as
\begin{align}
\label{NUM_SYS}
\max \limits_{\textbf{x}} \sum_{i}^{N}U_i(x_i) -C(\textbf{x})\\ \notag
\text{    s.t. } x_i \in [a_i,b_i], i=1,\dots, N,
 \notag
\end{align}
where $x_i$ is the resource units that cloud tenant $i$ receives, $\textbf{x}=(x_1,\dots,x_N)$ is the vector of resource allocation, $U_i(x_i)$ is the monotonically increasing utility function associated with tenant $i$ which is a strictly concave function of its resource allocation, $C(\textbf{x})$ is a strictly convex function, and $a_i$ and $b_i$ are the constants.

Since the cloud provider has no knowledge of the utility functions, and the cost function is unknown to the cloud tenants, centralized methods such as interior point methods \cite{bertsekas2003convex} which are typically applied to solve the NUM problems may not be applicable to the specific problem in (\ref{NUM_SYS}). Alternatively, pricing-based iterative solutions are often used. Assume that the cloud provider charges cloud tenant $i$ a price $p_i$ for using resource $x_i$. Given a price vector $\textbf{p}=(p_1,\dots,p_N)$, the problem in (\ref{NUM_SYS}) is equivalent to
\begin{equation}
\label{NUM_SYS_PRICE}
\max \limits_{\textbf{x}\in \Pi_i[a_i,b_i]} \sum_{i}^{N}(U_i(x_i)-p_ix_i) + (\textbf{p}^\top \textbf{x}-C(\textbf{x})),
\end{equation}
where $(U_i(x_i)-p_ix_i)$ is the surplus of cloud tenant $i$ for using $x_i$ resource units, and $(\textbf{p}^\top \textbf{x}-C(\textbf{x}))$ is the profit of the cloud provider. With the price vector $\textbf{p}$, each cloud tenant selects $x_i$ resource units to maximize its surplus as follows:
\begin{equation}
\label{NUM_SYS_PRICE_User}
x_i(p_i)= \arg \max \limits_{x_i\in [a_i,b_i]} (U_i(x_i)-p_ix_i) , i=1,\dots, N.
\end{equation}
Given the returned resource request $x_i$ and applying the dual decomposition and gradient methods, the cloud provider updates iteratively the price vector $\textbf{p}$ according to the rule: $p_i=p_i-\gamma(y_i(\textbf{p})-x_i(p_i))$, where $\gamma$ is an appropriate step size, and $y_i(\textbf{p})$ is taken from $\textbf{y(\textbf{p})}=(y_1(\textbf{p}),\dots,y_N(\textbf{p}))= \arg \max \limits_{x_i\in [a_i,b_i]}\textbf{p}^\top \textbf{x}-C(\textbf{x}) $. The process is repeated until the vector of resource allocation $\textbf{x}$ converges to the optimal resource allocation $\textbf{x}^*$. Since (\ref{NUM_SYS}) is a convex optimization problem, the solution $\textbf{x}^*$ is unique.

In practice, the above optimal solution is feasible only when utility functions of cloud tenants are concave. However, in delay-sensitive services, e.g., video and voice services, such utility functions vary with different types of services with inelastic flows. Therefore, the resource allocation may be a non-convex optimization problem \cite{lee2005non}.

\section{Applications of economic and pricing models for resource management in cloud data center networking}
\label{sec:cloud_data_center}
Cloud networking provides local network connections to servers \cite{abts2012guided} (Fig.~\ref{cloud_networking_architecture}) and remote links for data center to create a resource pool supporting a large number of users and applications with diverse resource demands and utilities. Therefore, resource management becomes one of the most important issues in cloud networking. This section reviews applications of economic and pricing models for the resource management in cloud data center networking. The major issues include:

\begin{itemize}
\item{\textit{Bandwidth allocation}: Bandwidth allocation involves reserving and allocating bandwidth to users or application service providers. Traditional bandwidth allocation algorithms often assume that the available resources do not change. However, in cloud networking, resources and demands can fluctuate randomly. Economic and pricing models have been used as the solutions in which all scarce resources can be best utilized with the variability of budget.}
\item{\textit{Request allocation}: Request allocation is to assign massive users' requests, e.g., the transaction and data processing, to data centers. Economic and pricing models provide efficient approaches to achieve the load balancing, latency and cost minimization taking network resource availability into account.}
\item{\textit{Workflow allocation}: The workflow allocation in the cloud data center networking is the computing task allocation among the data centers. Market-based approaches provide an efficient task assignment with the lowest cost and the fastest completion time through using negotiation mechanisms among network entities.}
\end{itemize}


\subsection{Bandwidth Allocation}
\label{sec:cloud_data_center_bandwidth}
As part of bandwidth allocation, the bandwidth reservation can be implemented to guarantee the bandwidth availability for users in future. This allows users to pay a lower price due to advance reservation. However, the bandwidth reservation may cause oversubscription or undersubsciption issues \cite{chasejoint}. This section reviews the applications of economic and pricing models for bandwidth reservation and allocation in cloud data center networking. Note that the bandwidth reservation is often implemented in advance compared with the bandwidth allocation which is done in an online basis.

\subsubsection{VCG auction}
\label{sec:cloud_data_center_bandwidth_VCG}
\begin{figure}[ht]
 \centering
\includegraphics[width=7.2cm, height=6cm]{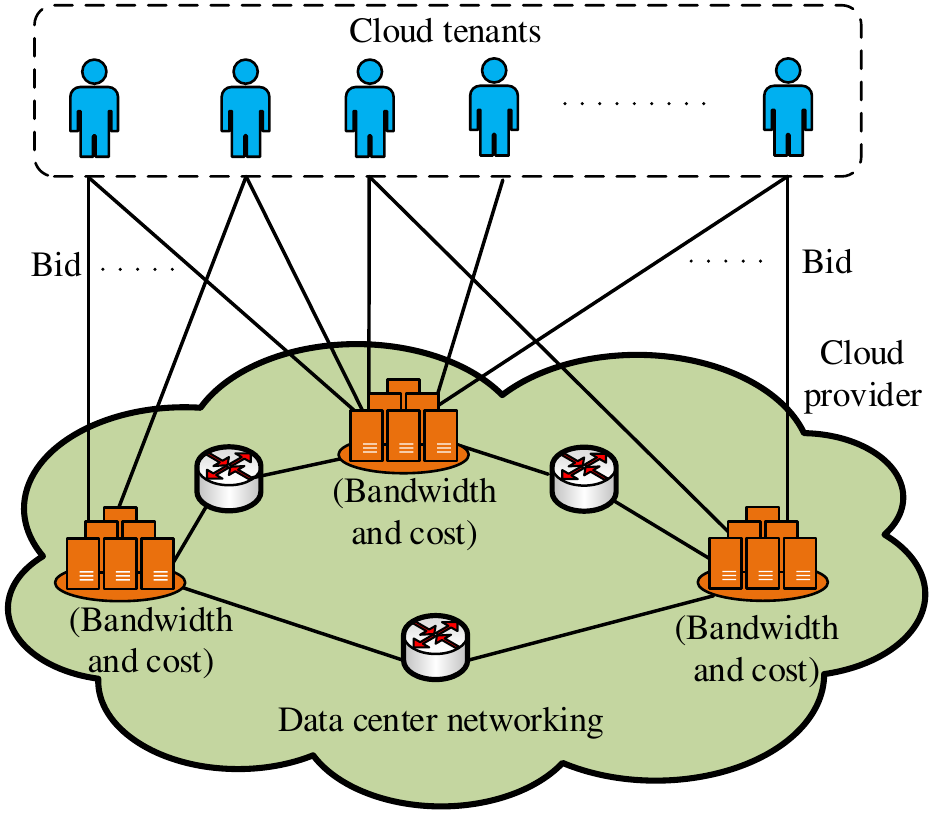}
 \caption{A market based cloud bandwidth reservation in cloud data center networks.}
 \label{cloud_data_center_bandwidth_auction}
\end{figure}

A typical bandwidth reservation model in cloud networking is shown in Fig. \ref{cloud_data_center_bandwidth_auction}. The model consists of a cloud provider, i.e., a seller, which owns a number of distributed data centers, and cloud tenants, i.e., buyers, which act as application and service providers. Cloud tenants rent bandwidth from the cloud provider to serve their subscribers. To avoid the high bandwidth reservation payment, the cloud tenants can lie about their revenues obtained by serving subscribers. Therefore, the authors in \cite{gui2014soar} adopted the VCG auction for the bandwidth reservation to achieve both optimal social welfare and strategy-proofness \cite{osborne1994course} that the cloud tenants have no incentive to lie about their revenue information. The cloud tenants simultaneously submit their bids to compete for bandwidth to the cloud provider. Each bid consists of bandwidth demands and the price per unit of bandwidth for which the buyer is willing to pay. To achieve the highest social welfare for the allocation, the winners are determined through a linear programming model which can be solved in polynomial time. The VCG mechanism was then applied to calculate the charge for each winner. The charge is the difference between the social welfare when the winner does not participate and when the winner participates in the auction. Since the proposed approach has an optimal allocation and calculates the charge based on the VCG auction, it was concluded to be a strategy-proof and optimal auction mechanism for cloud bandwidth reservation. The simulation results showed that both the social welfare and the bandwidth satisfaction ratio of the proposed approach exceed 90\% of the optimal solution when there are 200 tenants with 15 data centers.

\subsubsection{Shapley value based auction}
\label{sec:cloud_data_center_bandwidth_shapley}
The VCG mechanism mentioned in \cite{gui2014soar} can simultaneously guarantee the truthfulness and the optimal economic efficiency. However, when the underlying allocation problem is NP-hard, e.g., the traffic scheduling, the VCG auction becomes computationally intractable. Approximation algorithms can be used, but they cause the VCG auction to lose its truthfulness \cite{mu2008truthful}. To guarantee the truthfulness, approximation algorithms can be combined with suitable rules, e.g., by exploiting critical bids \cite{lehmann2002truth} or resorting to the linear programming decomposition technique \cite{lavi2011truthful}. In the context of bandwidth allocation for cloud users, the authors in \cite{shishapley} used the Shapley value method as the payment strategy instead of the VCG rule. In the auction, once receiving the data transfer requests (bids) from users, i.e., buyers, the cloud provider calculates a feasible traffic schedule to maximize social welfare based on linear programming. The Shapley value of each user is calculated as the average marginal charge by the cloud provider incurred by the user's traffic. Then, the cloud provider decides to reject or accept a user's request by comparing the user's bid price and its Shapley value. For example, if the user's bid price is larger than its Shapley value, the user's request is accepted, and its payment is equal to its Shapley value. Through the idea of cost sharing, the Shapley value approach was proved to be computationally efficient, budget balanced, individually rational, and truthful. However, the proposed approach only introduced the price. Other dimensions, e.g., path selection, need to be considered for each transfer task.
\subsubsection{Sealed-bid uniform price auction}
\label{sec:cloud_data_center_bandwidth_uniform}
The above approaches only considered either bandwidth reservation \cite{gui2014soar} or bandwidth allocation \cite{shishapley}. The authors in \cite{tan2014uniform} addressed both of them jointly through a two-tier pricing model with the aim of allocating efficiently bandwidth and maximizing the cloud provider's revenue. In the first tier, the reservation phase uses a premium price strategy to guarantee cloud tenants' minimum bandwidth ahead in time. A premium price, also called image pricing, is a strategy of keeping the price of a product or service artificially high to encourage favorable perception among buyers \cite{gittings2002advertising}. The unallocated bandwidth remained from the reservation phase is traded in the second tier through a sealed-bid uniform price auction. The sealed-bid uniform price auction is a multiunit auction where a fixed number of identical units of a homogeneous commodity are set at the same price. The uniform price auction is applied due to its fairness in charging identical price for identical goods. In a single auction round in the second tier, the cloud provider formulates allocation functions based on the tenants' demand requests and bidding prices. Then, the cloud provider determines the market clearing price and allocates bandwidth based on these allocation functions. Finally, all winners pay the market clearing price for their respective allocated quantities. However, the proposed approach did not consider the future demands and the uncertainty in resource utilization of the reservation requests.

\subsubsection{Bargaining game}
\label{sec:cloud_data_center_bandwidth_bargaining}
\begin{figure}[ht]
 \centering
\includegraphics[width=7.8cm, height=5.6cm]{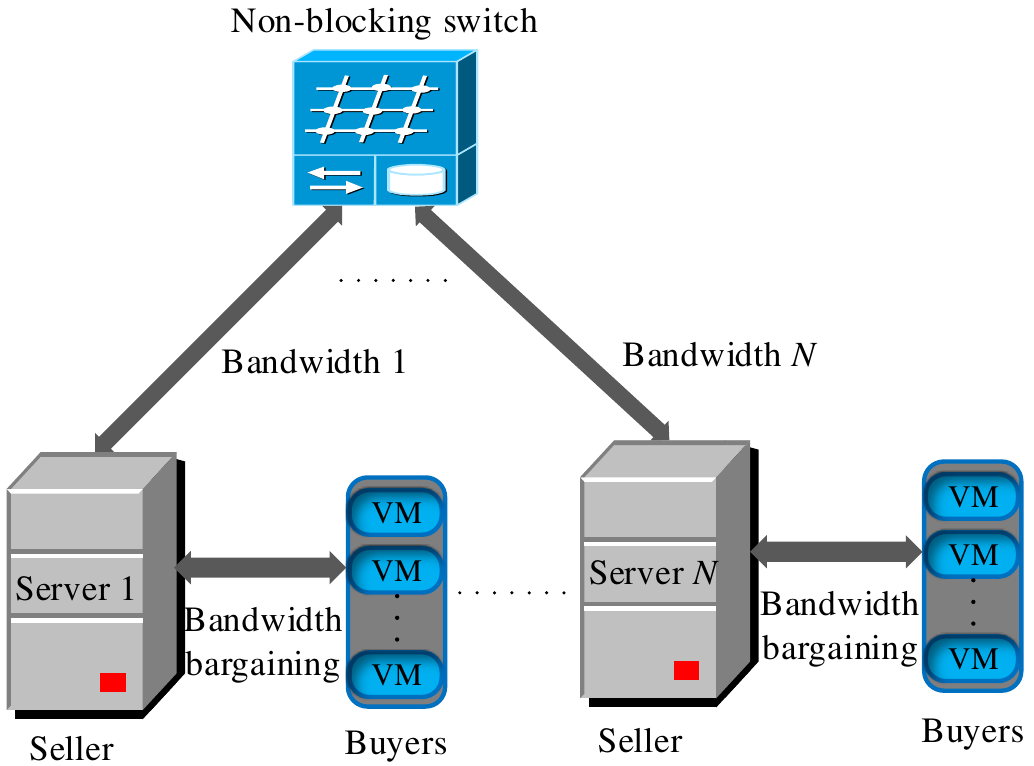}
 \caption{A Nash bargaining game based rate allocation in cloud data center networks.}
 \label{cloud_data_center_bandwidth_bargain}
\end{figure}

The aforementioned auctions only satisfy either the buyer's or the seller's objective. To get a win-win solution for both, a cooperative game such as a bargaining game can be used. The authors in \cite{guo2013cooperative}, \cite{guo2013falloc} employed the bargaining game to address the rate allocation for VM-pairs in data centers as shown in Fig.~\ref{cloud_data_center_bandwidth_bargain}. The VM-pairs, i.e., buyers, participate in an iterative bargaining game to negotiate the rates with the servers, i.e., sellers. Each VM-pair is associated with a utility gain which is assumed to be a convex function. The optimization problem is formulated to maximize the joint profit which is the product of buyers' utility gains. The problem is solved by the dual-based decomposition with the Lagrange multiplier method with the interpretation of rates and prices that buyers are willing to pay. The rate constraints are assumed to be linear \cite{guofair}. Using the Karush-Kuhn-Tucker (KKT) conditions \cite{boyd2004convex}, the approach achieves a unique Nash bargaining solution of the rate allocation. The Nash bargaining solution ensures the Pareto optimality and achieves the fairness in resource allocation \cite{kelly1997charging}, \cite{yaiche2000game}. The simulation results showed that the proposed approach can satisfy bandwidth demands of VM-pairs up to 99\%. However, using the subgradient method results in slow convergence speed.

\subsubsection{Stackelberg game}
\label{sec:cloud_data_center_bandwidth_stackelberg}
Auction approaches, e.g., sealed-bid auctions, allow the cloud provider to determine the price of bandwidth based on their bids. Typically, the cloud provider sets the bandwidth price to maximize its own revenue, and then the tenants decide their payments to obtain more bandwidth without paying too much. Thus, the interactions between the cloud providers and tenants can be modeled as a two-stage Stackelberg game as proposed in \cite{liidaas}. In the first stage, based on tenants' bandwidth demands, the cloud providers, i.e., leaders, cooperate with each other in a Nash bargaining game to optimize their pricing strategies. Different from \cite{guo2013cooperative}, \cite{guofair} which employed the subgradient method, found to have slow convergence to the optimal price, the authors in \cite{liidaas} proposed to combine the advantage of the demand segmentation method and the geometrical Nash bargaining solution \cite{feng2012bargaining}. Given the price and the amount of bandwidth that the cloud providers are willing to allocate, each selfish tenant, i.e., a follower, optimizes its bandwidth reservation through the weighted fair and max-min fair bandwidth reservation algorithms. The simulation results indicated that the proposed algorithms ensure the high average revenue of cloud providers and guarantee the bandwidth requirement for tenants.

In practice, the tenants' VMs of the same applications in a data center can be partitioned into the same virtual networks through network slicing \cite{nikaein2015network}. The authors in \cite{yuan2012game}, \cite{wang2014lease} adopted the Stackelberg game for allocating bandwidth to the tenants' virtual networks from the cloud provider. The cloud provider acts as the leader which announces a rental price per unit of bandwidth. Unlike \cite{liidaas}, the leader's pricing strategy attempts to drive tenants' virtual networks, i.e., followers, to the social optimal solution and to make the highest profit for itself simultaneously. Due to the non-cooperative nature in terms of demand for network resources, virtual networks can be considered to be players in a non-cooperative game in which their strategies are to choose bandwidth demands to maximize their own utilities. Assume that the utilities are strictly concave functions of bandwidth and additive, the optimal bandwidth allocation among followers can be formulated as an optimization problem with multiple constraints. Through the distributed implementation, it was proved that the problem has a unique optimal solution which is the Nash equilibrium of the game. Given the set of optimal strategies of the followers, the leader then defines an objective function to be its own profit. Optimizing this function using the first-order conditions, there exists a unique Stackelberg equilibrium at which the leader's profit is maximized, and the bandwidth allocation for the followers reaches the Nash equilibrium \cite{yuan2012game}. Theoretically, the outcome of the Stackelberg game can achieve the better utilities compared with Nash equilibrium strategies which are obtained by assuming a simultaneous choice for leaders and the follower \cite{huck2001stackelberg}. However, the experiment results to demonstrate this important comparison were not given.

\subsubsection{Differential pricing}
\label{sec:cloud_data_center_bandwidth_diffe}
As stated earlier, the differential pricing applies different prices of the same product to different buyers. The authors in \cite{divakaran2014bandwidth}, \cite{divakaran2013probabilistic} adopted the differential pricing for the allocation of bandwidth to tenants with the aim of maximizing the cloud provider's revenue. Specifically, the tenants, i.e., buyers, submit their requests to the cloud provider, i.e., the seller. Each request has a bandwidth demand and a deviation factor. The deviation factor represents the flexibility in bandwidth allocation available to the provider. If two requests have different deviation factors, the tenant with a higher value gives higher flexibility to the provider than that with a lower value. The cloud provider gives more discounts, i.e., a lower price, to the tenant with the higher deviation factor. This flexibility introduces more degree of freedom of bandwidth allocation, leading to a higher efficiency. Simulation results showed that compared with the deterministic bandwidth allocation algorithm from \cite{kamiyama1998efficient}, the proposed approach increases the allocated bandwidth up to 28\% and the provider's revenue up to 12\%. However, as shown in \cite{divakaran2015towards}, the difference of the provider's revenue between the two approaches is slight when more discounts are given.
\subsubsection{Smart data pricing}
\label{sec:cloud_data_center_bandwidth_smart}
Smart data pricing adapts the resource price according to network congestion. This pricing strategy is used to create proper economic incentives for users. The authors in \cite{shen2014new} adopted this strategy for allocating communication bandwidth to tenants' VMs when the VMs communicate with each other via multiple links as shown in Fig.~\ref{cloud_networking_architecture}. The objectives are to achieve the min-guarantee, i.e., guaranteeing the minimum bandwidth that the tenants expect for each VM, high utilization, i.e., maximizing network utilization, and network proportionality, i.e., the fairness among tenants. The experimental results showed that the proposed approach can satisfy up to 98\% of tenants' VM demand compared with 76\% of the baseline pricing from \cite{popa2012faircloud} which sets price based on only bandwidth. However, the proposed approach does not explain how to set the unit price of the min-guarantee bandwidth optimally.

The schemes to calculate the unit price of the min-guarantee bandwidth can be found in \cite{zhan2014distributednet}, \cite{zhan2015adaptive}. The price is proportional to the number of serving VMs and is inversely proportional to the total sharing bandwidth. However, this pricing strategy is simple and does not ensure that the price is optimal.


\subsubsection{Other pricing models}
\label{sec:cloud_data_center_bandwidth_other}
Apart from the common schemes described above, others pricing models have also been applied to the resource allocation in cloud networking.

\textbf{Dominant resource pricing:} Typically, cloud tenants rent both VM and network resource, i.e., a data transfer service. Therefore, the price that a tenant pays the data center owner, i.e., a cloud provider, depends on the time of VM occupancy and the size of the data transferred among VMs. In particular, the time of the VM occupancy depends on the location of the tenant because if the tenant is far from VMs' locations, the transfer time is longer. The authors in \cite{ballani2011price}, \cite{ballani2011towards}, \cite{stefani2013trust} proposed dominant resource pricing that applies prices independently of the tenants' locations, and thus reducing the tenants' costs. Since only the VM occupancy price depends on the tenant's location, the key technique is to enable the network price to dominate the VM occupancy price. Since the network price can be larger or smaller than the VM occupancy price, a bandwidth baseband threshold was introduced to guarantee that the bandwidth requirement is always larger than the base bandwidth. Therefore, the network price always dominates the VM occupancy price, and the tenants' payments are considered to be flat, i.e., a location independent price. The simulation results showed that compared with the baseline pricing which sets price based on the task completion time, the tenants pay 70-80\% less with the proposed pricing. However, the proposed approach does not specify how to set the bandwidth baseband threshold optimally.

\textbf{Ramsey pricing:} The authors in \cite{wanisefficient} exploited the Ramsey pricing \cite{baumol1970optimal} to regulate the demand of cloud tenants' bandwidth requests, achieving higher network utilization. The model consists of a cloud provider which owns a number of geographically distributed data centers connected with each other through a high-capacity network. Each virtual link between any two data centers is divided into several virtual pipes corresponding to several service classes. Each service class has a specific expected latency and a corresponding price which is set based on the Ramsey problem. The Ramsey problem pricing is a strategy in which a monopolist sets the price to maximize social welfare subject to the constraint on profit \cite{oum1988ramsey}. Applying this strategy to the proposed model, the price is set to maximize the welfare of both the cloud tenants and the cloud provider subject to the preset profit threshold. Solving this problem requires to have knowledge of the profit threshold and the expected demand for each service class. The threshold can be defined according to the market competition, and the expected demand can be obtained by cloud resource monitoring methods, e.g., a semi-centralized monitoring mechanism \cite{alizadeh2011data}. Generally, increasing the price leads to a decrease in the demand of that service class, or it will move the demand of that service class to other classes. The simulation results showed that the cloud provider's network utilization of the proposed approach is higher, e.g., 60\% with the supply and demand pricing \cite{mihailescu2010dynamic} or 40\% with the static pricing. However, multiple cloud providers participating in the market need to be considered in the future work.

\textbf{Pricing model based on dynamic programming:} The above approaches did not consider the workload fluctuation of cloud tenants. Some events, e.g., releasing of a new movie, cause a high workload fluctuation. These applications require a continuous resource allocation elasticity to accurately adapt to the time-varying application's needs \cite{li2013elasticity}. Therefore, it is important to predict the workload fluctuation. The authors in \cite{wanis2015modeling} proposed an approach to tackle this problem. The model consists of a cloud application owner, i.e., a buyer, which rents the inter-data center networking resources from a service provider, i.e., a seller. The application's requirements can be predicted by using a Markov chain model, utilizing the temporal variability of workload fluctuations \cite{pacheco2011markovian}. The prediction allows the service provider to dynamically re-size the inter-data centers bandwidth pool to ensure the availability of network resources. Given the traffic workload fluctuation information, the service provider sets the resource prices at each time slot to maximize its long-term expected revenue. A dynamic programming algorithm \cite{bertsekas1995dynamic} was adopted to search for the optimal prices. The simulation results showed that the mean square error between the estimated and the actual fluctuations can be marginal. Moreover, when the elasticity level, i.e., the bandwidth adaptation, is higher, the service provider receives a higher accumulated revenue since more bandwidth demands are met at a higher price. However, more advanced techniques should be applied to accurately forecast workload fluctuations.

\subsection{Request Allocation}
\label{sec:cloud_data_center_request}
The term ``request'' in this section represents the resource/workload/data processing requests from cloud users or tenants. A simple technique can be implemented by assigning each request of a user to the closest data center \cite{wong2006closestnode}. However, such a method can overload the data center during peak time and further degrade application performance as well as the revenue of the provider. Therefore, economic and pricing models have been developed to optimize the benefits of users and providers.

\subsubsection{Cost-based pricing}
\label{sec:cloud_data_center_resource_request_cost}
To maximize the service provider's profit while satisfying Service-Level Agreements (SLAs) of the users, the authors in \cite{prasad2010resource} considered charging incoming user requests through the cost-based pricing. Cost-based pricing is a strategy to set the price of a request based on the cost of executing the request. In particular, whenever a new request from a user arrives at the cloud site, the service provider, i.e., the seller, estimates necessary parameters including network parameters (e.g., the bandwidth, control overhead, session throughput, and session lifetime), resource parameters (average inflow and processing speed), and data processing parameters (e.g., a job size, block size, and total time). These parameters allow the service provider to determine the SLA of the user and the resources allocated for the request. Then, the price of the request is defined according to the costs of resources. To maximize the service provider's profit, the consumer perceive pricing is also employed through considering the money for which the user is willing to pay. However, other market factors, e.g., the market competition, need to be considered.

\subsubsection{Bargaining game}
\label{sec:cloud_data_center_resource_request_bargain}

To direct the users' requests to the appropriate data centers, mapping nodes, e.g., authoritative DNS servers, can be deployed at different regions. The objective of this deployment is to optimize the general system performance. The mapping nodes cooperate for receiving the cloud resources from data centers. The authors in \cite{xu2012general} modeled the data selection problem for users' requests as the bargaining game among mapping nodes. Each mapping node, i.e., a player, has its own average utility as the objective which can be a convex function of the latency or throughput. Each node also has an initial utility which represents the minimum utility requirement for the users based on the SLAs with the cloud provider. The goal is to maximize the average utility cooperatively. Similar to \cite{guo2013cooperative}, the optimization problem is formulated by maximizing the joint utility which is the product of nodes' average utilities. However, unlike \cite{guo2013cooperative}, to solve the non-linear problem, the sequential-linear-approximation algorithm \cite{avriel2003nonlinear} was employed for reducing computational complexity. The linear problem is then solved by the dual-based decomposition with the Lagrange multiplier method with the interpretation of load balancing, capacity, and cost constraints. Using the subgradient method, the load balancing and capacity constraints from data centers serve as price signals. When the total traffic routed to a data center exceeds its capacity, the data center increases its price for the next round to suppress the excessive demand. This process continues until the algorithm converges to the optimal resource allocation. However, the slow convergence from using the subgradient method can impact the real-time applications.

\subsubsection{Non-cooperative game}
\label{sec:cloud_data_center_resource_request_compete_game}
In a competition market with multiple service providers, a non-cooperative game can be used. The authors in \cite{zhang2012dynamic} adopted the game for the request allocation among service providers with the aim of maximizing the social welfare of the system. The market consists of service providers, i.e., sellers, and a user, i.e., a buyer. Each service provider has its own control strategy which involves deciding on the number of servers placed in each data center and routing a user's request to an appropriate server. The objective of each service provider is to minimize its operational costs while satisfying the users' SLA and data center capacity constraints. Assume that each service provider's strategy is kept private from other service providers. The set of optimal strategies yields a unique Nash equilibrium in which no service provider can
optimize its cost by unilaterally changing its allocation strategy
over time. The price of anarchy and the price of stability are then determined as the metrics to measure the best-case and worst-case efficiency loss of the game, respectively.

To tackle the worst-case where a service provider behaves selfishly in an uncoordinated manner, the authors in \cite{zhang2013dynamic} extended the market with the participation of the cloud provider by taking into account a penalty function. The cloud provider must pay a penalty cost for the service provider when the cloud provider rejects the resource request from the service providers. This is reasonable since the request rejection can hurt the service satisfaction of users, resulting in the loss in service provider's revenue. The experimental results showed that the proposed approach can improve the social welfare from 10\% to 20\% compared with the outcome of the competition game in \cite{zhang2012dynamic}.

\subsection{Workflow Allocation}
\label{sec:cloud_data_center_workflow}

Workflow allocation in cloud network is to map required tasks from users to the resources at multiple network locations and order their executions so that task-precedence requirements are satisfied \cite{wu2015workflow}. Due to the large resource distribution in the cloud network, the requirements for the workload allocation include minimizing the cost and delay of the task execution as well as adapting dynamically to the resource demand fluctuation from users. The traditional approaches exploited the diversity in local electricity prices, e.g., \cite{gupta2014cost}, \cite{gupta2015cost}, \cite{kantarciinter}. However, they focused only on minimizing the cost, i.e., Operational expenditures (Opex), of the operators rather than satisfying the users. Besides, the static approaches, e.g., \cite{blythe2005task}, \cite{lopez2006analysis}, may not achieve the objectives due to the lack of interactions among the entities in the cloud network. Therefore, the market based schemes with negotiation mechanisms among network entities have been developed to optimize the workflow allocation, resource utilization, and profit of the operators.
\subsubsection{Profit maximization}
\label{sec:cloud_data_center_workflow_profit}
The authors in \cite{zhao2014dynamic} proposed a task scheduling algorithm across data centers through the economic model based on profit maximization as described in Section~\ref{subsec:Profit_maximization}. Once receiving workload requests from users including the types of VMs and the number of time slots, the cloud provider solves the profit maximization problem. The total revenue depends on the number of tasks and their prices, and the costs are the Opex in the data center. The solution of the optimization problem allows the cloud provider to choose (i) an appropriate price for each type of tasks at each data center, (ii) the best number of servers to provision each type of VMs in each data center, and (iii) the optimal number of tasks of each type to schedule and to drop. Moreover, to enable the cloud provider to achieve a high time-average overall profit, the authors adopted the drift-plus-penalty framework in Lyapunov optimization \cite{neely2010stochastic}. It is a classical approach for translating a long-term time-average optimization problem into a series of similar one-shot optimization problems. The simulation results showed that the proposed pricing algorithm outperforms the static pricing, e.g., the pricing strategy in Amazon EC2's on-demand instance market, in terms of the profit. Moreover, the proposed approach achieves the stable profit over time.

\subsubsection{Spot instance pricing}
\label{sec:cloud_data_center_workflow_spot_instance}
\begin{figure}[ht]
 \centering
\includegraphics[width=8 cm, height = 8.6 cm]{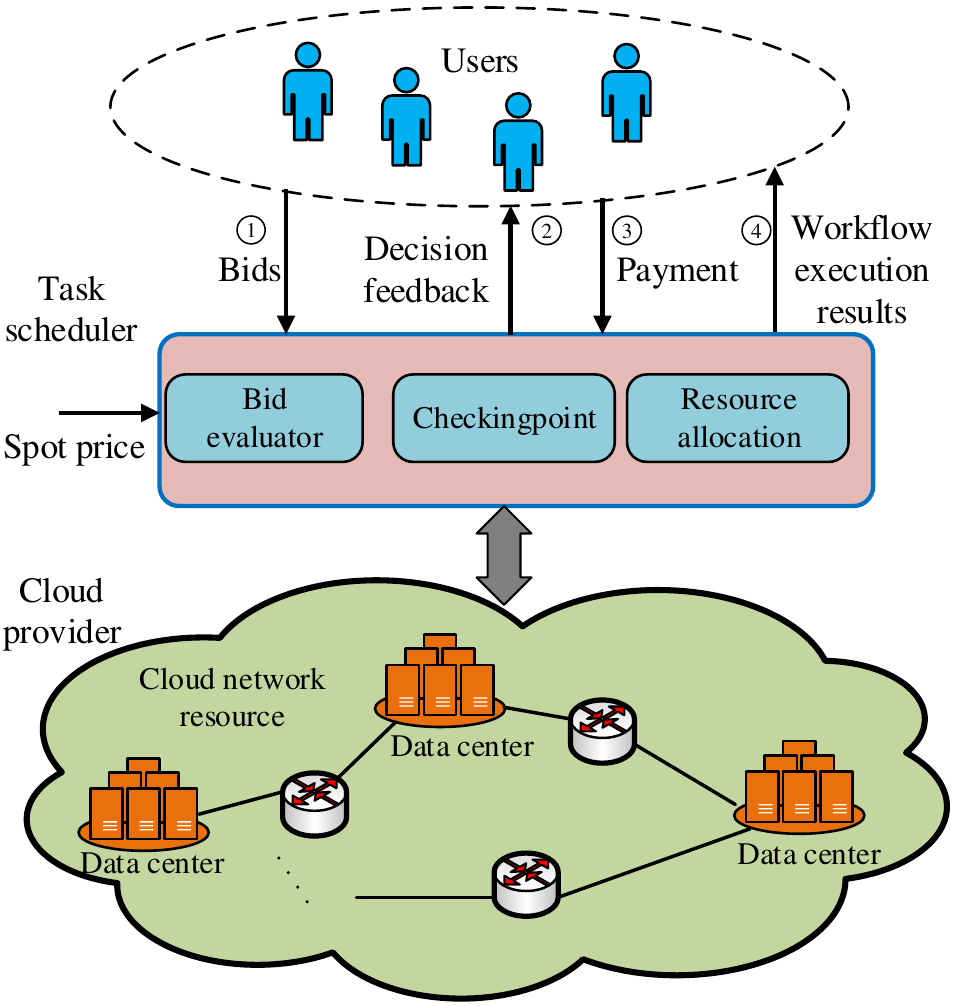}
 \caption{Spot instance pricing based workflow allocation mechanism.}
 \label{cloud_data_center_workflow_spot}
\end{figure}

The approach in \cite{zhao2014dynamic} aims at maximizing the cloud provider's profit. To reduce costs for users and achieve high utilization, the spot instance pricing can be used. Spot instance pricing which is practically used to set the price of the Amazon EC2 instances (https://aws.amazon.com/ec2/spot/pricing/) allows users to bid for idle or unused cloud instances. The users receive the instances immediately as long as their bids are higher than the spot price (i.e., the price of the spot instance) \cite{tang2012towards}, which is often a discounted price. Different from the auction approaches, e.g., the first-price sealed-bid auction, the user only pays the spot price instead of its bidding price. The authors in \cite{poola2014fault} addressed the workload scheduling in cloud networks by combining the spot instance pricing and the on-demand instance pricing \cite{shen2013scheduling} to reduce the cost of workload execution while guaranteeing the workflow deadline. The model is shown in Fig. \ref{cloud_data_center_workflow_spot}. Each user submits its bid to the task scheduler including the application task, current spot price, on-demand price, the current time, and factors that specify the weights of the types of pricing. The scheduler calculates the bid values and employs the checkpointing technique \cite{yi2010reducing} which is known as an efficient fault tolerant strategy to find a suitable cloud resource for every task. If the spot instance is not enough for accomplishing the task, the scheduler adaptively switches to on-demand instances to meet workflow deadline. Experimental results showed that the proposed approach can reduce up to 70\% execution cost compared with the task scheduling algorithm using only the on-demand instance pricing.

\subsection{Applications of economic and pricing models for resource management in federated cloud networking}
\label{sec:cloud_data_center_federated_cloud}
\begin{figure}[ht]
 \centering
\includegraphics[width=7.5 cm, height = 8.9 cm]{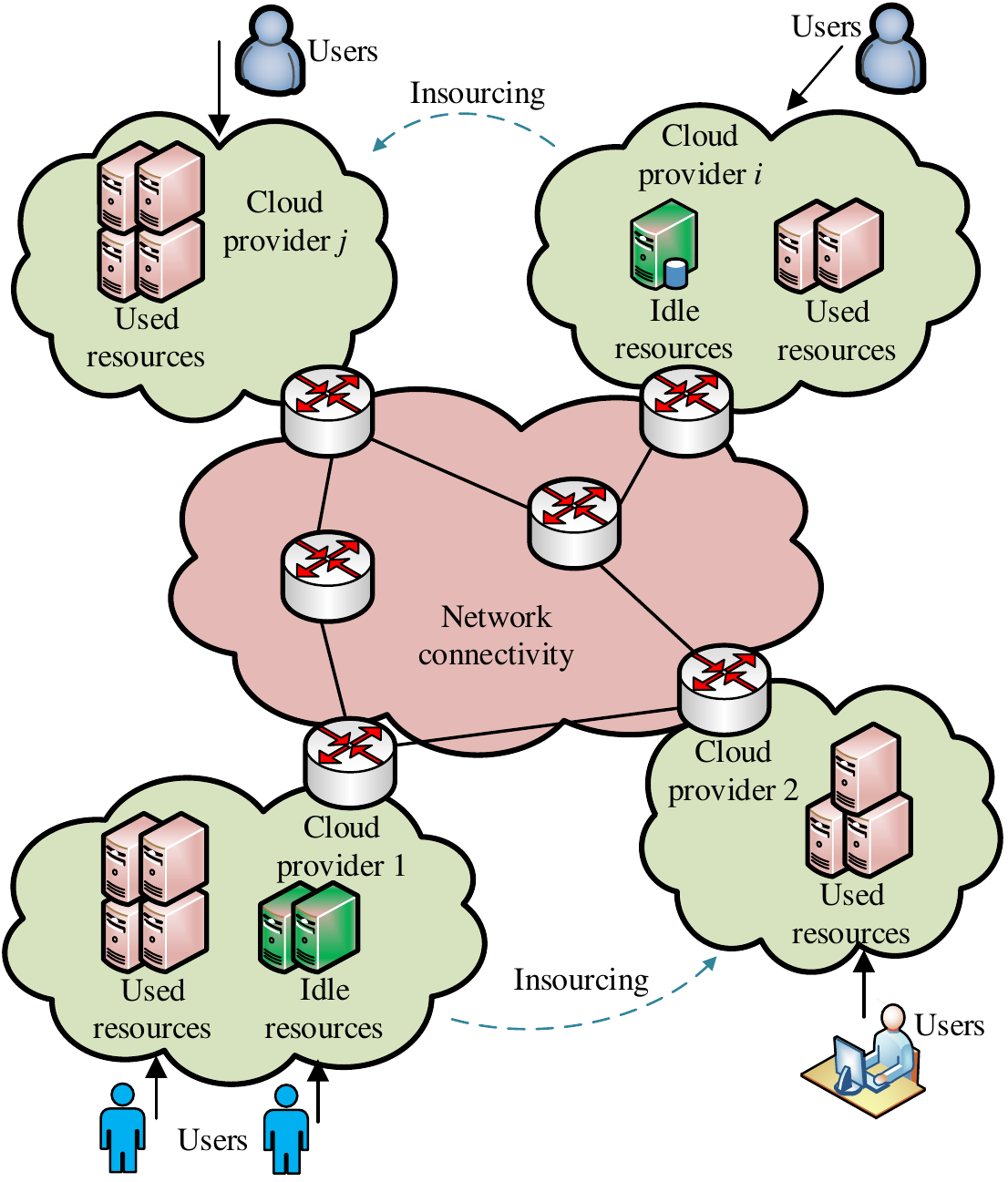}
 \caption{Federated cloud network.}
 \label{federated_cloud_networking}
\end{figure}

To guarantee SLAs for users' composite services, i.e., multi-cloud applications, cloud and network resources need to be distributed across multiple cloud providers \cite{bardhan2012mechanism}. Federated cloud networking (Fig.~\ref{federated_cloud_networking}) enables interconnecting the cloud providers to form cloud federation resources which can be shared to increase capacity, availability, and resilience at multiple network locations \cite{qiang2016cloud}. The federated cloud network can be considered to be an extension of the cloud data center network belonging to the providers. Mathematically, it was also proved in \cite{abdo2014cloud} that the Return On Investment (ROI), i.e., the earned amount of money for each unit of investment, of a cloud provider in the federated cloud networking is higher than that in the stand-alone cloud data center network. In what follows, we discuss economic and pricing models which provide incentives to cloud providers to pool their resources in the federated cloud network to satisfy users' requests.

\subsubsection{Profit maximization}
\label{sec:cloud_federation_request_profit}
To stimulate cloud providers within the federated cloud network for sharing resources, their profits need to be maximized. Therefore, the authors in \cite{rebaiimproving} adopted the economic model based on profit maximization for allocating resource requests from users to cloud providers in a federated cloud network. The model consists of cloud providers interconnected through high capacity links. As shown in Fig.~\ref{federated_cloud_networking}, each cloud provider receives resource requests from its local users and other providers. A cloud provider can insource, i.e., rent out, its unused resources to serve other providers, but it can also outsource, i.e., borrow, resources from other providers. To maximize cloud providers' profits and guarantee load balancing among them, each cloud provider determines the insourcing price as well as the available quota of resources. Using pricing policy proposed in \cite{toosi2011resource}, the insourcing price is set according to the VM costs, the fixed price to local users, and the maximum hosting capacity and idle capacity. Given the insourcing price, outsourcing prices (set by other providers), and networking costs (from the network providers), each provider solves the integer linear programming problem with the aim of maximizing the total profit of the providers in the federated cloud network. The solution of this problem helps each provider to partition received requests into subsets that will be hosted locally or outsourced to other providers as well as to select the insourcing requests from other providers to accept. The simulation results showed that compared with the case without federation, the proposed approach can improve the profits of some providers up to 42\% and the request acceptance rate up to 48\%.

\begin{figure}[ht]
 \centering
\includegraphics[width=7.4 cm, height = 7.1 cm]{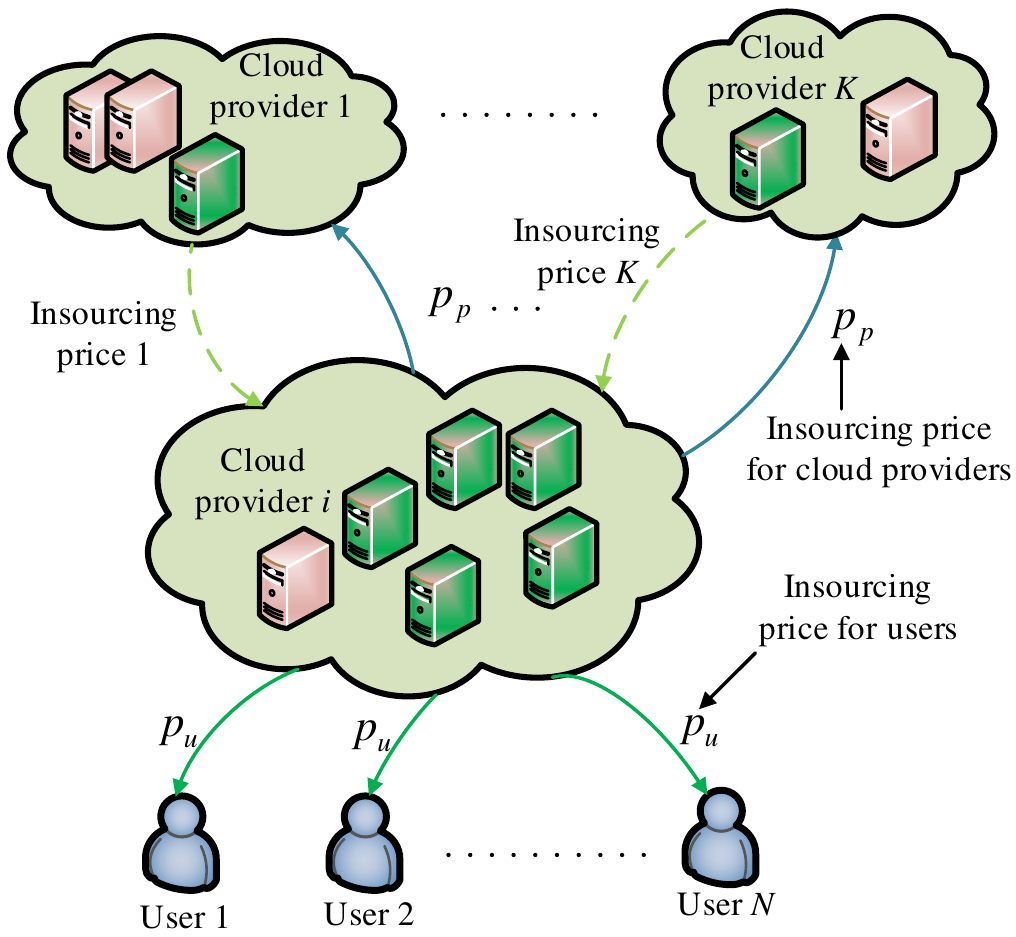}
 \caption{Differential pricing based request allocation for federated cloud network.}
 \label{federated_cloud_networking_differential_pricing}
\end{figure}

 The approach in \cite{rebaiimproving} did not specify how to set the resource prices. The authors in \cite{hadjimathematical} determined upper and lower bounds for resource prices for both cloud providers and users. The model is shown in Fig.~\ref{federated_cloud_networking_differential_pricing}. First, each cloud provider calculates costs, e.g., energy costs, networking costs, deployment costs, and outsourcing costs. The lower bound for the insourcing price charging to the other providers is determined based on the condition to ensure that the revenue is higher than the overall cost. The upper bound for the insourcing price is obtained based on the proposed insourcing prices by the other providers, and particularly the upper bound is the minimum insourcing price among them. This enables one cloud provider to compete with the other cloud providers. Otherwise, the proposed resource price to users needs to be higher than the insourcing price to improve revenues of the providers within the federated cloud network. This condition allows to determine the lower and upper bounds for the proposed price to the users. These lower and upper bound prices are used as inputs for the revenue maximization problem of each cloud provider. The branch and bound algorithm \cite{land1960automatic} is used to solve the integer problem to find the optimal resource amount for serving the users, the amount for outsourcing and for insourcing. Nonetheless, the low complexity algorithm is needed.

The same approach can be found in \cite{darzanos2015model} which determines a portion of job requests transferred from one provider to the other provider so that their total profit is maximized while guaranteeing the user utility. The model consists of two cloud service providers and their servers interconnected with each other through Internet links. First, each provider defines a profit function of its revenue and energy consumption cost. The revenue is based on the total incoming requests at the queue of the provider and the price per request. The total request of a provider includes the requests from its local users and the other provider's users. The price per request is set based on a QoS-dependent pricing scheme which should be convex and decreasing in the average delay of job execution. This delay is calculated according to the M/M/1 queueing model \cite{abate1987transient}, taking into account the Internet transfer delay. The total profit maximization of the two providers is a non-linear optimization problem that is then solved with the method of Lagrangian and the KKT conditions. The solution is the optimal portions of the requests at each provider to be routed to the other provider \cite{boyd2004convex}. The simulation results indicated that the proposed approach improves the total profit up to 100\% compared with the case when each provider serves only its own users. However, the case with more than two providers that will make the problem more general was not considered.

\subsubsection{Differential pricing}
\label{sec:cloud_federation_differential}
The aforementioned approaches did not provide any prediction techniques which address the workload elasticity of users. The future resource consumption of current users in the federated cloud network can be estimated through their behavior and probability of using the resource as proposed in \cite{aazam2014advance}, \cite{aazam2014broker}. The resources considered here include virtual servers, computing and storage resources, virtual networks, and network bandwidth. The prediction can be implemented by a cloud broker which acts as an intermediary between the cloud providers and users. Basically, the resource estimation for the current user is based on the current resource price, the average of service relinquishing probability, and the history of relinquishing probability of the user. Relinquishing probability of a user is the probability of giving up the resource for which the user has requested. Given the relinquishing probability, the differential pricing scheme was introduced to set different resource prices depending on the behavior of users \cite{aazam2014broker}. Specifically, the proposed resource price is proportional to the service relinquishing probability of the user. For example, if the user has the high probability, the proposed price is set at a high value. This policy discourages the users with a high relinquishing probability from participating in the market because they may degrade the profit of providers. However, the proposed approach did not explain how to obtain the relinquishing probability for each user.

The same approach can be found in \cite{aazamfog}. However, the authors in \cite{aazamfog} investigated refunding the remaining amount to a user when its service consumption will be discontinued. The authors also considered that fog computing which is located between underlying Internet of Things (IoT) devices and the cloud computing can implement this task. The total refund is the sum of the refund of unutilized resources and the refund to be paid on quality degradation, i.e., not satisfying SLA. The refund is determined based on the amount of unutilized resources and a depreciation index. A high depreciation index allows the users which have used more services, e.g., more than 60\%, to receive more refunds. The refund on quality degradation is based on the ratio of the acquired quality of service and the provided QoS.

\subsubsection{Cost-based pricing}
\label{sec:cloud_federation_costpricing}
\begin{figure}[ht]
 \centering
\includegraphics[width=\linewidth]{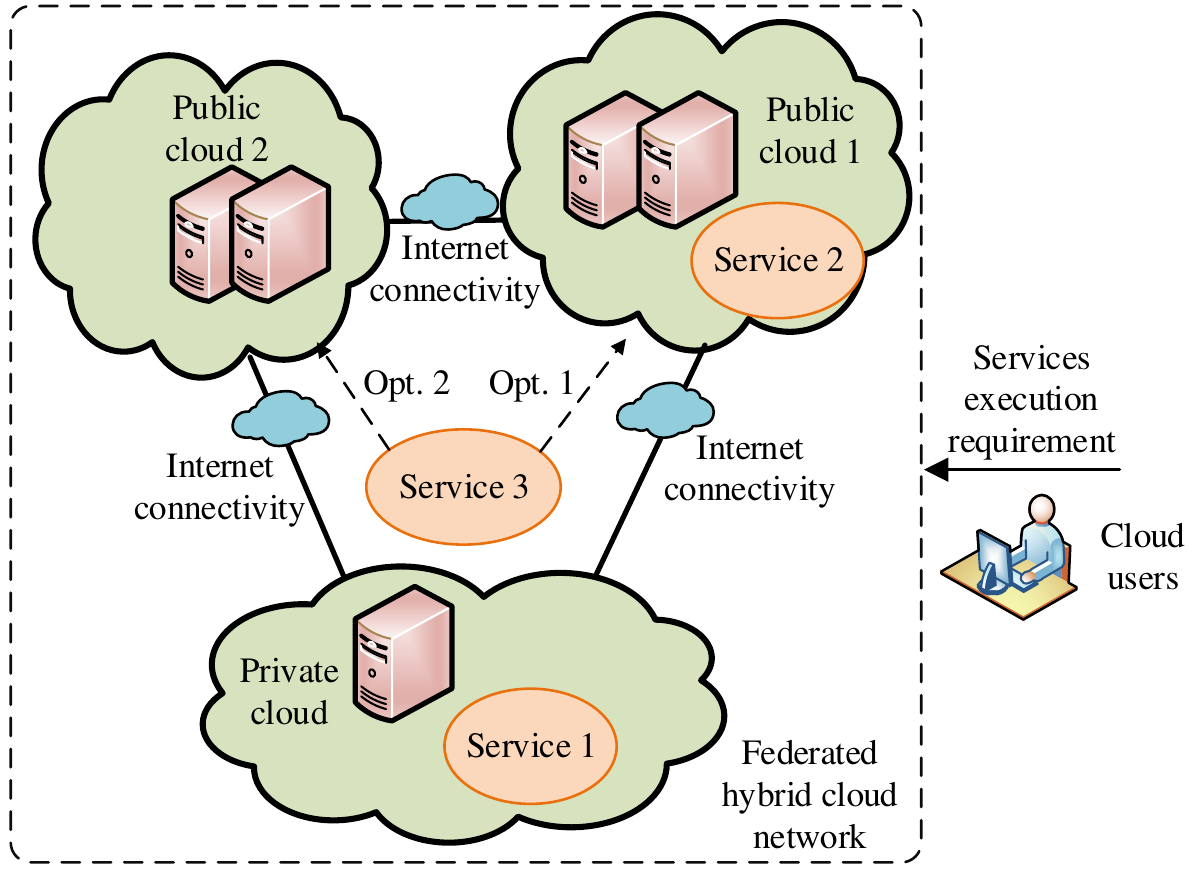}
 \caption{Service placement in hybrid federated cloud network.}
 \label{hybrid_federated_cloud_networking}
\end{figure}

In the federated cloud network, users have to decide where to place their services on the federated cloud such that the service cost is minimized. The authors in \cite{altmann2014cost} addressed the service placement optimization based on the cost model in federated hybrid clouds to guarantee the cost minimization for cloud users. A federated hybrid cloud (Fig.~\ref{hybrid_federated_cloud_networking}) is a composition of interoperable private and public cloud networks. Given a set of cloud services from the cloud providers, the total cost for each possible service placement option is evaluated by calculating the sum of the fixed costs and the variable costs. The fixed costs include the costs for hardware, e.g., servers and network devices, and software licenses. The variable costs comprise, e.g., the electricity cost, Internet connectivity cost, and cost of data transfer between different clouds. Finally, the cloud user selects the best option which has the minimum total cost. However, more performance factors, e.g., a service latency, need to be considered to guarantee user SLAs.

\subsubsection{Double auction}
\label{sec:cloud_federation_double_auction}
Most of the above approaches consider the request allocation. The authors in \cite{zheng2015star} addressed the bandwidth reservation in federated cloud networking by using the double auction. The model consists of multiple cloud providers, i.e., sellers, multiple cloud tenants, i.e., buyers, and an auctioneer. The double auction used in this work is similar to that presented in Section~\ref{subsec:Auction_theory}. However, there are some slight modifications in the winner determination and payment stages. Specifically, upon receiving asks from sellers and bids from buyers, the auctioneer sorts sellers in the first list by the selling prices in a non-decreasing order. Then, the auctioneer sorts buyers in the second list by their bids in a non-increasing order. The water filling method \cite{proakis2001digital} is used to fill the sellers one by one following the order in the first list, with buyers by the order in the second list. The auctioneer finds the largest indexes $k$ and $l$ in the two lists to satisfy two constraints: (i) the first $k$ sellers in the first list have sufficient bandwidth to satisfy the demands of the first $l$ buyers in the second list, and (ii) the total charge to $l$ buyers is less than total payment to $k$ sellers. The second constraint is to guarantee the ex-post budget balance of the auction, i.e., the profit of the auctioneer is non-negative. The winners are the first $k$ sellers and the first $l$ buyers. The auctioneer pays each winning seller by the $(k+1)$th selling price in the first list and charges each winning buyer by the $(l+1)$th bidding price in the second list. Such payment and charging mechanisms are similar to those of the Vickrey auction that guarantee the truthfulness of the auction. However, the cheating behaviors of the bidders such as false-name bidding and collusion need to be considered in the future work.

\textbf{Summary:} In this section, we have discussed three major issues in cloud data center networks, i.e., the bandwidth allocation, request allocation, and workload allocation. We have then reviewed the applications of economic and pricing models for these issues which are summarized along with references in Table~\ref{table_data_center_networking_1} and Table~\ref{table_data_center_networking_2}. We observe that in the cloud data center networking, economic and pricing models for bandwidth allocation get more attentions than other issues. In the federated cloud networking, the models aim at improving the total profit for cloud providers. In the next section, we review economic and pricing models in mobile cloud networking for efficient management of cloud and radio resources.

\begin{table*}
\caption{Applications of economic and pricing models for resource management in cloud data center networking}
\label{table_data_center_networking_1}
\scriptsize
\begin{centering}
\begin{tabular}{|>{\centering\arraybackslash}m{0.2cm}|>{\centering\arraybackslash}m{0.4cm}|>{\centering\arraybackslash}m{1.6cm}|>{\centering\arraybackslash}m{1cm}|>{\centering\arraybackslash}m{1cm}|>{\centering\arraybackslash}m{1.2cm}|>{\centering\arraybackslash}m{5cm}|>{\centering\arraybackslash}m{2.4cm}|>{\centering\arraybackslash}m{1.2cm}|}
\hline
\multirow{2}{*} {\textbf{}} & \multirow{2}{*} {\textbf{Ref.}} & \multirow{2}{*} {\textbf{Pricing model}} & \multicolumn{3}{c|} {\textbf{Market structure}} & \multirow{2}{*} {\textbf{Mechanism}} & \multirow{2}{*} {\textbf{Objective}} & \multirow{2}{*} {\textbf{Solution}} \tabularnewline
\cline{4-6}
 & & & \textbf{Seller} & \textbf{Buyer} & \textbf{Item} & & &\tabularnewline
\hline
\hline
\parbox[t]{2mm}{\multirow{9}{*}{\rotatebox[origin=c]{90}{ \hspace{-8cm} Bandwidth allocation}}}

& \cite{gui2014soar} & VCG auction & Cloud provider&Cloud tenants&Bandwidth & Based on buyers' bids, the seller determines winners by solving a linear program in polynomial time and charges each winner with the VCG
payment & Resource efficiency, and welfare social
maximization&Nash equilibrium\tabularnewline \cline{2-9}

&\cite{shishapley} &Shapley value based auction& Cloud provider & Cloud users&Bandwidth& Based on buyers' bids, the seller compares each buyer's price with its own Shapley value to accept or reject the buyer. The accepted buyers pay their Shapley values for the seller&Social
welfare maximization, computational efficient, individually rational, and truthful&Nash equilibrium\tabularnewline \cline{2-9}

& \cite{tan2014uniform} & Premium price and sealed-bid uniform price auction & Cloud provider &Users&Bandwidth&Buyers reserve bandwidth based on the premium pricing. The unallocated bandwidth remaining is traded using the sealed-bid uniform price auction&Maximum social welfare, and revenue maximization&Nash equilibrium\tabularnewline \cline{2-9}

& \cite{guo2013cooperative} \cite{guofair} \cite{guo2013falloc}& Bargaining game & Servers&VM-pairs&Transmission rate&The unique optimal solution for the rate and price is determined by using the method of Lagrange multipliers and the karush-kuhn-tucker conditions &Min-bandwidth guarantee, high utilization, and fair allocation&Nash bargaining solution\tabularnewline \cline{2-9}

& \cite{liidaas}& Stackelberg game & Cloud provider&Application provider&Bandwidth&Sellers set the bandwidth price based on the geometrical Nash bargaining game. The buyer optimizes its bandwidth reservation through the weighted fair and max-min fair reservation algorithms&High revenue, and bandwidth guarantee&Stackelberg
equilibrium solution\tabularnewline \cline{2-9}

& \cite{yuan2012game} \cite{wang2014lease}& Stackelberg game & Cloud provider&Tenants' virtual networks&Bandwidth&Seller uses the first-order conditions to set price, and buyers compete the bandwidth in a non-cooperative game&Efficient and fair allocation, and seller's profit maximization&Stackelberg equilibrium solution\tabularnewline \cline{2-9}

& \cite{divakaran2014bandwidth} \cite{divakaran2013probabilistic}& Differential pricing & Cloud provider&Tenants&Bandwidth&Seller sets lower prices for buyers which accept to receive bandwidth allocation with high flexibility&Payment minimization, and high utilization&Pareto
efficiency\tabularnewline \cline{2-9}

& \cite{shen2014new}&Smart data pricing & Cloud provider&Tenants' VMs&Bandwidth&Seller sets price of links between VMs depending on the congestion degree of the links to regulate VMs' demands&Min-bandwidth guarantee, high utilization, and network proportionality&Market
equilibrium\tabularnewline \cline{2-9}

& \cite{zhan2014distributednet} \cite{zhan2015adaptive}&Smart data pricing & Cloud provider&Tenants' VMs&Bandwidth&Same as \cite{shen2014new}, but the seller sets price of bandwith reservation according to the number of serving VMs, and the total sharing bandwidth&Min-bandwidth
guarantee, high utilization&Market equilibrium\tabularnewline \cline{2-9}

&\cite{ballani2011price} \cite{ballani2011towards}&Dominant resource pricing& Cloud provider&Cloud tenants&Data transfer service&Seller introduces a bandwidth threshold which the buyers must buy so that the network price dominates the VM occupancy price &Min-bandwidth
guarantee, buyers' cost reduction& Flat-rate equilibrium\tabularnewline \cline{2-9}

&\cite{wanisefficient}&Ramsey pricing & Cloud provider&Cloud tenants&Bandwidth &Seller sets different prices for different virtual pipes based on Ramsey problem & High utilization, social welfare maximization& Second-best equilibrium \tabularnewline \cline{2-9}

&\cite{wanis2015modeling}&Dynamic programming based pricing& Service provider&Application provider&Bandwidth &Seller predicts the buyer's needs using a Markov chain and sets price for the network resource based on a dynamic programming algorithm& High resource adaptation, and high revenue& Optimal solution \tabularnewline \cline{2-9}
\hline
\parbox[t]{2mm}{\multirow{9}{*}{\rotatebox[origin=c]{90}{ Workload allocation\hspace{-1 cm} }}}

&\cite{zhao2014dynamic}& Profit maximization & Cloud provider &Users&Workload& Once receiving workload requests from buyers, the seller solves the profit maximization through the drift-plus-penalty framework in the Lyapunov optimization & Time-averaged overall profit maximization, and optimal accepted tasks&Optimal solution\tabularnewline \cline{2-9}
&\cite{poola2014fault}&Spot instance & Cloud provider &User&Workload& Seller employs the checkpointing technique for scheduling the buyer's workload. The buyer pays the seller the spot instance & Buyer's cost minimization&Spot market equilibrium\tabularnewline \cline{2-9}
\hline
\end{tabular}
\par\end{centering}
\end{table*}

\begin{table*}
\caption{Applications of economic and pricing models for resource management in cloud data center networking (cont.)}
\label{table_data_center_networking_2}
\scriptsize
\begin{centering}
\begin{tabular}{|>{\centering\arraybackslash}m{0.2cm}|>{\centering\arraybackslash}m{0.4cm}|>{\centering\arraybackslash}m{1.6cm}|>{\centering\arraybackslash}m{1cm}|>{\centering\arraybackslash}m{1cm}|>{\centering\arraybackslash}m{1.2cm}|>{\centering\arraybackslash}m{5cm}|>{\centering\arraybackslash}m{2.4cm}|>{\centering\arraybackslash}m{1.2cm}|}
\hline
\multirow{2}{*} {\textbf{}} & \multirow{2}{*} {\textbf{Ref.}} & \multirow{2}{*} {\textbf{Pricing model}} & \multicolumn{3}{c|} {\textbf{Market structure}} & \multirow{2}{*} {\textbf{Mechanism}} & \multirow{2}{*} {\textbf{Objective}} & \multirow{2}{*} {\textbf{Solution}} \tabularnewline
\cline{4-6}
 & & & \textbf{Seller} & \textbf{Buyer} & \textbf{Item} & & &\tabularnewline
\hline
\hline
\parbox[t]{2mm}{\multirow{9}{*}{\rotatebox[origin=c]{90}{ \hspace{-7cm} Request allocation}}}
&\cite{prasad2010resource} & Cost-based pricing & Service provider&User&Request execution & Seller sets price of the request execution based on the total cost of resources & Seller's profit maximization&Pareto efficiency\tabularnewline \cline{2-9}

&\cite{xu2012general} & Bargaining game & Data centers&Mapping nodes&Request directing service& A bargaining game among buyers is solved by the dual decomposition with the Lagrange multipliers. The balancing and capacity act as price signals to regulate the buyers' needs& Social welfare maximization&Nash bargaining equilibrium \tabularnewline \cline{2-9}

&\cite{zhang2012dynamic} & Non-cooperative game& Cloud provider&Service providers&Request execution& Sellers participate in the non-cooperative game in which their strategies are to decide the number of servers placed in each data center and routing users' requests to appropriate servers& Social welfare maximization&Nash equilibrium \tabularnewline \cline{2-9}

&\cite{zhang2013dynamic} & Non-cooperative game& Cloud provider&Service providers&Request execution& Same as \cite{zhang2012dynamic}, but sellers will receive a penalty cost if the cloud provider rejects the buyers' resource demands & Social welfare maximization&Nash equilibrium \tabularnewline \cline{2-9}

& \cite{rebaiimproving} & Profit maximization & Cloud providers&Users&Resource requests &Each seller sets resource insourcing price to other sellers based on its maximum hosting capacity and idle capacity. Maximizing total profit of federation providers is solved via the integer linear problem & Profit maximization&Optimal solution\tabularnewline \cline{2-9}

&\cite{hadjimathematical} & Profit maximization & Cloud providers&Users&Resource requests &Each seller proposes upper and lower bounds for resource prices to other sellers and buyers. Then, the branch and bound algorithm is used to determine optimal resource allocation& Profit maximization, and optimal request allocation&Optimal solution\tabularnewline \cline{2-9}

& \cite{aazam2014advance} \cite{aazam2014broker}& Differential pricing & Cloud providers&Users&Service requests &A broker considers behaviors of buyers through defining their relinquishing probabilities to set different prices & Efficient resource allocation, and profit improvement&Pareto
efficiency\tabularnewline \cline{2-9}

&\cite{altmann2014cost} &Cost-based pricing & Cloud providers&Cloud user&Service requests &Buyer evaluates the total cost of all possible resource options for its service execution from different sellers through calculating fixed costs and variable costs. The option with minimum total cost is selected& Buyer's cost minimization&Pareto
efficiency\tabularnewline \cline{2-9}

& \cite{zheng2015star} & Double auction & Cloud providers&Cloud tenants&Bandwidth &The auctioneer uses the water filling method to select the winning buyers and winning sellers. The charing and payment policies are based on the Vicrkey auction &Truthfulness, and ex-post budget balance&Market equilibrium\tabularnewline \cline{2-9}
\hline
\end{tabular}
\par\end{centering}
\end{table*}

\begin{table*}[h]
\caption{A summary of advantages and disadvantages of major approaches for the resource management in cloud data center networking.}
\label{table_sum_advantage_data_center}
\scriptsize

\begin{centering}
\begin{tabular}{|>{\centering\arraybackslash}m{2cm}|>{\centering\arraybackslash}m{5.5cm}|>{\centering\arraybackslash}m{5.5cm}|}
\hline
\cellcolor{myblue} &\cellcolor{myblue} &\cellcolor{myblue} \tabularnewline
\cellcolor{myblue} \multirow{-2}{*} {\textbf{Major approaches}} &\cellcolor{myblue} \multirow{-2}{*} {\textbf{Advantages}} &\cellcolor{myblue} \multirow{-2}{*}{\textbf{Disadvantages}} \tabularnewline
\hline
\hline
 \cite{gui2014soar} &\begin{itemize} \item Support multiple bandwidth units \end{itemize} & \begin{itemize}  \item  Have high computational complexity \end{itemize}\tabularnewline \cline{2-3}
 \hline
\cite{shishapley}  & \begin{itemize} \item Achieve computational efficiency  \end{itemize}& \begin{itemize} \item Evaluate bids based on only a single attribute  \end{itemize} \tabularnewline \cline{2-3}
 \hline
 \cite{tan2014uniform}  & \begin{itemize} \item Support both bandwidth reservation and bandwidth allocation  \end{itemize} &\begin{itemize} \item Support a single bandwidth unit  \item Have uncertainty in future demands and resource utilization \end{itemize}\tabularnewline \cline{2-3}
  \hline
 \cite{guo2013cooperative} &  \begin{itemize} \item Achieve win-win solution \end{itemize}& \begin{itemize} \item Have slow convergence  \end{itemize}\tabularnewline \cline{2-3}
  \hline
\cite{divakaran2014bandwidth} & \begin{itemize} \item Adapt to the high flexible requests of users \end{itemize} & \begin{itemize} \item Not adapt time-guarantee requests \end{itemize} \tabularnewline \cline{2-3}
  \hline
\cite{wanisefficient}  &\begin{itemize} \item  Achieve high network utilization  \end{itemize}&\begin{itemize} \item Support only a single cloud provider \end{itemize} \tabularnewline \cline{2-3}
  \hline
\cite{wanis2015modeling}  &\begin{itemize} \item Adapt to the time-varying application's needs  \end{itemize} &\begin{itemize} \item Have high computational complexity \end{itemize} \tabularnewline \cline{2-3}
  \hline
\cite{zhao2014dynamic}  & \begin{itemize} \item Achieve stable
profit over time  \end{itemize}& \begin{itemize} \item Have high computational complexity  \end{itemize} \tabularnewline \cline{2-3}
  \hline
\cite{poola2014fault}  & \begin{itemize} \item  Achieve low computational complexity \item Guarantee workflow deadline  \end{itemize} & \begin{itemize} \item  Have unreliability in spot instances \end{itemize} \tabularnewline \cline{2-3}
  \hline
\cite{prasad2010resource}  & \begin{itemize} \item Be easy to be implemented \end{itemize}& \begin{itemize} \item Not use market factors \end{itemize} \tabularnewline \cline{2-3}
  \hline
\cite{xu2012general}  & \begin{itemize} \item Achieve win-win solution \end{itemize} &\begin{itemize} \item Have slow convergence \end{itemize}  \tabularnewline \cline{2-3}
  \hline
 \cite{aazam2014advance}  &\begin{itemize} \item Address the workload elasticity of users  \end{itemize} & \begin{itemize} \item Be challenging to determine of users' behavior  \end{itemize} \tabularnewline \cline{2-3}
  \hline
 \cite{zheng2015star}  & \begin{itemize} \item Support multiple cloud providers and multiple cloud tenants  \end{itemize}&\begin{itemize} \item Have unstable equilibrium \item Have slow convergence  \end{itemize} \tabularnewline \cline{2-3}
\hline
\end{tabular}
\par\end{centering}
\end{table*}

\section{Applications of economic and pricing models for resource management in mobile cloud networking}
\label{sec:mobile_cloud_networking}
As stated in Section \ref{sec:cloud_general}, the Mobile Cloud Networking (MCN) is structured by several segments including cloud computing infrastructures (i.e., data centers), Radio Access Network (RAN), cloud mobile core network, and mobile platform services \cite{jamakovic2013mobile}. Since the wireless environment in MCN is dynamic, distributed, and heterogeneous, traditional static methods are not possible to adapt to achieve optimal resource management \cite{gizelis2011survey}. Economic and pricing models have been recently employed to dynamically and efficiently manage resources, e.g., the bandwidth and energy, in MCN, which are reviewed in the following.


\subsection{Bandwidth Allocation}
\label{sec:mobile_cloud_networking_spectrum}
In this section, we review dynamic pricing schemes using auction mechanisms to address bandwidth allocation issues in MCN.
\subsubsection{Combinatorial clock auction}
\label{sec:mobile_cloud_networking_spectrum_clock_auction}
The authors in \cite{forde2011exclusive} studied the method to assign efficiently wireless bandwidth to cloud users in MCN. The cloud users typically reserve the bandwidth for both uplink and downlink transmission within a specific time period. Thus, the combinatorial clock auction which is described in Section \ref{subsec:Auction_theory} is applied. The model consists of a mobile cloud network owner which acts as a seller (auctioneer), and cloud users which act as buyers (bidders). The auctioneer owns a set of spectrum bands. Each band may have several bandwidth channels. The channels are auctioned within a specified time period and location. Bidders submit their package bids to the auctioneer. Each bid is a combination of channels in different bands, locations, and time periods and the corresponding price for which the bidder is willing to pay. The winner determination problem was formulated to identify the bids as winning or losing. The objective is to maximize the sum of the accepted bidding prices. The constraints include spectrum availability and duplex spacing between the uplink and downlink spectrum channels, i.e., paired spectrum channels. In particular, the duplex spacing constraint is to avoid the interference between the uplink and downlink channels. The spacing can be the amount of unpaired spectrum allocated to the bidder. The winner determination problem is NP-hard, and the anytime search algorithm \cite{sandholm2002algorithm} can be used which leads to a feasible solution at any time for the bandwidth allocation. However, the joint allocation of bandwidth and other network resources, e.g., base stations and backhaul links, needs to be investigated.

\subsubsection{Conventional auction}
\label{sec:mobile_cloud_networking_spectrum_Dutch_auction}
\begin{figure}[ht]
 \centering
\includegraphics[width=7.8 cm, height=9cm]{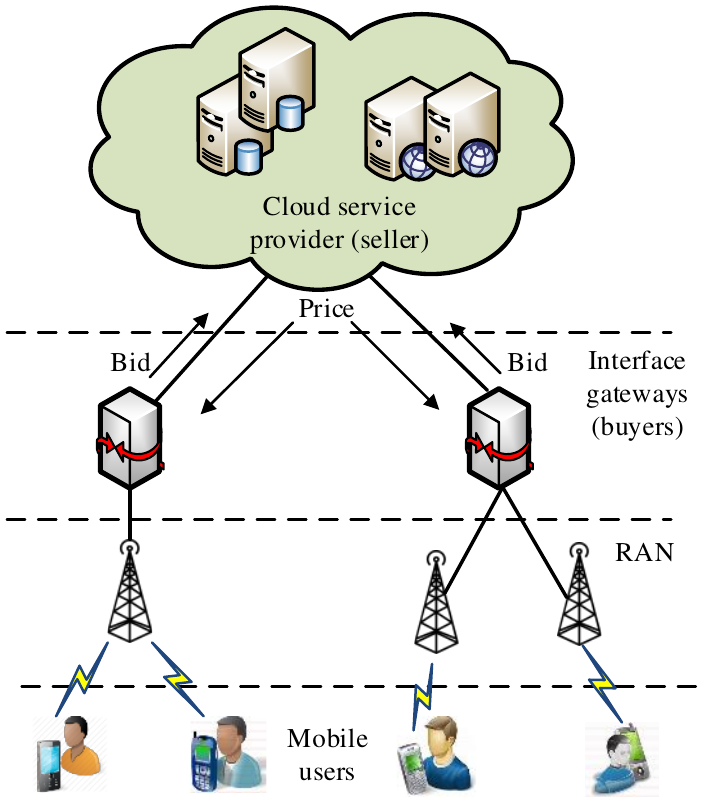}
 \caption{Bandwidth allocation based on auction in MCN.}
 \label{mobile_cloud_networking_conventional_auction}
\end{figure}

The auction mechanism as proposed in \cite{forde2011exclusive} requires a high communication overhead. Alternatively, the Dutch auction (see Section \ref{subsec:Auction_theory}) can be used for the bandwidth redistribution as proposed in \cite{misra2014qos}. The model is shown in Fig.~\ref{mobile_cloud_networking_conventional_auction} in which there are one cloud service provider, i.e., a seller, and multiple interfacing gateways, i.e., buyers. At a time slot, each gateway has a number of mobile users connected with it with the QoS guarantee. Initially, the service provider sets the ceiling price per unit of bandwidth and broadcasts the price to all gateways. Each gateway submits a bid to the service provider including the minimum required bandwidth which guarantees the QoS, i.e., the service delay, for its mobile users. The service provider compares its maximum bandwidth availability and the total demand of the gateways. If the available bandwidth is greater than or equal to the total demand, the service provider terminates the auction and allocates the bandwidth proportionally to the demands of gateways. Otherwise, the service provider decreases the price and broadcasts it in the next time slot. The auction continues until the total demand is greater than or equal to available bandwidth. Each gateway then pays the service provider the price of bandwidth based on the prepay mechanism as in \cite{wang2010spectrum}. In general, a gateway tries to maximize its utility which is the difference between the revenue that it receives from serving mobile users and the price that the gateway pays to the service provider for the allocated bandwidth. The utility is a concave function with respect to the gateway's bid. Using the second-order derivative of the utility function, it was proved that there exits a Nash equilibrium at which the gateways' utilities are maximized. However, the assumption that each gateway knows the bid value of others is not realistic. 

One shortcoming of the approach in \cite{misra2014qos} is that the service provider may redistribute bandwidth for all gateways even if only one gateway is overloaded by many users. The authors in \cite{dasquality} investigated sharing users among an overloaded gateway and its neighbors. The English auction is used. Initially, the overloaded gateway, i.e., the seller, broadcasts the information of the user's location to neighbors, i.e., buyers. If the user is within the region of a neighbor, this gateway sends the overloaded gateway an average QoS index which is determined based on its allocated bandwidth and the service delays of current users. The overloaded gateway also estimates a QoS index for the user based on the gateway's average QoS index and the information received from the service provider including the minimum delay threshold and the service delay. The overloaded gateway accepts the interested gateways as participants in the auction if the difference between its estimated index and the neighbors' average index is smaller than a threshold. At the initial state of the auction, the overloaded gateway broadcasts the minimum price for sharing the user. In each iteration, the gateway increases the price until its current utility value exceeds the initial value. The overloaded gateway allocates the user to the neighbor which accepts the price. The overloaded gateway then gets the payoff from the winning neighbor. However, the assumption about the information provided by the service provider for estimating the QoS index is not always possible.



\subsection{Resource management in Cloud-RAN}
\label{sec:mobile_cloud_networking_cloud_RAN}
\begin{figure}[ht]
 \centering
\includegraphics[width=\linewidth]{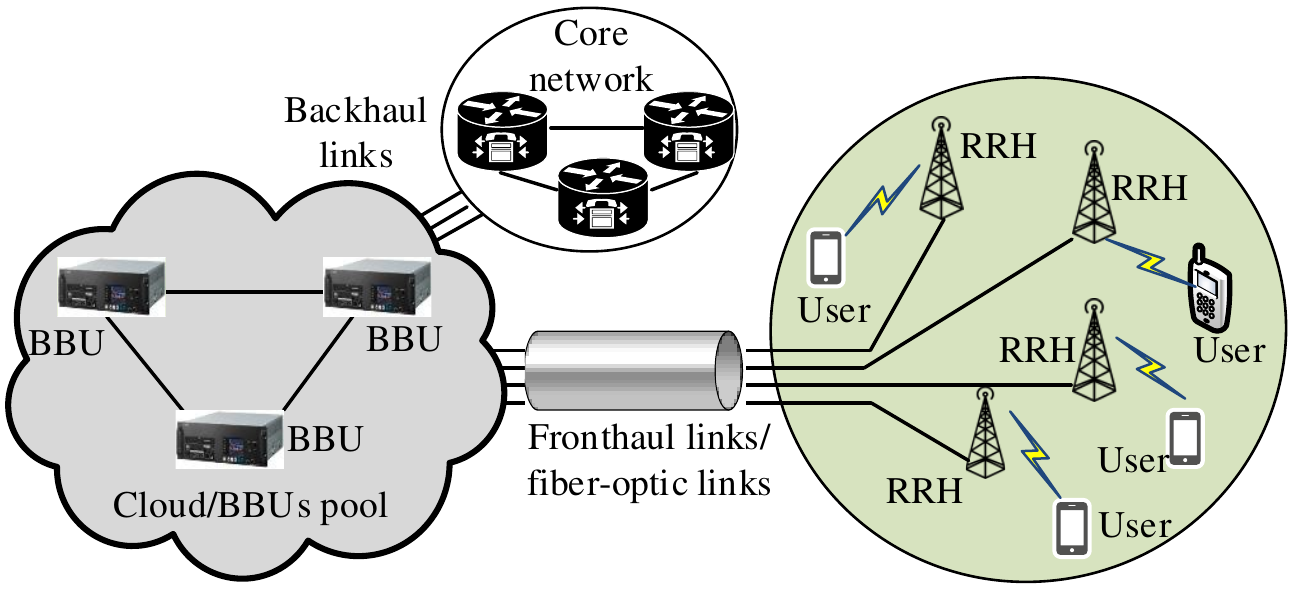}
 \caption{Cloud-RAN architecture.}
 \label{mobile_cloud_networking_CRAN}
\end{figure}

Cloud-Radio Access Network (Cloud-RAN) is a centralized, cloud computing-based architecture for radio access networks \cite{mobile2011c}. In Cloud-RANs, signal processing functions of a base station is performed in the cloud, i.e., centralized BaseBand processing Units (BBUs) or BBUs pool, as shown in Fig. \ref{mobile_cloud_networking_CRAN}. Then, the transmissions of radio signals to users are performed by Remote Radio Heads (RRHs) based on the baseband signals received from the cloud. To connect BBUs and RRHs, fronthaul links are used. One of the design goals in Cloud-RAN is to minimize the total downlink transmission power from RRHs to users while maintaining the fronthaul capacity and user QoS constraints.

 To achieve the goal, the authors in \cite{ha2014energy} studied two problems. The first problem aims to determine a   set of RRHs serving each mobile user and the precoding vectors for RRHs to minimize the total transmission power from RRHs to mobile users with the constraints on fronthaul capacity. The second problem is to minimize the total transmission power from RRHs to mobile users and total fronthaul capacity between BBUs and RRHs. The first problem is addressed by iteratively solving the second problem while adjusting pricing coefficients associated with RRHs. A pricing coefficient for each RRH refers to the price per unit of fronthaul capacity for the link between the cloud and RRH. The fronthaul capacity of the RRH is a decreasing function of its pricing coefficient. Solving the first problem is implemented via the binary searching method which adjusts the pricing coefficients so that the fronthaul capacities of RRHs are equal to their maximum allowable limits. In the second problem, the objective function is concave, and the feasible region corresponding to all the constraints is also convex. Thus, this problem can be solved by using the gradient method. Simulation results showed that the total transmission power is smaller when the maximum number of RRHs serving one user is larger. However, this also results in the high computational complexity.

 The model in \cite{ha2014energy} consists of a single cloud serving multiple users via RRHs. Multiple clouds can be used to satisfy the processing demands of users, called M-CRAN (MultiCloud RAN) \cite{dhifallah2015decentralized}. The authors in \cite{dahrouj2015distributed} addressed the issue of assigning users, i.e., buyers, to clouds, i.e., sellers, such that the overall net benefit of each cloud is maximized. Each cloud solves the knapsack problem \cite{chu1998genetic} the objective and constraint of which are the net benefit and the resource budget, respectively. The resource budget is defined as the maximum number of users that the cloud can serve. The cloud also pays its users penalty costs if the QoS service cannot be guaranteed. Therefore, the net benefit function of the cloud serving a user is the difference between the price that the user pays and the penalty cost. The optimization problem is NP-hard. However, a full polynomial time approximation scheme \cite{kellererknapsack} can be used to find an optimal set of users. The cloud can increase the penalty cost for attracting more users and iteratively perform the algorithm to maximize its overall net benefit. However, the high computational complexity will be incurred.

\begin{table*}
\caption{Applications of economic and pricing models for resource management in mobile cloud networking}
\label{table_mobile_cloud_networking}
\scriptsize
\begin{centering}
\begin{tabular}{|>{\centering\arraybackslash}m{0.2cm}|>{\centering\arraybackslash}m{0.4cm}|>{\centering\arraybackslash}m{1.6cm}|>{\centering\arraybackslash}m{1cm}|>{\centering\arraybackslash}m{1cm}|>{\centering\arraybackslash}m{1.2cm}|>{\centering\arraybackslash}m{5cm}|>{\centering\arraybackslash}m{2.4cm}|>{\centering\arraybackslash}m{1.2cm}|}
\hline
\multirow{2}{*} {\textbf{}} & \multirow{2}{*} {\textbf{Ref.}} & \multirow{2}{*} {\textbf{Pricing model}} & \multicolumn{3}{c|} {\textbf{Market structure}} & \multirow{2}{*} {\textbf{Mechanism}} & \multirow{2}{*} {\textbf{Objective}} & \multirow{2}{*} {\textbf{Solution}} \tabularnewline
\cline{4-6}
 & & & \textbf{Seller} & \textbf{Buyer} & \textbf{Item} & & &\tabularnewline
\hline
\hline
\parbox[t]{2mm}{\multirow{9}{*}{\rotatebox[origin=c]{90}{ \hspace{-2cm} Resource management}}}
& \cite{forde2011exclusive} & Combinatorial clock auction& Mobile cloud network owner&Cloud users&Bandwidth &Based on buyers' bids, the winner determination problem is solved by the anytime search algorithm&Efficient allocation&Optimal solution\tabularnewline \cline{2-9}

& \cite{misra2014qos} & Dutch auction&Cloud service provider&Gateways&Bandwidth &Based on buyers' bids, the seller compares its bandwidth availability and the total demand of buyers to either allocate resources or decrease the price such that the demand and the supply are equal &Buyers' utility maximization&Nash equilibrium\tabularnewline \cline{2-9}

& \cite{dasquality}& English auction&Overloaded gateway&Neighboring gateways&Sharing user &Seller initially broadcasts the minimum price for sharing user to buyers, and then increases the price until its current utility exceeds the initial value. The buyer which accepts this price is selected as the winner for serving the sharing users &QoS guarantee, and seller's utility maximization &Nash equilibrium\tabularnewline \cline{2-9}

& \cite{ha2014energy}& Generic pricing&Cloud/BBU&RRHs&Fronhaul capacity&Seller sets the price per unit of fronthaul capacity. The binary searching method is used to adjust the price to solve the problem which determines a set of RRHs serving each mobile user and the precoding vectors for RRHs&Buyers' power minimization, and optimal trade-off between transmission power and required fronthaul capacity &Optimal solution\tabularnewline \cline{2-9}

& \cite{dhifallah2015decentralized}& Knapsack problem&Clouds&Users&Service&Each seller solves the knapsack problem to select an optimal set of buyers which
maximize its overall net benefit. The problem is solved by the full polynomial time approximation method&Seller's benefit maximization, and QoS guarantee &Pareto efficiency\tabularnewline \cline{2-9}
\hline
\end{tabular}
\par\end{centering}
\end{table*}

\begin{table*}[h]
\caption{A summary of advantages and disadvantages of major approaches for the resource management in mobile cloud networking.}
\label{table_sum_advantage_mobile_cloud}
\scriptsize

\begin{centering}
\begin{tabular}{|>{\centering\arraybackslash}m{2cm}|>{\centering\arraybackslash}m{5.5cm}|>{\centering\arraybackslash}m{5.5cm}|}
\hline
\cellcolor{myblue} &\cellcolor{myblue} &\cellcolor{myblue} \tabularnewline
\cellcolor{myblue} \multirow{-2}{*} {\textbf{Major approaches}} &\cellcolor{myblue} \multirow{-2}{*} {\textbf{Advantages}} &\cellcolor{myblue} \multirow{-2}{*}{\textbf{Disadvantages}} \tabularnewline
\hline
\hline
\cite{forde2011exclusive} &\begin{itemize} \item Require little global information \item Achieve economic efficiency \end{itemize} & \begin{itemize}  \item Have high computational complexity \end{itemize}\tabularnewline \cline{2-3}
\hline
 \cite{misra2014qos}  &\begin{itemize} \item Have fast convergence \end{itemize} & \begin{itemize}  \item Require a centralized allocation algorithm \end{itemize}\tabularnewline \cline{2-3}
 \hline
 \cite{dhifallah2015decentralized}&\begin{itemize} \item Support multiple cloud providers \end{itemize} & \begin{itemize}  \item Have high computational complexity    \end{itemize}\tabularnewline \cline{2-3}
  \hline
\end{tabular}
\par\end{centering}
\end{table*}

\textbf{Summary:} In this section, we have reviewed the existing literature of pricing-based resource management approaches in MCN. These approaches with their references are summarized in Table~\ref{table_mobile_cloud_networking}. As seen, auction-based approaches are well suited for the bandwidth allocation in MCN. However, the pricing models developed for the resource management in Cloud-RAN are relatively few. Further research is required to extend the preliminary results reviewed in this section. In the following section, we review the existing economic and pricing models for resource allocation in edge computing which includes a variety of network technologies, i.e., the cloudlet, cloudIoT, social cloud, and self-organization cloud.

\section{Applications of economic and pricing models for resource management in edge computing}
\label{sec:social_edge_computing}
\begin{figure}[ht]
 \centering
\includegraphics[width=7.7 cm, height=7.7cm]{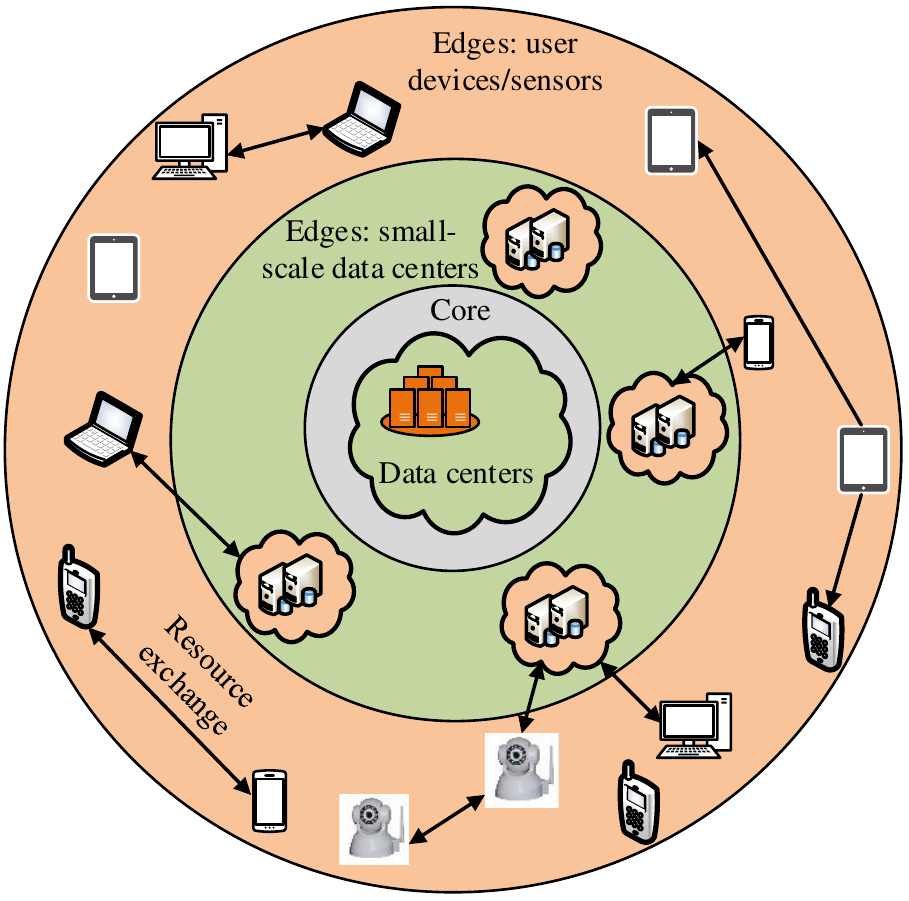}
 \caption{Edge-centric computing model.}
 \label{edge_computing_model}
\end{figure}

The cloud data center and mobile cloud networking are considered to be centralized paradigms, with storage and processing resources hosted within large data centers belonging to cloud providers. However, such paradigms face issues such as peak usage, high operational costs, bandwidth bottlenecks, and service interruption due to natural disasters (e.g., fire, earthquake, and power outage) \cite{greenberg2008cost}. Edge computing models that exploit distributed ``edge'' devices can solve the issues. As shown in Fig.~\ref{edge_computing_model}, edge devices can be small-scale data centers, volunteered computers, users devices (e.g., laptops, smartphones, and iPads) and sensors that are at the periphery of the network. Therefore, edge computing pushes the frontier of computing applications, data, and services away from the core of the data center network to the edges \cite{garcia2015edge}. To attract users to contribute their resources, incentive mechanisms using pricing and payment strategies have been adopted in order to guarantee the stable scale of participants and QoS. Thus, this section reviews economic and pricing models for the resource management in some common models of edge computing. 

\subsection{Cloudlet}
\label{sec:mobile_cloud_networking_IoT_Edge_cloudlet}
\begin{figure}[ht]
 \centering
\includegraphics[width=6.5 cm, height=8.3cm]{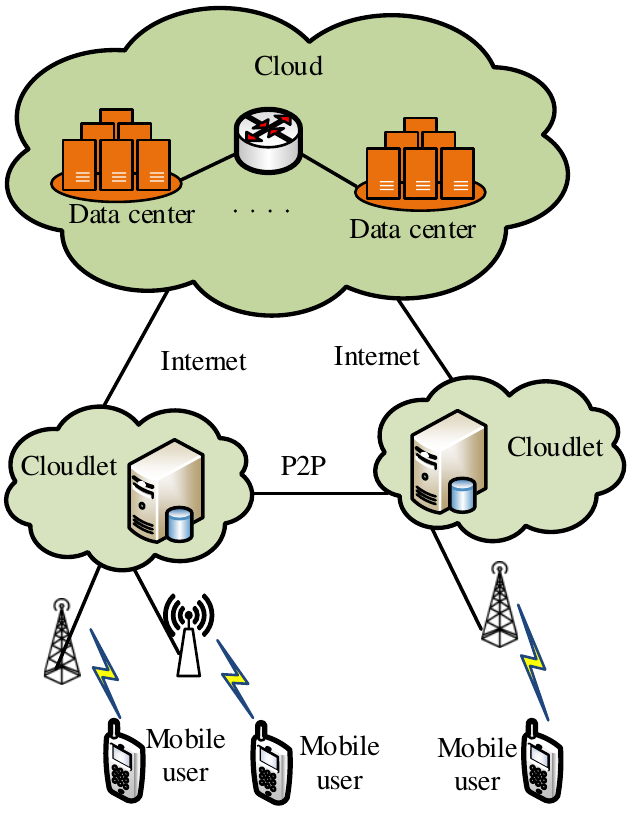}
 \caption{A 3-tier model with mobile devices, cloudlets, and cloud.}
 \label{cloudlet_architecture}
\end{figure}

Cloudlet, also known as mobile edge computing \cite{patel2014mobile} or mobile micro-cloud \cite{wang2015emulation}, is a mobility-enhanced small-scale cloud data center. It can be located at the edge of the network, e.g., at the base station to which the mobile users connect, with the aim of providing cloud services to mobile users at lower latency \cite{satyanarayanan2009case}. Thus, the cloudlet is considered to be the middle tier of a 3-tier model, i.e., the mobile device-cloudlet-cloud as shown in Fig.~\ref{cloudlet_architecture}. Resources at the cloudlet tier are limited \cite{kumar2010cloud}. Therefore, competition-based pricing models such as auction, non-cooperative game, or supply-demand model are effectively used for resource allocation to mobile users.

\subsubsection{Double auction}
\label{sec:mobile_cloud_networking_IoT_Edge_double_auction}
To motivate resource pooling, cloudlets can form cloudlet groups as illustrated in Fig.~\ref{cloudlet_architecture}. The authors in \cite{zhang2014community} adopted the real-time group-buying auction for the cloudlet group to offer its services, i.e., mobile videos, to nearby mobile users with lower prices while maximizing the profit of the group. The group-buying auction is a type of double auction in which buyers get more discounts from sellers if more buyers participate\cite{chen2002bidder}. The model consists of the cloud (i.e., the supplier) connected to the cloudlet group (i.e., the retailer) through the Internet and the mobile users (i.e., buyers). One of the cloudlets starts the auction with an initial auction price, a specified supply quantity, and an auction period. Users with bidding prices higher than the auction price are selected as successful bidders, i.e., winners. Potential bidders including unsuccessful bidders and new incoming users are sorted in a descending order of bidding prices. The cloudlet then finds the first bidder with the bidding price not less than the auction price in the price curve. The price curve is a non-increasing sequence of auction prices obtained via maximizing the expected profit function of the cloudlet. Potential bidders, the bidding prices of which are higher than that of the first bidder, are the winners. When the auction ends, all winners buy the services at the final deal price, which is lower when more winners are selected. This strategy stimulates more users to choose the service from the cloudlet group.

The double auction discussed in \cite{zhang2014community} can achieve the individual rationality and budget balance, but it cannot guarantee the truthfulness. The authors in \cite{Jin} addressed this problem by charging users according to the payment policy of the Vickrey auction. The model consists of: (i) mobile users, i.e., buyers, (ii) cloudlets, i.e., sellers, and (iii) a central controller, i.e., an auctioneer. The cloudlet serves only its nearby mobile users to reduce communication latency. The auctioneer sorts buyers in an ascending order of bids and sellers in a descending order of asks. The ask of the median seller is selected as a threshold to determine the winning buyer and seller candidates. For each winning seller candidate, the auctioneer selects a winning buyer with the highest price and charges it a price of the second highest bid. If the buyer wins two or more sellers, the auctioneer can select only one seller such that the buyer's utility is the highest. The simulation results showed that when a buyer bids a truthful price, its utility is improved. However, the system efficiency in terms of the number of final matchings between winning buyers and winning sellers only achieves around 50\% of that of the optimal strategy.


Using the same model as in \cite{Jin}, the authors in \cite{jinauction} considered the randomness and the uncertainty in the auction to improve the system efficiency. Specifically, the auctioneer sorts sellers randomly as a list. To determine a winning buyer for each seller, the auctioneer defines the ask vector excluding the ask of that seller and then calculates the median ask of this vector. Among buyers with the bids higher than the ask of the seller, the buyer with the highest bid wins the service of the seller. Then, the winning buyer and the seller are inserted to the sets of winning buyers and winning sellers, respectively. The clearing price charging to the winning buyer and the price paid to the seller are set to be the same. More specifically, the price is the maximum of the median ask and the second highest bid of all buyers for the seller. Since any winning buyer in the set of the winners does not compete with other buyers for the remaining sellers, the candidate elimination algorithm as in \cite{Jin} is not necessary, and the system efficiency is thus improved. The simulation results showed that the proposed solution achieves the system efficiency up to 80\% of that of the optimal strategy. However, the proposed solution cannot guarantee strong truthfulness for buyers.

\subsubsection{Non-cooperative game}
\label{sec:mobile_cloud_networking_IoT_Edge_non_cooperative}
\begin{figure}[ht]
 \centering
\includegraphics[width=7.3 cm, height=9.8cm]{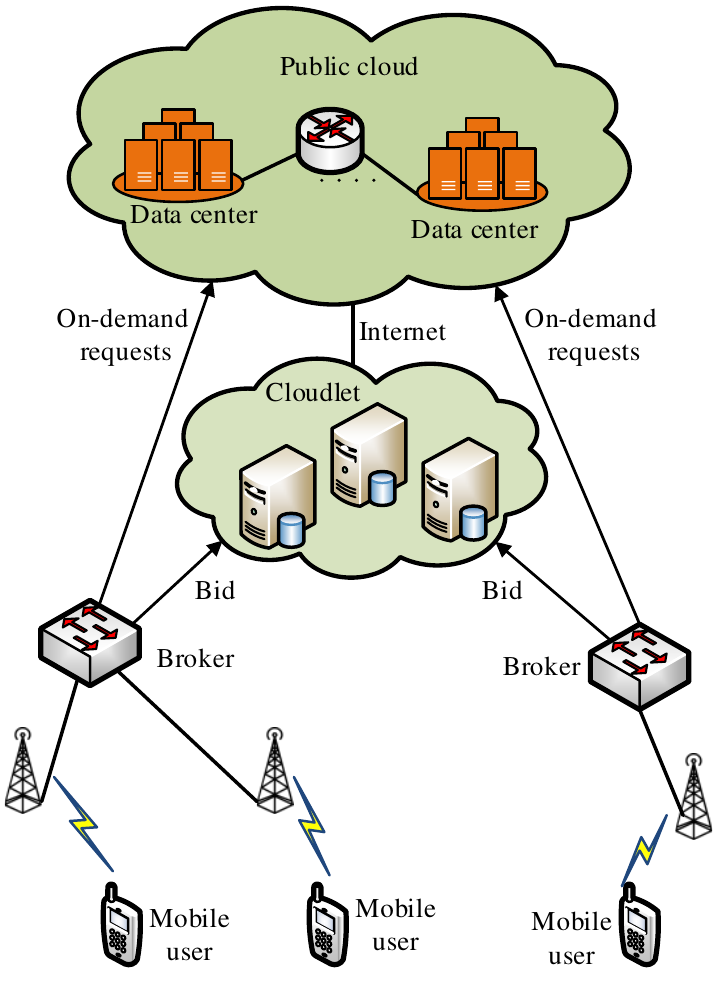}
 \caption{Resource allocation in presence of brokers.}
 \label{cloudlet_non_cooperative_game}
\end{figure}

The authors in \cite{guanvalue} considered a model with multiple brokers which assign cloud resources, i.e., the computation resource and network bandwidth, reserved from the cloudlet and the public cloud to mobile users as shown in Fig.~\ref{cloudlet_non_cooperative_game}. Long-term reservation and on-demand request are applicable at the public cloud, but the bid proportion policy should be implemented at the cloudlet due to its limited resources. The bid proportion policy \cite{teng2010resource} allocates resources to buyers proportionally to their purchasing prices. Each broker, i.e., each buyer, decides its bidding price and on-demand request such that its average cloud price is minimized, given other brokers' strategies. The average price is a convex function of the bidding price, and the brokers are selfish to minimize the brokers' average prices. Therefore, the non-cooperative game is used to determine their optimal decisions. The Jacobi best-response algorithm \cite{scutari2014decomposition} is then adopted to iteratively achieve an approximation of the Nash equilibrium at which bidding prices of all brokers are optimal. Simulation results showed that the proposed solution can reduce the price around 23\% compared with the case that mobile users submit their requests to the public cloud directly. However, information sharing schemes among brokers to enforce the truthfulness need to be developed.
\subsubsection{Supply and demand model}
\label{sec:Edge_computing_cloudlet_supply_demand}
The above models are for static environments in which users are not on the move. When they are moving, pricing strategies relying on central entities, e.g., the auction, are not appropriate. In such a system, the authors in \cite{wu2015cloudlet} addressed the issue of exchanging cloudlet resources including CPU cycles, storage, and broadband network, among mobile users through the supply and demand model. Each user owns a cloudlet and acts as a buyer and a seller during its mobility. Given the budget, the objective of each user is to maximize the individual payoff which is the difference between the utility of buying resources and the cost of selling resources. This is a convex problem and can be solved by the primal-dual algorithm \cite{williamson2002primal}. The solution allows the user to decide an optimal amount of resources to sell. To clear the resource market while maximizing each user's utility, an aggregate excess demand function is introduced. This function is the difference between the total demand and the total supply over all users in network. The classical tatonnement process \cite{uzawa1960walras} is used for the price adjustment to achieve the market equilibrium where the total demand equals the total supply. In particular, if the total demand exceeds the total supply, the seller increases the unit resource. Otherwise, the unit price should be reduced. The process continues until the function equals zero. However, the equilibrium may not be stable, and the computational cost for converging to the equilibrium can be high.

\begin{figure}[ht]
 \centering
\includegraphics[width=6.2 cm, height=9.3cm]{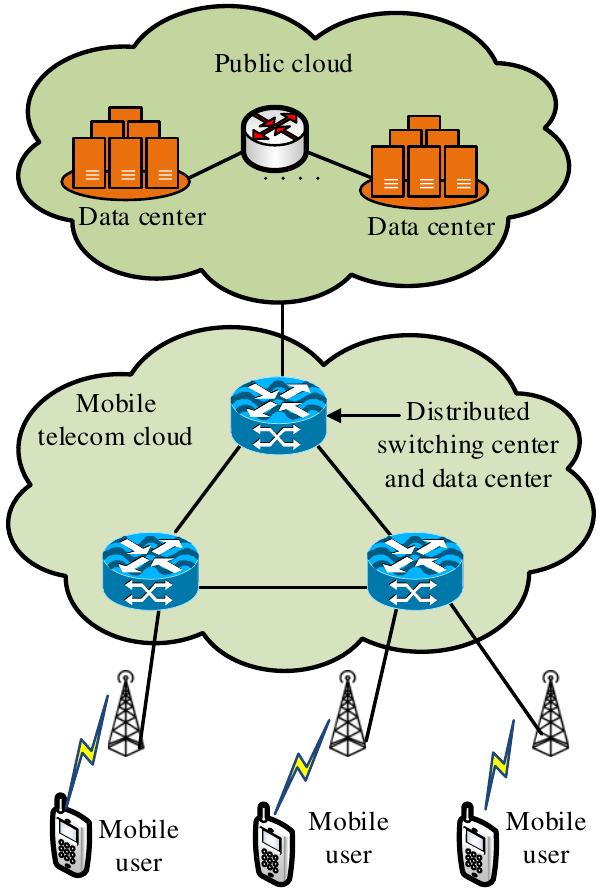}
 \caption{Mobile telecom cloud model.}
 \label{cloudlet_MTC_cost_minimization}
\end{figure}

Apart from the aforementioned cloudlet models, a similar model, called Mobile Telecom Cloud (MTC), is found in \cite{vaezpour2015mobile}. Edge cloud services are given by mobile network operators which provide the last-mile Internet access to mobile users as shown Fig.~\ref{cloudlet_MTC_cost_minimization}. The network operators act as brokerages which use discount from cloud providers, e.g., Amazon, to offer better and cheaper cloud services to their users. Specifically, when receiving users' cloud requests, the brokerage formulates the resource reservation as the total cost minimization problem. The total cost depends on the users' cloud requests, costs of cloud services offered by the brokerage or cloud providers, and the discount threshold from the cloud providers. This problem is then solved by either the linear programming combined with the rounding technique \cite{raghavan1987randomized} or the min-cost greedy. The optimal solution allows the brokerage to set its price range using two conditions: (i) its offer price is less than the proposed price of the cloud provider, and (ii) the total cost of the brokerage is less than the sum of charged prices to users. These conditions aim to guarantee a high profit for the brokerage while attracting more users. The simulation results indicated that the cost of brokerage with the min-cost greedy algorithm is smaller than that obtained from with the linear programming. Moreover, the min-cost greedy algorithm runs much faster than the linear programming with rounding which requires multiple iterations to converge.

\subsection{Volunteer Computing Systems}
\label{sec:Volunteer_system}
\begin{figure}[ht]
 \centering
\includegraphics[width=\linewidth]{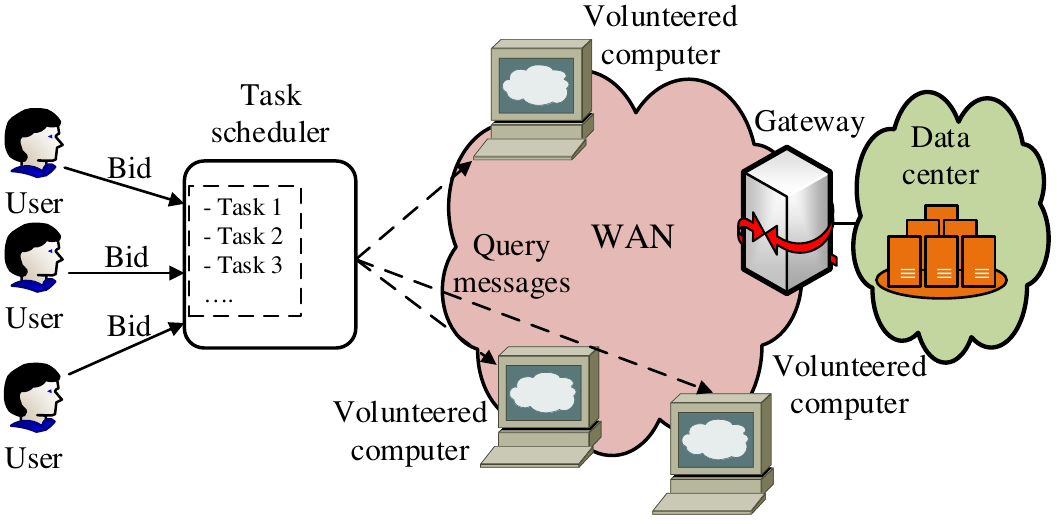}
 \caption{Task scheduling in volunteer computing system.}
 \label{Volunteer_computing}
\end{figure}

This section reviews economic approaches for resource allocation in volunteer computing systems. Similar to the cloudlet model, volunteer computing systems allow computer owners to contribute their computing resources for processing users' tasks. However, instead of using small-scale data centers as in the cloudlet model, a large number of distributed volunteered computers are used. They are connected with each other over WANs as illustrated in Fig.~\ref{Volunteer_computing}. Some projects are recently designed based on the framework, e.g.,~BOINC~\cite{anderson2004boinc} with 65,000 computers, and~Cloud@Home (http://clouds.gforge.inria.fr/pmwiki.php). In such a network, any volunteered computer, also called a host or a node, can act as a task scheduler and a resource contributor to schedule and execute users' tasks.

\subsubsection{Reverse Vickrey auction}
\label{sec:Volunteer_system_reverse}
Like the cloudlet model, due to the limited resource at the volunteered computers, the reverse Vickrey auction is usually applied for allocating users' tasks to the volunteered computers in volunteer computing systems. Such an approach was proposed in \cite{di2011social}. More specifically, when receiving a task request from a user, the task scheduler sends a query message including the expected qualified resource demand of the task, e.g., the CPU, memory, and network bandwidth, to its neighbors using the distributed range query protocol \cite{sheu2007distributed}. This searching continues until the number of hops is greater than a time-to-live threshold. Resource nodes (sellers) reply the task scheduler (buyer) the response messages (asks) containing the resource nodes' identifiers (e.g., IP address), resource availability states, and the corresponding prices. The resource node with the lowest price is selected for executing the user's task, and the payment for this node is implemented according to the Vickrey auction policy. Compared with the random-node-selection approach, the simulation results indicated that the proposed approach improves the resource node's payoff up to five times. However, the finished task ratio of the proposed approach is slightly lower than that of the random selection approach. A multi-attribute reverse auction can be used to improve the performance.

\subsubsection{Profit maximization}
\label{sec:Volunteer_system_profit}
Since the task scheduler acts as a broker to map task requirements to the resources of nodes, the profit of the broker needs to be considered. The authors in \cite{mukherjeerisc} employed the profit maximization-based pricing to determine the costs paid for the volunteered computer owners and the resource prices offered to the users to maximize the broker's profit. Each user submits to the scheduler a task execution request including the required resources (i.e., the computation and network resources), the number of time slots, the deadline, and a desired success probability of the reservation in the case of resource failures. The scheduler determines the probability distributions of price acceptance for users and the probability distributions of cost acceptance by the owners. These distributions can be learned in an online fashion. Based on these distributions, the broker defines its profit by computing the total revenue minus the total cost. The optimization problem is then solved by a sequential optimization algorithm in which each price and cost for a request type is optimized sequentially. Simulation results showed that the total profit of the broker at the low demand is significantly higher than that at the high demand. The reason is that even at high demand, the broker cannot increase the price offered to users to ensure that they still accept the prices.


\subsubsection{Demand-based pricing}
\label{sec:Volunteer_system_demand}

The approaches in \cite{di2011social}, \cite{mukherjeerisc} may not satisfy some users which are willing to pay proportionally to the QoS. Hence, the authors in \cite{dilip2014priority} adopted the demand-based pricing for allocating users' tasks to resource nodes. The demand-based pricing charges users according to their demand. This pricing strategy is practically adopted by some cloud service providers, e.g., CloudTweaks (http://cloudtweaks.com/), to maximize the resource utilization and guarantee low costs. Specifically, given users' resource requirements and nodes' budgets for executing tasks, the node sets the resource price based on the total resource demand of tasks across the network. If the total demand is less than 50\% of the total capacity of the node, the price is charged according to the base price. Otherwise, the higher price is applied. Also, the node uses the $k$-means clustering algorithm to classify the tasks into high, medium, and low priority levels. The node checks the resource requirement of the task with the highest priority. If the node has available resource to serve the task and the price for executing the task is within the budget constraint, the task is given with the preference for execution in the node. Compared with the first-come first-served approach, the proposed approach improved significantly the throughput ratio while reducing the average payment of users. However, this may lead to revenue loss of resource owners.

\subsubsection{VCG auction}
\label{sec:Volunteer_system_VCG}
In practice, users require not only resources within an internal system, but also external resources, e.g, data center resources. They need to access the external resources via a gateway of a bandwidth provider \cite{danniswara2015stream}. The authors in \cite{khantowards} considered allocating bandwidth to users so as to maximize the provider's revenue. Additionally, it must ensure the social welfare maximization for the users even if they can lie about their priority to get higher utility. The VCG auction is used to achieve this goal. Users, i.e., buyers, submit to a provider, i.e., a seller, their bids, each of which specifies the priority class, the bandwidth demand, the valuation, and the price. Given these information, the provider defines each user's utility. The provider finds an optimal schedule to maximize the sum of utilities of all users through the use of the greedy approach. Each user is then charged according to the VCG payment policy. The simulation results highlighted that when varying the probability of user lying, the social welfare of the VCG auction is always significantly higher than that of the first-price and the Vickrey auctions.

\subsection{Client-Assisted Cloud Systems}
\label{sec:social_P2P_networking_social_Client_assisted}
Client-assisted cloud models are distributed cloud paradigms which form resource pooling by exploiting resources of clients \cite{garcia2015edge}. Here, a client or user is referred to as an ``edge'' device belonging to the external network environment of the edge-centric computing as shown in Fig.~\ref{edge_computing_model}. Such paradigms aim to reduce the network traffic and resource burden at servers in volunteer computing systems, cloudlets, and data centers \cite{chu2013user}. In particular, in what follows, we review economic approaches which have been used to incentivize users/clients to contribute their local resources in the different distributed cloud models.

\subsubsection{Client-assisted cloud storage system}
\label{sec:social_P2P_networking_social_SOC_storage}
\begin{figure}[ht]
 \centering
\includegraphics[width=7.8 cm, height=7.5cm]{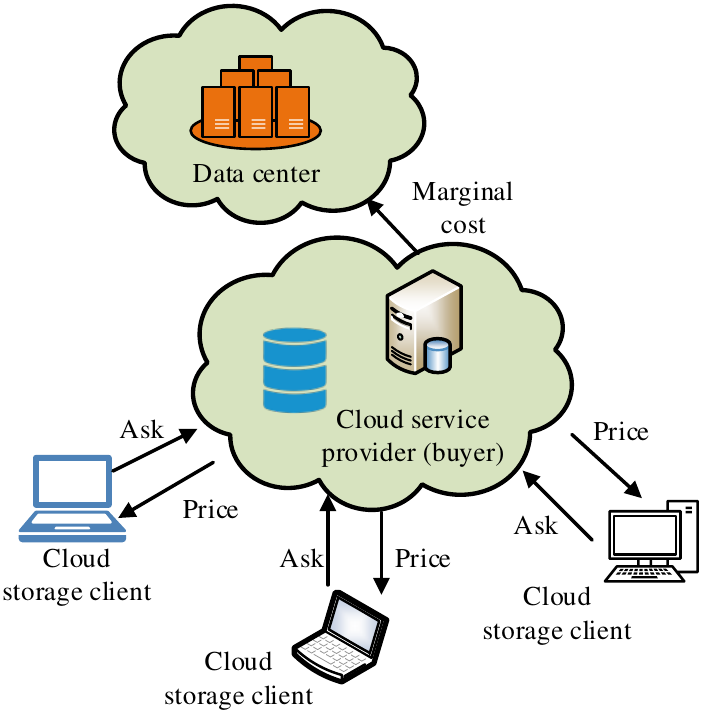}
 \caption{Client-assisted cloud storage allocation using online reverse auction.}
 \label{client_assisted_cloud_storage_online_auction}
\end{figure}

The authors in \cite{zhao2015online} addressed the issue of constructing a storage pool for storage service providers, e.g., Amazon S3, using under-untilized storage and network bandwidth resources of cloud users as illustrated in Fig.~\ref{client_assisted_cloud_storage_online_auction}. Due to asynchronous arrivals of users and service provider \cite{davoli2014triton}, the online reverse auction can be applied. Note that the auction is commonly used for the online e-commerce, e.g., eBay (http://www.ebay.com/rpp/live-auctions). Accordingly, users, i.e., sellers, submit their asks to the service provider, i.e., the buyer. Each ask contains information about the amount of resources that the user can contribute, the time window when it is available, and money remuneration. Upon receiving the asks, the service provider determines a completeness ratio, which is the ratio of the total resource from users and its resource demand. If the ratio is less than one, the service provider uses the storage and bandwidth from servers in data centers. The resource pooling cost is thus the sum of the payments to the users plus the marginal resource cost from the servers. The optimization problem for the service provider determines the allocation rule and payment to minimize the resource pooling cost, given the constraints ensuring the individual rationality of users and the truthfulness of the mechanism. To achieve the goals, the allocation rule and payment are designed according to a marginal pricing function, which is a non-increasing function of the completeness ratio. It was then proved that the allocation rule is monotone, and the online auction scheme is truthful. Simulation results indicated that the social cost in the online auction is always less than that of the offline VCG auction when varying the resource pooling demand, the number of asks, and the ratio between the server cost and the average asking price.

The client-assisted cloud model as in \cite{zhao2015online} still requires the communication among service providers and users. Therefore, when another user needs to use the resource, it makes a request to the service provider which results in latency. Moreover, to guarantee the availability of resources, the interaction between demand users and supply users needs to be considered. Distributed cloud models such as self-organization cloud and social cloud can be adopted as discussed below.

\subsubsection{Self-organization cloud}
\label{sec:social_P2P_networking_social_SOC_CNC}
Self-organization cloud allows a number of host machines of users to be connected by a P2P overlay network on the Internet \cite{di2013dynamic}. Since each user may act as a resource provider or a resource requester, resource exchange between them is typically modeled by the double auction. Such an approach was presented in \cite{wu2016scalable} where the double auction is adopted for the task allocation among users. The model consists of request users, i.e., buyers, which require resources, i.e., computational resources and network bandwidth, for executing their tasks from provision users. Provision users act as sellers to contribute their resources for the task execution from the request. Before submitting bids, buyers have the rough estimation about the price of the required resources by using a price-setting mechanism, e.g., SpotCloud (http://www.spotcloud.com). The buyers then submit to a cloud planner, i.e., an auctioneer, their own bids including task descriptions, resource specifications, and the prices that they estimate. The cloud planner selects the buyer with the highest price as the winner. Then, the cloud planner sends the request of the winner to all sellers in the network. Interested sellers return the cloud planner by their asks. The seller with the lowest price is selected to provide resources to the buying winner. The lowest price is also the payment of the buying winner. However, how to discover users with available resource in the network was not given.

\begin{figure}[ht]
 \centering
\includegraphics[width=6 cm, height=5.5cm]{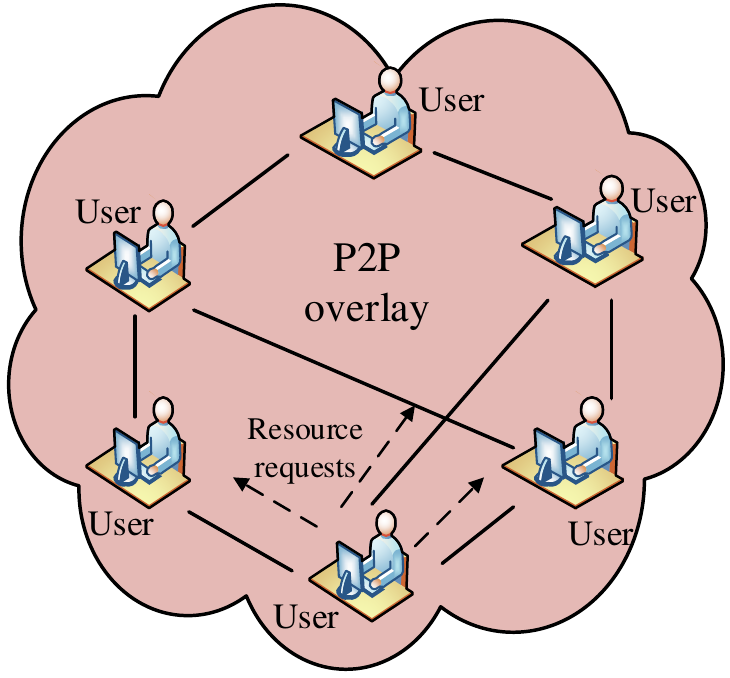}
 \caption{A general system of self-organization cloud.}
 \label{self_organization_cloud}
\end{figure}

The authors in \cite{khethavath2013introducing}, \cite{praveen2014game} proposed to use the kademlia protocol \cite{maymounkov2002kademlia} which allows a buyer to perform a \textit{resource discovery} process to find candidate sellers with available resource in the network before the reverse auction is executed. Indeed, the candidate sellers can submit to the buyer their asks, each of which includes the information about available resources (i.e., CPU, memory, and bandwidth), QoS (i.e., a latency), a participation factor, and an incentive value. The participation factor is defined based on the historical resource contribution of a seller. The buyer calculates its utility value corresponding to each ask, taking into account weights assigned to each component in the ask. The buyer selects a seller whose ask enables it to achieve the highest utility as the winner. The winner receives the incentive value representing the resources that it will receive in future. Such a non-money reward (along with coupons \cite{albers2013coupons} or reputation score \cite{li2015reputation}) reduces the incentive costs for the buyer. This cost is significantly low as the number of sellers increases as shown in the simulation results. The resource allocation is thus fair and achieves the system stability. However, the reverse auction cannot be applied when the resource is insufficient.

\subsubsection{Social Cloud}
\label{sec:social_P2P_networking_social}
A social cloud is \textit{``a resource and service sharing framework utilizing relationships established between members of a social network''} \cite{chard2012social}. This model is thus similar to the self-organization cloud. However, if users in the self-organization cloud are anonymous and are not accountable for their actions, then accountability can be established through existing friend relationships in the social cloud.

\begin{figure}[ht]
 \centering
\includegraphics[width=\linewidth]{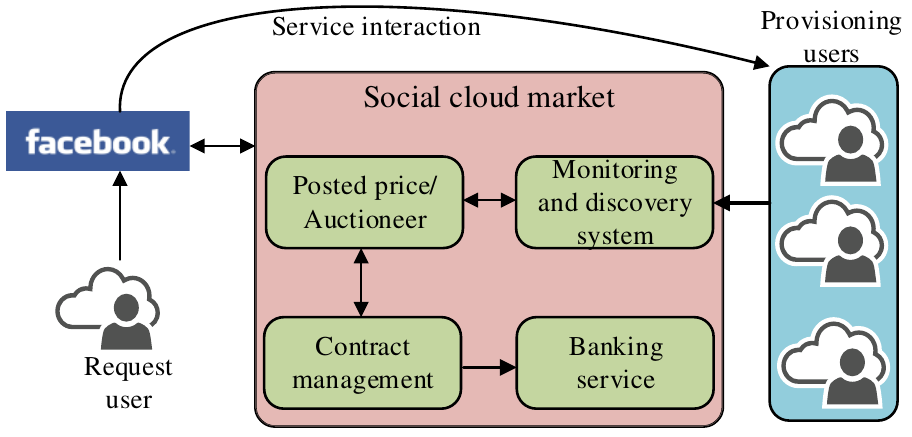}
 \caption{Social cloud market infrastructure.}
 \label{social_cloud}
\end{figure}

The social marketplace is at the core of the social cloud which is similar to online procurement markets. Provision users in the social network arrive in a sequential manner to offer their services. Thus the posted-price mechanism is usually used. The authors in \cite{chard2010social} adopted the posted-price model for storage service marketplaces in the social cloud as shown in Fig.~\ref{social_cloud}. To enable the accountability, information of users such as user ID and credit balance, is managed by a bank. When each request user, i.e., a buyer, requests posted price offers for specific services, the cloud application checks the available balance of the buyer. The cloud application gives a list of provision users, i.e., sellers, their resource availability and the pricing information. These information is periodically updated and stored in a monitoring and discovery system. When the request user selects a service from a provision user, the cloud application creates SLA. If both parties accept the agreement, it is then passed to the bank for transferring credits between them. However, the provision users can lie about service costs to gain higher payoffs.

Therefore, the reverse Vickrey auction can be used as proposed in \cite{chard2012social}. Upon receiving the storage service description of the request, an auction manager, i.e., an auctioneer, solicits asks from provision users within the social network. The asking prices are determined based on pre-defined metrics, and a seller with the lowest asking price is selected as the winner. Similar to \cite{chard2010social}, SLA between the buyer and the winner is created, and the credit transfer is done through a bank. The credit is determined according to the payment policy of the reverse Vickrey auction which prevents misreports of provision users. The auction-based model is generally more complex than the posted-price model due to soliciting asks from sellers. To be appropriate in the social cloud environment with a large number of bidders, the auction manager needs to execute a number of concurrent auctions. As shown in the simulation results, a small scale auction manager may complete 65 concurrent auctions each with 50 bidders per minute.

To adapt to more requirements of users, the multi-attribute reverse Vickrey auction can be used. Multi-attribute auctions allow buyers and sellers to negotiate multiple attributes in addition to the price such as service quality and service deadline. The authors in \cite{khan2014service} adopted such an auction for the task allocation among users in the social network. The request user initially sends its task description including a set of task attributes and the corresponding weights to a nearby auctioneer. The auctioneer broadcasts the request in the publishing area which may be described by the friends' community reachable through, e.g., a web-portal. Upon receiving the task request, potential provision users which meet the minimum attribute requirements submit their asks to the auctioneer. The auctioneer evaluates the utility score of each ask based on the available attributes of the ask and the request user's weights. The provision user with the highest utility score will be selected as the winner. The auctioneer then creates SLA involving the service price, the availability of resources, and the agreed attribute values between the winner and the buyer. The winner will get the price, which is increased by the amount, such that the utility of the transaction equals the utility score of the second-highest ask. However, how to define the request user's weights is not specified.
\subsubsection{Cloud-Centric Internet of Things}
\label{sec:mobile_cloud_networking_IoT_incentive}
This subsection considers an edge computing model in which the edge devices are sensing devices, e.g., smartphones of users, as shown in Fig.~\ref{edge_computing_model}. Since these devices are equipped with various types of sensors, they are capable of generating sensing data. A large number of smartphones allow to gather sensing data from interest areas, and such a paradigm is called a crowdsensing network \cite{guo2014participatory}. To provide the data to interested individuals/organizations, the crowdsensing networks and the cloud platforms are integrated. Since the crowdsensing network is also a main component of IoT \cite{sundmaeker2010vision}, the integration can be called Cloud-centric Internet of Things (Cloud-IoT) \cite{kantarci2014mobility} or Cloud of Things \cite{aazam2014cloud}. The main challenge in such models is to provide incentive mechanisms to users such that they perform sensing tasks and provide sensing data with the lowest cost. Typically, the reverse auction mechanisms are used.

The authors in \cite{kantarci2014mobility} employed the reverse auction scheme to assign required tasks to phone users. The model has three entities: (i) cloud users which act as buyers, (ii) phone users, i.e., sellers, and (iii) a sensing server, i.e., an auctioneer. Once receiving a set of sensing tasks from the cloud users, the sensing server broadcasts the task requests to all phone users in a specific area. Interested phone users reply with asks, and the server sorts these users based on their marginal value contributions. The marginal value contribution of a user is defined as an additional value introduced to the set by including the sensing task of the user \cite{kantarci2014trustworthy}. Users with the highest marginal value contributions are selected for the tasks. The payments for the selected users are not less than their asking prices and determined based on their marginal value contributions to the set of sensing tasks.

In practice, mobility of phone users can degrade the utility of the platform since the users can move out a service region. The lightweight triangulation method \cite{kim2006extracting} can be adopted for the server to keep track of the mobility pattern of each phone user. The utility of the platform also decreases due to the malicious phone users aiming at sending altered sensing data to the cloud users. Reputation of users is introduced to overcome this issue as proposed in \cite{kantarci2014trustworthyconf} and \cite{kantarci2014reputation}. The model and the reverse auction mechanism used in these approaches are similar to those in \cite{kantarci2014mobility}. However, upon receiving sensing data from each phone user, the server runs an outlier detection (anomaly detection) algorithm \cite{zhang2010outlier} to classify the sensing data as an outlier or a non-outlier. Since the trustworthiness of a user can be time-varying, the user's instantaneous reputation is stored and updated in the database of the cloud platform. The current reputation of a user is determined as the function of the number of its non-outliers and outliers. When selecting winners, the server uses the ratio of its bidding price and current reputation to evaluate the bid. This selection process is to increase the selection probability of a phone user with a lower price and higher reputation, i.e., a smaller ratio. The payments for the winners are then applied similar to those in \cite{kantarci2014mobility}. The simulation results showed that with the proposed approach, the payments of malicious phone users can be reduced around 55\% compared with those in \cite{kantarci2014mobility}.

\begin{table*}
\caption{Applications of economic and pricing models for resource management in edge computing}
\label{table_edge_computing_1}
\scriptsize
\begin{centering}
\begin{tabular}{|>{\centering\arraybackslash}m{0.2cm}|>{\centering\arraybackslash}m{0.4cm}|>{\centering\arraybackslash}m{1.6cm}|>{\centering\arraybackslash}m{1.2cm}|>{\centering\arraybackslash}m{1cm}|>{\centering\arraybackslash}m{1.2cm}|>{\centering\arraybackslash}m{5cm}|>{\centering\arraybackslash}m{2.4cm}|>{\centering\arraybackslash}m{1.2cm}|}
\hline
\multirow{2}{*} {\textbf{}} & \multirow{2}{*} {\textbf{Ref.}} & \multirow{2}{*} {\textbf{Pricing model}} & \multicolumn{3}{c|} {\textbf{Market structure}} & \multirow{2}{*} {\textbf{Mechanism}} & \multirow{2}{*} {\textbf{Objective}} & \multirow{2}{*} {\textbf{Solution}} \tabularnewline
\cline{4-6}
 & & & \textbf{Seller} & \textbf{Buyer} & \textbf{Item} & & &\tabularnewline
\hline
\hline
\parbox[t]{2mm}{\multirow{9}{*}{\rotatebox[origin=c]{90}{ \hspace{-6cm} Cloudlet}}}
& \cite{zhang2014community} & Real-time group-buying auction& Cloudlet&Mobile users&Mobile videos &Seller forms a price curve of auction prices, and buyers with bidding prices higher than the auction price are the winners. The final deal price is the final auction price & Payment minimization, and expected profit maximization&Subgame perfect
equilibrium \tabularnewline \cline{2-9}

&\cite{Jin}& Double auction& Cloudlets&Mobile users& Processing, storage, and networking&Based on sellers' asks, the auctioneer determines buyer and seller winning candidates. The buyer candidate with the highest price is the winner and is charged with a price of the second highest bid & Individual rationality, budget balance, and truthfulness&Market equilibrium\tabularnewline \cline{2-9}

&\cite{jinauction} & Double auction& Cloudlets&Mobile users& Processing, storage, and networking&Winning buyer for each seller is determined based on sellers' asks. The clearing price charged to the winning buyer and the price paid to the winning seller are set at the same price&Individual rationality, budget balance, truthfulness, and system efficiency&Market equilibrium\tabularnewline \cline{2-9}

&\cite{guanvalue}& Non-cooperative game& Cloudlet&Brokers&Cloud resources&The Jacobi best-response algorithm is used to optimize the bidding prices of buyers& Cost minimization for mobile users&Nash equilibrium\tabularnewline \cline{2-9}

& \cite{wu2015cloudlet}& Supply and demand model& Mobile user&Mobile users&Cloudlet servers &Seller uses an aggregate excess demand function to define the total demand and total supply. The classical tatonnement process is used to adjust the price depending on the resource demand to clear the market resource& Payoff maximization, and resource efficiency&Market equilibrium\tabularnewline \cline{2-9}

& \cite{vaezpour2015mobile}&Cost minimization& MTC brokerage&Mobile users&Cloud services &Given buyers' requests, the seller formulates the resource reservation as the total cost minimization. The linear programming and the min-cost greedy are used to solve the problem & Seller' profit maximization, and buyers' payment minimization&Optimal solution\tabularnewline \cline{2-9}
\hline
\parbox[t]{2mm}{\multirow{9}{*}{\rotatebox[origin=c]{90}{ \hspace{-2cm} Volunteer computing system}}}
& \cite{di2011social} & Reverse Vickrey auction& Volunteered computers&Task scheduler&CPU, memory, and network bandwidth &Buyer searches sellers with available resources via the distributed range query protocol. The buyer selects the seller with the lowest price as the winner, and the payment is based on the Vickrey auction policy& Payoff improvement, and truthfulness&Nash
equilibrium \tabularnewline \cline{2-9}

& \cite{mukherjeerisc} & Profit maximization& Volunteered computer owner&Users&CPU, memory, and network bandwidth &A broker determines the costs paid for the seller and the prices offered to the buyers to maximize the broker's profit. A sequential optimization is then introduced to solve this problem& Profit maximization&Optimal solution \tabularnewline \cline{2-9}

&\cite{dilip2014priority} &Demand-based pricing& Volunteered computer&Users&Task execution service&Seller assigns priority levels to tasks based on the $k$-means clustering algorithm and then executes tasks with the highest priority levels. The price is set based on the total resource demand of buyers& Throughput ration improvement, and payment reduction &Value
optimization \tabularnewline \cline{2-9}

&\cite{khantowards} &VCG auction& Bandwidth provider&Users&Bandwidth&Given buyers' requests, the seller finds an optimal schedule to maximize the sum of utilities of all buyers by using the greedy approach. The price is then set according to the VCG payment policy& Social welfare maximization & Nash equilibrium \tabularnewline \cline{2-9}
\hline
\end{tabular}
\par\end{centering}
\end{table*}

\begin{table*}
\caption{Applications of economic and pricing models for resource management in edge computing (cont.)}
\label{table_edge_computing_2}
\scriptsize
\begin{centering}
\begin{tabular}{|>{\centering\arraybackslash}m{0.2cm}|>{\centering\arraybackslash}m{0.4cm}|>{\centering\arraybackslash}m{1.6cm}|>{\centering\arraybackslash}m{1.2cm}|>{\centering\arraybackslash}m{1cm}|>{\centering\arraybackslash}m{1.2cm}|>{\centering\arraybackslash}m{5cm}|>{\centering\arraybackslash}m{2.4cm}|>{\centering\arraybackslash}m{1.2cm}|}
\hline
\multirow{2}{*} {\textbf{}} & \multirow{2}{*} {\textbf{Ref.}} & \multirow{2}{*} {\textbf{Pricing model}} & \multicolumn{3}{c|} {\textbf{Market structure}} & \multirow{2}{*} {\textbf{Mechanism}} & \multirow{2}{*} {\textbf{Objective}} & \multirow{2}{*} {\textbf{Solution}} \tabularnewline
\cline{4-6}
 & & & \textbf{Seller} & \textbf{Buyer} & \textbf{Item} & & &\tabularnewline
\hline
\hline
\parbox[t]{2mm}{\multirow{9}{*}{\rotatebox[origin=c]{90}{ \hspace{-6cm} Client-assisted cloud systems}}}

& \cite{zhao2015online} &Online reverse auction& Cloud storage users &Storage service provider&Storage service&Based on sellers' asks, buyer uses a marginal pricing function to determine the allocation rule and the payment to minimize the resource pooling cost& Social cost minimization, truthfulness, and individual rationality & Nash equilibrium \tabularnewline \cline{2-9}

& \cite{wu2016scalable} &Double auction& Provision users &Request users&Resources&Given buyers' bids, a planner selects a buyer with the highest price and then finds a seller with the lowest price to provide resources to the buyer & Resource efficiency & Market equilibrium \tabularnewline \cline{2-9}

& \cite{khethavath2013introducing} \cite{praveen2014game} &Reverse auction& Provision users &Request user&Resources&Based on sellers' asks, the buyer calculates the corresponding utility values. The seller whose ask maximizes the buyer' utility is selected as the winner. The winner then gets an incentive value & System stability, and low incentive cost & Nash equilibrium \tabularnewline \cline{2-9}

& \cite{chard2010social} &Posted-price& Provision users &Request user&Storage service&Based on a list of sellers along with their posted price offers, the buyer selects a seller and the cloud application creates an SLA. The buyer then pays the seller through the bank &Utility maximization for buyer & Nash equilibrium \tabularnewline \cline{2-9}

& \cite{chard2012social} &Reverse Vickrey auction& Provision users &Request user&Storage service&Given sellers' asks, the auction manager selects a seller with the lowest price as the winner. The payment follows the payment policy of the reverse Vicrkey auction &Truthfulness, and payment minimization& Nash equilibrium \tabularnewline \cline{2-9}

& \cite{khan2014service} &Multi-attribute reverse Vickrey auction& Provision users &Request user&Task execution service&The auctioneer evaluates the utility scores corresponding to sellers' asks. The seller with the highest utility score is selected as the winner. The payment is determined based on the payment policy of the reverse Vicrkey auction &Utility maximization of buyer, truthfulness& Nash equilibrium \tabularnewline \cline{2-9}

& \cite{kantarci2014mobility} &Reverse auction& Phone users&Cloud users&Sensing task&The server selects the sellers with the highest marginal contributions as the winners. Payments for the winners are determined based on their marginal contributions &Incentive cost minimization& Nash equilibrium \tabularnewline \cline{2-9}

& \cite{kantarci2014trustworthyconf} \cite{kantarci2014reputation} &Reverse auction& Phone users&Cloud users&Sensing task&Same as \cite{kantarci2014mobility}, but for each seller, the server defines a ratio of its asking price and reputation. The reputation is determined using the outlier detection algorithm. Sellers with smaller ratios are selected as the winners &Utility maximization, and incentive cost reduction& Nash equilibrium \tabularnewline \cline{2-9}

\hline
\end{tabular}
\par\end{centering}
\end{table*}

\begin{table*}[h]
\caption{A summary of advantages and disadvantages of major approaches for the resource management in edge computing.}
\label{table_sum_advantage_edge_computing}
\scriptsize

\begin{centering}
\begin{tabular}{|>{\centering\arraybackslash}m{2cm}|>{\centering\arraybackslash}m{5.5cm}|>{\centering\arraybackslash}m{5.5cm}|}
\hline
\cellcolor{myblue} &\cellcolor{myblue} &\cellcolor{myblue} \tabularnewline
\cellcolor{myblue} \multirow{-2}{*} {\textbf{Major approaches}} &\cellcolor{myblue} \multirow{-2}{*} {\textbf{Advantages}} &\cellcolor{myblue} \multirow{-2}{*}{\textbf{Disadvantages}} \tabularnewline
\hline
\hline
\cite{zhang2014community} &\begin{itemize} \item Achieve good economic properties \end{itemize} & \begin{itemize}  \item Have slow convergence \item Have unstable equilibrium  \end{itemize}\tabularnewline \cline{2-3}
\hline
\cite{jinauction}  &\begin{itemize} \item Reduce communication latency  \item Improve the system efficiency \end{itemize} & \begin{itemize}  \item Have slow convergence \item Have unstable equilibrium  \end{itemize}\tabularnewline \cline{2-3}
 \hline
\cite{guanvalue}&\begin{itemize} \item Support both long-term reservation and on-demand request  \end{itemize} & \begin{itemize}  \item Require buyers to submit their requests simultaneously \item Have unstable equilibrium       \end{itemize}\tabularnewline \cline{2-3}
 \hline
 \cite{mukherjeerisc}&\begin{itemize} \item Guarantee reliability to buyers  \end{itemize} & \begin{itemize}  \item Be challenging to learn probabilities of price acceptance for buyers \item Have high computational complexity \end{itemize}\tabularnewline \cline{2-3}
 \hline
 \cite{zhao2015online}&\begin{itemize} \item Support asynchronous arrivals of both buyers and sellers \end{itemize} & \begin{itemize}  \item  Require high communication   \end{itemize}\tabularnewline \cline{2-3}
  \hline
\cite{chard2010social}&\begin{itemize} \item Support anonymous buyers \item Achieve high reliability and resource availability \end{itemize} & \begin{itemize}  \item Require frequently monitoring and discovering the resource availability and prices of sellers\end{itemize}\tabularnewline \cline{2-3}
  \hline
\cite{kantarci2014mobility}&\begin{itemize} \item Support the mobility of sellers \end{itemize} & \begin{itemize}  \item Require frequently calculating and updating instantaneous reputation of sellers \end{itemize}\tabularnewline \cline{2-3}
  \hline
\end{tabular}
\par\end{centering}
\end{table*}

\textbf{Summary:} In this section, we have discussed four major models of the edge computing, and for each model we have reviewed the related economic and pricing approaches. We summarize the approaches along with references in Table~\ref{table_edge_computing_1} and Table~\ref{table_edge_computing_2}. From the two tables, we observe that more client-assisted cloud systems have been recently studied. This is reasonable because the distributed cloud models reduce costs and service latency for users. In the next section, we review economic and pricing approaches for the resource management in cloud-based Video on Demand (VoD) systems. Cloud-based VoD systems are new video content delivery models in the development of cloud networking. 
\section{Applications of economic and pricing models for resource management in cloud-based VoD systems}
\label{sec:VoD_system}
This section describes and reviews the related work of cloud-based Video on Demand (cloud-based VoD), which is one important service supported by the cloud networking \cite{Murray_2012_tech}. VoD, e.g., Internet Protocol TeleVision (IPTV) \cite{newslog2006iptv}, is a system that enables users/clients to select and watch video contents whenever they want instead of watching at a specific broadcast time \cite{hua1998patching}. Traditional VoD services are based on the client-server or P2P architectures which have a major drawback of high costs and low bandwidth utilization, especially with the large and imbalanced demands of users. Therefore, to reduce the costs and improve resource utilization, VoD providers, i.e., cloud tenants, can use cloud platforms from cloud providers to form the cloud-based VoD systems. The cloud-based VoD system can also provide flexibility to support users with various requirements \cite{wu2011cloudmedia}. However, the bandwidth cost is still significant in cloud-based VoD systems since video contents typically consume a large amount of bandwidth. The ultimate goal of VoD providers is to minimize the bandwidth cost and maximize their profit while satisfying users' demand, and thus pricing strategies are applied. In particular, the economic and pricing models have been used to address the following issues.

\begin{itemize}
\item{\textit{Bandwidth allocation}: In cloud-based VoD systems, VoD providers store their video contents in the cloud servers belonging to cloud providers. Therefore, the VoD providers need to reserve bandwidth which allows them to upload the video contents to the cloud servers as well as to guarantee the access for their users. Economic and pricing models have been used as solutions in which the bandwidth reservation cost is minimized while still satisfying users' demand.}
\item{\textit{P2P caching}: To reduce the reservation cost from the cloud servers, the VoD providers may use local resources, e.g., the storage and upload bandwidth, of peers or users to cache video data. Since the users or peers are naturally selfish, pricing strategies have been adopted to incentivize the users to contribute their resources while minimizing the cost.}
\end{itemize}

\subsection{Cloud-Based VoD Models}
\label{sec:VoD_system_model}

This section presents economic and pricing approaches for bandwidth allocation in Cloud-based VoD systems. A major issue in such models is to allocate the bandwidth owned by cloud providers to VoD providers for delivering video contents. Depending on the specific scenario, a pricing model is applied to address the issue. In particular, if the model consists of a VoD provider and multiple cloud providers or multiple VoD providers and a cloud provider, competition-based pricing schemes such as auctions and non-cooperative game will be used. On the contrary, if the objective of the resource allocation is to maximize social welfare of all VoD providers, the network utility maximization can be adopted.
\subsubsection{Combinatorial auction}
\label{sec:VoD_system_combinatorial_auction}
\begin{figure}[ht]
 \centering
\includegraphics[width=\linewidth]{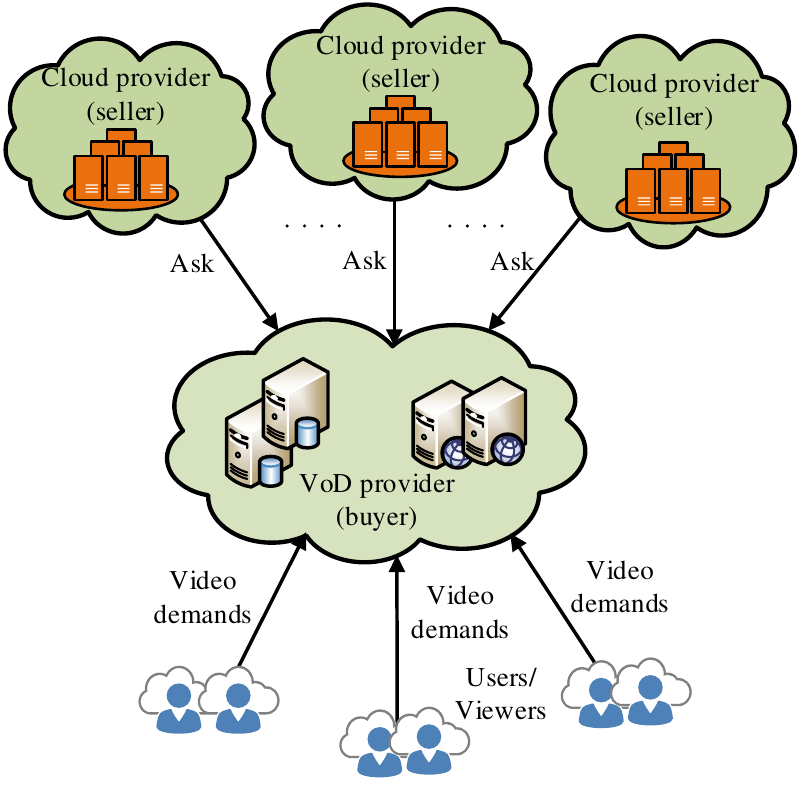}
 \caption{Bandwidth allocation in cloud-VoD system based on combinatorial auction.}
 \label{cloud_VoD_combinatorial_auction}
\end{figure}

The authors in \cite{cong2014lbas} considered the cloud-based VoD system with a VoD provider and several cloud providers as shown in Fig. \ref{cloud_VoD_combinatorial_auction}. The VoD provider delivers/allocates groups of videos from its local servers to the cloud providers. Since there are several combinations of videos for trading, the combinatorial auctions are adopted. The VoD provider as a buyer classifies videos into groups, and each group consists of the same user demands. The VoD provider also evaluates price for each group, which is generally a decreasing function of user demands. Then, the VoD provider sends the requests to the cloud providers. Depending on available bandwidth and memory, each cloud provider, i.e., a seller, determines the number of groups and the number of videos in each group that it can serve. Cloud providers respond to the VoD provider with their asks including the information related to the number of groups and videos along with the corresponding prices. The VoD provider computes a \textit{distance} value associated with each ask. The distance value is the price evaluated by the VoD provider minus the price offered by the cloud provider. The VoD provider selects a cloud provider with the largest distance value as the winner to allocate its video groups. To guarantee the truthfulness of the auction, the payment policy from the Vickrey auction is adopted. The Approximate Efficiency Maximization (AEM) algorithm \cite{wang2012cloud} is used to avoid the collusion among cloud providers. The simulation results showed that the proposed approach saves up to 10\% of the cost compared to the video migration strategy in \cite{li2011cost}. However, multiple VoD providers need to be considered in the future work.

\subsubsection{Generic pricing mechanism}
\label{sec:VoD_system_generic}
The authors in \cite{niu2012theory} extended the market model in \cite{cong2014lbas} involving multiple public cloud providers, i.e., sellers, VoD providers (cloud tenants), i.e., buyers, and a broker. In this setting, the broker reserves the actual bandwidth from the cloud providers and sells \textit{probabilistic bandwidth guarantee services} to tenants. The broker determines the lower and upper bounds of the price to maximize its profit and minimize the bandwidth reservation cost. The broker then determines a \textit{load direction matrix} which directs tenants' bandwidth demands to the cloud providers. To find the lower bound of the price, the broker defines its profit which is the sum of prices offered to tenants minus the total cost for reserving bandwidth from the cloud providers. The price offered to the tenant is a concave function of its demand, and the demand is assumed to follow the Gaussian distribution\cite{niu2012quality}. The lower bound of the price is determined by using the gradient ascent algorithm for the profit maximization. For the upper bound of the price, the broker needs to set the price lower than that offered by the cloud providers to attract more tenants. Therefore, the upper bound of the price is actually the cloud providers' pricing scheme \cite{niu2012theory_tech}. The simulation results showed that the broker can save the bandwidth reservation cost by more than 30\% on average compared with the case that each tenant reserves bandwidth individually.

\subsubsection{Non-cooperative game}
\label{sec:VoD_system_non_cooperative}
For the market model in \cite{niu2012theory}, the pricing scheme of the broker is affected by the cloud providers, and such a market is considered to be a \textit{controlled market}. The authors in \cite{niu2012pricing} considered a \textit{free market} which only consists of tenants, i.e., VoD providers and a broker. The tenants competes with each other for the cloud bandwidth by submitting their pricing strategy to the broker. The interactions among selfish tenants are modeled as a non-cooperative game in which the strategy of each tenant is to set the price so as to maximize its own utility. The utility is inversely proportional to the price that the tenant pays. Using the Cauchy-Schwarz inequality and the proof by contradiction, it was shown analytically that if the broker decides the \textit{load direction matrix} to maximize its profit as mentioned in \cite{niu2012theory}, the tenants' prices will converge to a unique Nash equilibrium. This equilibrium still holds even if multiple brokers exist in the market since the game is played by the tenants. However, the competition among brokers may lead to a zero profit of brokers.

\subsubsection{Utility maximization}
\label{sec:VoD_system_utility_maximization}

Unlike the above market models, the authors in \cite{niuasynchronous} and \cite{niu2013efficient} considered multiple VoD providers (cloud tenants), which reserve the egress network bandwidth, i.e., the upload speed, from a data center of a cloud provider to guarantee delivering smoothly videos to their users. The objective is to find the resource allocation that maximizes the tenants' social welfare. The social welfare is the sum of tenants' utilities minus the total cost at the cloud provider, and thus the utility maximization problem as presented in Section \ref{subsec:Utility_maximization} can be applied. In particular, the tenant utility is a strictly concave and monotonically increasing function of the allocated resource while the total cost is a strictly convex function that is also monotonically increasing with the allocated resource. Such concavity of the utility function is to guarantee that the optimization problem has a unique optimal solution. Since the utility function may not be known by the cloud provider, and the total cost function may not be known to the tenants, the centralized algorithms, e.g., the Newton's method \cite{bertsekas2003convex}, cannot be applied. The authors in \cite{niuasynchronous} and \cite{niu2013efficient} adopted two distributed iterative algorithms based on price updates.

In \cite{niuasynchronous}, at each iteration, the cloud provider updates the price charged to each tenant in the next iteration using the first-order partial derivative of the total cost with respect to the tenant's resource allocation and the price at the current iteration. Given the price at the current iteration, each tenant determines its resource allocation so as to maximize its utility minus the cost of using resources. Unlike \cite{niuasynchronous}, the price in \cite{niu2013efficient} is updated by the tenants while the resource allocation is updated by the cloud provider. At each iteration, each tenant sets the price in the next iteration based on the first-order derivative of its utility function and the price at the current iteration. In both \cite{niuasynchronous} and \cite{niu2013efficient}, it was proved that if the price and the resource allocation reach fixed points, then the resource allocation will be the optimal solution of the optimization problem.

In general, compared with the algorithms such as the Newton's method and the Alternating Direction Method of Multiplier (ADMM)\cite{boyd2011distributed}, the proposed approaches minimize the message passing overhead. This is because the approaches use only their local information, i.e., the cost and utility functions, and feedback update variables. Moreover, the approaches remove the assumption on the cooperativeness of entities in the market. For example, in \cite{niuasynchronous} once the cloud provider sets a price vector, the tenants will find their own resource allocation to maximize their utilities, which are natural strategies of selfish tenants. This property is in contrast to the ADMM \cite{boyd2011distributed} which requires more complicated variable updates. The simulation results showed that the average number of iterations needed to converge of the proposed approach in \cite{niuasynchronous} is 10 while those of the primal gradient descent algorithm and the gradient descent-based consistency pricing method \cite{tan2006distributed} are 100 and 50, respectively. However, a more general market with multiple cloud providers needs to be investigated in the future work.

\subsection{P2P-Assisted Cloud-Based VoD Models}
\label{sec:VoD_system_P2P}
\begin{figure}[ht]
 \centering
\includegraphics[width=\linewidth]{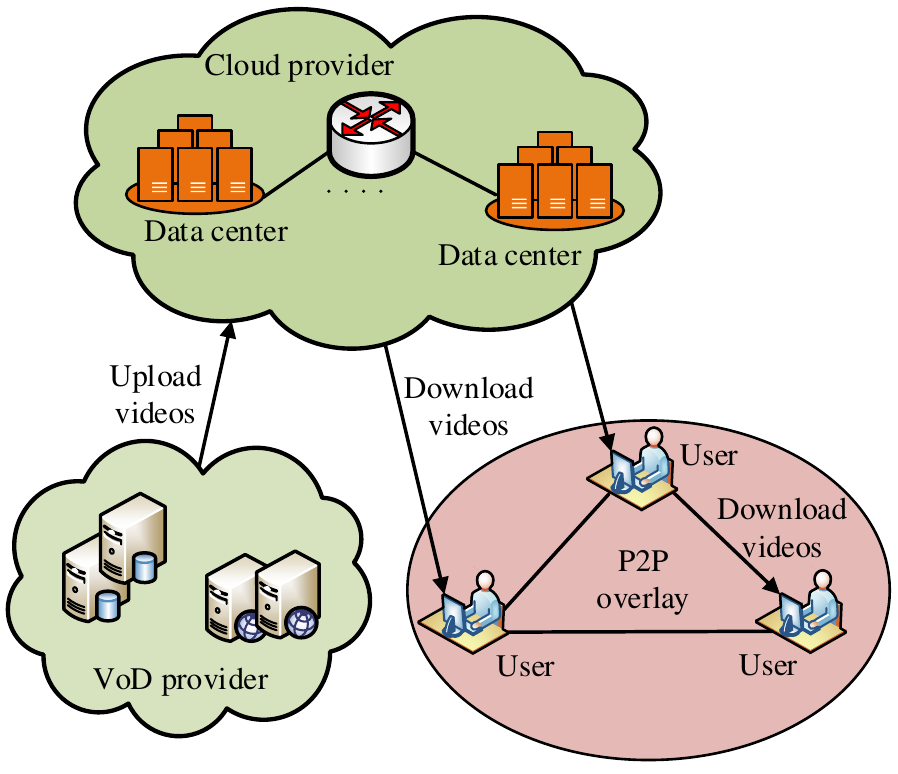}
 \caption{Overview of a P2P-assisted cloud-based VoD
model.}
 \label{P2P_assisted_cloud_based_VoD}
\end{figure}

As stated earlier, an important challenge in cloud-based VoD systems is the bandwidth cost on the servers in the cloud. For example, YouTube is estimated to spend \$470 million a year while Facebook spends \$500,000 a month on bandwidth (http://www.slate.com/articles/technology/technology/2009/). To improve scalability and save bandwidth costs at the cloud, the cloud-based VoD systems can be combined with P2P networks as shown in Fig.~\ref{P2P_assisted_cloud_based_VoD}. Peers, i.e., users, can download video contents from both the cloud and other peers in the P2P network. However, this requires the VoD provider to provide the peers with an incentive for (i) downloading videos from other peers rather than the cloud and (ii) caching the video contents by using their resources, e.g., the memory and upload bandwidth. Pricing models were introduced to achieve these goals as discussed in what follows.

\subsubsection{Stackelberg game}
\label{sec:VoD_system_P2P_stackelberg}
Typically, to reduce the access of users to servers in the cloud, the VoD provider can apply smart pricing strategies, e.g., the usage-based pricing, to the bandwidth consumption. To optimize both utilities of the VoD provider and users, the Stackelberg game can be adopted as proposed in \cite{lin2015autotune}. First, the VoD provider which acts as the leader estimates the cloud bandwidth usage of users by using the exponentially weighted moving average method \cite{lucas1990exponentially}. Then, it sets the VoD service price proportionally to the estimated cloud bandwidth usage. Given the price, users, i.e., followers, select bit rates to maximize their utility. The utility is the difference between the satisfaction degree and the price that the user pays to the VoD provider. The satisfaction degree is a concave function of the bit rate. Upon receiving users' responses, the VoD provider sets the price so as to maximize its revenue. Based on the new price, each user recalculates the optimal bit rate that maximizes its utility. The simulation results showed that the cloud bandwidth consumption of the proposed approach is significantly lower than that of the server-side bit rate adaptation approach in \cite{mansy2011analysis}. The reason is that the proposed approach encourages users to download video chunks from peers through the usage-based pricing. However, how the users determine their optimal bit rates was not specified.

\subsubsection{Double auction}
\label{sec:VoD_system_double_auction}
Another solution which reduces cloud bandwidth consumption is to share the surplus upload bandwidth among peers, called the prefetching strategy \cite{huang2008challenges}. As mentioned in Section \ref{sec:social_P2P_networking_social_SOC_CNC}, the resource sharing among peers is typically implemented in a decentralized manner since any peer may be a buyer or a seller. Thus, the resource sharing can be modeled as the double auction-based market as proposed in \cite{feng2010peer}. The market is divided into several sub-markets, each of which only trades one video segment. When a peer acts as a buyer, it broadcasts a bid to its neighbors. A seller compares its ask with the bid. If the ask is less than the bid, the seller will propose the buyer to trade the segment at a transaction price, which is the average of the ask and the bid. Since the transaction price is less than the bid and larger than the ask, the proposed approach is the ex-post individual rationality, i.e., the expected utility of participants is non-negative. This incentivizes peers to exchange video segments with each other rather than to download from the server. The simulation results showed that the proposed approach outperforms the near-sequential prefetching strategy \cite{choe2007improving} and the popularity-based prefetching strategy \cite{he2009optimal} in terms of server bandwidth costs. However, since the buyer only receives a video segment in a sub-market, it may need to participate in other sub-markets to satisfy its demand. This increases the latency of delivering the video.

\subsubsection{Generic pricing mechanisms for the P2P caching}
\label{sec:VoD_Caching}
In practice, peers may not be satisfied due to the lack of some video contents in the market. One possible reason is that the other peers may clear their local storage after finishing watching the videos. The authors in \cite{wu2012incentivizing} designed a reward price-based incentive mechanism for peers to cache the video replicas to satisfy all peers' demand. Videos in the system are categorized into different classes, each consisting of videos of similar popularity associated with a price. The VoD provider determines these prices such that the number of supplied video replicas equals the demand. First, it defines the \textit{refreshing probability} of each peer, i.e., the probability that the peer clears its local storage. Second, it defines the \textit{storage probability} which can be considered to be the \textit{popularity} of a video. Since each peer decides whether to cache a video based on the price of the video, the price of each video is proportional to the refreshing probability, the desired number of replicas of the video in the system, and inversely proportional to the storage probability. For example, videos which are more popular have higher prices to incentivize more peers to cache them so as to meet the greedy cache requirement \cite{wu2012exploring}.

In reality, the VoD provider always considers its profit when setting prices for videos. The authors in \cite{wu2014distributed} addressed this problem by investigating \textit{strategic pricing} to minimize the VoD provider's operation cost. The VoD provider calculates the operation cost involving the upload cost to the cloud and the reward price paid to peers. The upload cost is proportional to the difference amount between the desired videos and the supplied videos. The reward price is the total cost paid to all peers due to their video contributions. The operation cost is a continuous and convex function of the reward cost. Thus, given a budget for reward prices, it was proved that there exists a unique solution of the reward price to the cost minimization problem. The simulation results revealed that compared with the approaches without using any incentive scheme, the proposed approach reduces significantly the operation cost, especially when the upload cost of servers increases. In practice, there is a fraction of peers which may not be sensitive to the reward prices. Thus, the cost may be reduced further if the VoD provider can learn the real sensitivity of the peers.

\subsubsection{Pricing models for resource allocation in cloud-assisted P2P streaming systems}
\label{sec:VoD_system_P2P_video_streaming}
 This section reviews a few pricing approaches to provide incentive to peers in cloud-assisted P2P streaming systems. Cloud-assisted P2P streaming systems are similar to the P2P-assisted cloud-based VoD models \cite{su2010incentive}. However, in the cloud-assisted P2P streaming systems, there is no specific VoD provider, and thus stakeholders involve only peers and streaming servers in the cloud.

As presented in \cite{chakareski2015cost}, the cloud as a buyer rents resources including network bandwidth, storage space, and CPU, from peers as sellers, namely \textit{helpers}, to contribute video contents to its customers. The cloud offers the service price, and then the peers decide their resource contributions. Thus, the Stackelberg game can be used. Each peer, i.e., a follower, computes its payoff which is the difference between the price offered by the cloud, i.e., the leader, and the cost incurred to offer the resource. The strategy of the peers is to find the amount of resources to maximize their own payoffs. The optimal amount can be determined by using the first-order derivative. Given the total amount of the resources from peers, the cloud determines the offered price to maximize its revenue.

The approach in \cite{chakareski2015cost} did not consider the budget of the cloud. The authors in \cite{mostafavi2015game} adopted the Stackelberg game to analyze the upload bandwidth sharing between helpers and a streaming server resided in the cloud, taking into account the budget of the server. In the first stage, the server, i.e., the leader, announces its budget. The second stage can be considered to be a non-cooperative game among selfish helpers, i.e., followers, which decide the number of bandwidth units to maximize their own utilities. The utility is the difference between the reward that the helper receives and its video sharing cost. Generally, the utility is a strictly concave function of the number of bandwidth units. By using the second derivative, it was proved that there exists a unique Nash equilibrium involving optimal strategies of helpers. Given the helpers' strategies, the server determines the budget to maximize its utility, which is the gain from the total transmitted video minus the reward paid to the helpers. In particular, the gain is characterized by the two-parameter rate-distortion model which expresses the trade-off between the allocated bandwidth and the video content distortion. The utility is a strictly concave function of the budget. It was also proved that there is a unique budget value which maximizes the utility.

The server can determine an optimal value of budget only if it has full information about the helpers' utility functions. In real scenarios, this assumption may not be valid since the helpers autonomously decide their upload bandwidth contributions. Thus, the reverse auction can be used. The server, i.e., the buyer, broadcasts a vector of bandwidth demands to helpers, i.e., sellers. Interested helpers submit to the server their asks including the number of bandwidth units and the prices that they are willing to pay. Given the asks, the server computes its utility. The utility is the difference between the value that the server would pay if it did not receive helps from helpers and the sum of asking prices. The server needs to select a set of winners to maximize the utility within its budget. The optimization problem is NP-hard. However, the server's utility was proved to have the submodularity property. Thus, the winner determination and the payment rules can be implemented by using the results on the budget feasible mechanism design in \cite{singer2010budget} for submodular functions. The experiment results showed that the reverse auction-based approach outperforms the greedy algorithm in terms of server utility and truthfulness. However, the number of shared bandwidth units achieved in the reverse auction-based is less than that of the Stackelberg game-based approach, which can access to full information about the helper utility function.

\subsection{Cloud-based Wireless Multimedia Social Networks (CWMSN)}
\label{sec:CWMSN}
\begin{figure}[ht]
 \centering
\includegraphics[width=8cm, height=10cm]{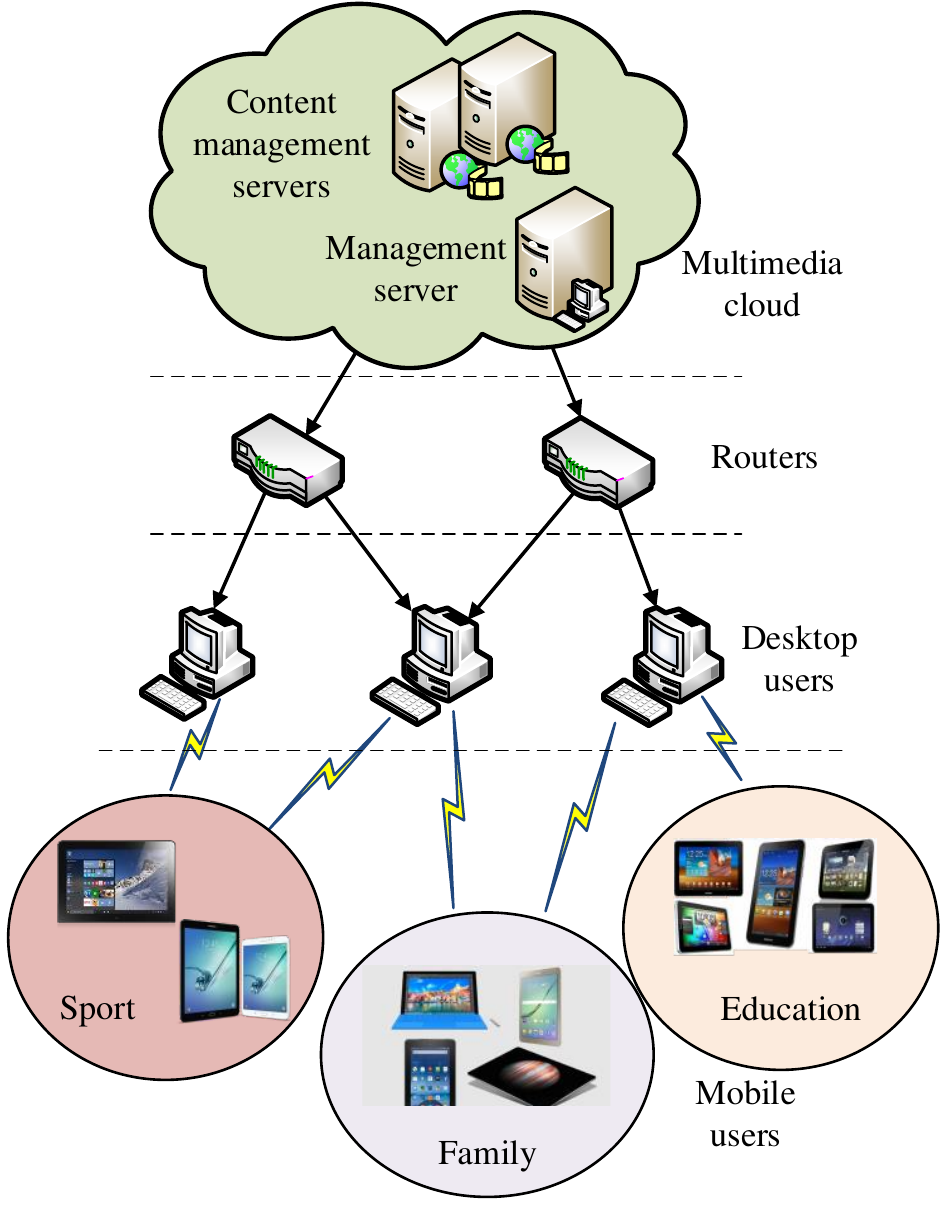}
 \caption{Cloud-based multimedia social network architecture.}
 \label{VoD_model_CWMSN}
\end{figure}

Cloud-based Wireless Multimedia Social Networks (CWMSN) are proposed to support heterogeneous services to users. The CWMSN model is shown in Fig.~\ref{VoD_model_CWMSN} which is essentially a combination of the multimedia cloud and different subnetworks. These subnetworks are based on the various social contexts, e.g., family, interest, and hobby. The system includes content providers, desktop users, and mobile users. The content providers deliver their live programs such as live-streaming and VoD, video conferences, VoIP, photo sharing and editing to the desktop users through distributed servers and gateways. Desktop users have higher computational capability and more resources than those of the mobile users. Therefore, desktop users can share their resources, i.e., the bandwidth, with mobile users which want to obtain live programs. Since the mobile users are selfish for resource competition, game theory-based pricing schemes are appropriate solutions to analyze the bandwidth sharing.

\subsubsection{Stackelberg game}
\label{sec:CWMSN_stackelberg}
In the context of CWMSN, the authors in \cite{nan2014stackelberg} addressed the bandwidth sharing issue between the desktop users and the mobile users. The desktop users decide the amount of bandwidth and the corresponding price, and then each mobile user selects the specific desktop user to connect to. The interactions between them can be formulated as the Stackelberg game in which the desktop users are leaders and the mobile users are followers. Mobile users are not fully rational when making their decisions, and their behaviors are thus modeled by the evolutionary game. Initially, each mobile user in a group randomly connects to a desktop user and calculates its utility based on the allocated bandwidth and the price offered by the desktop user. The mobile user also computes the average utility of the group. If the average utility is greater than its own utility, the mobile user changes its connection to another desktop user to possibly receive higher utility. Otherwise, it keeps the current connection. This process is repeated until all mobile users in the same group achieve equal utility at the equilibrium. Given the result of the mobile users' evolution, desktop users compete with each other in a non-cooperative game by deciding the amounts of bandwidth and corresponding prices to maximize its utility. The solution of the non-cooperative game is the Nash equilibrium which is computed by finding a fixed point of the best response functions of all desktop users \cite{rasti2009pareto}. However, the proof for the existence and uniqueness of the Nash equilibrium was not given.

The bandwidth allocation in \cite{nan2014stackelberg} is the proportional allocation mechanism, meaning that different mobile users receive the same amount of bandwidth if the ratio of their bids to their minimum bandwidth requirements is a constant. This is unfair to the desktop users since the mobile users may pay less while they still receive the requested amount of bandwidth, as long as the ratio is constant. The authors in \cite{nan2014distributed} introduced a punishment coefficient to evaluate the bandwidth of the mobile user obtained from the desktop user. The punishment coefficient is the ratio of a mobile user's bid and its minimum bandwidth requirement. The mobile user is considered to have a cheating behavior if its punishment coefficient is less than one since its bid cannot be lower than its minimum bandwidth requirement. Then, the bandwidth obtained by the mobile user will be reduced by the corresponding punishment. The simulation results highlighted that with the proposed solution, i.e., the cheat-proof strategy, each mobile user must honestly bid the amount of bandwidth which is equal to its minimum bandwidth requirement to obtain its desired amount. By contrast, with the non-cheat-proof method, the mobile users can obtain the same amount of bandwidth, but they only pay less.

In practice, the Nash equilibrium as mentioned in \cite{nan2014stackelberg} and \cite{nan2014distributed} cannot be obtained if the desktop users do not have the information about each other's strategies, i.e., the amount of shared bandwidth and the price. Learning-based algorithms that adjust strategies toward increasing the utility of the desktop users can be used to update this information.

\subsubsection{Generic pricing mechanism}
\label{sec:CWMSN_generic_pricing}

The approaches in \cite{nan2014stackelberg} and \cite{nan2014distributed} did not consider the characteristic of mobile users. Using the same model, on the contrary, the authors in \cite{nan2015pricing} divided the mobile users into two categories which are price-sensitive users and QoS-sensitive users. Each desktop user determines its two pseudo-demand functions corresponding to these two types of mobile users. In general, these functions depend on the possible maximum number of mobile users and the bandwidth price. The desktop user formulates its benefit maximization problem based on the demand functions. The problem is then solved by using the first-order derivative and the Girolamo Cardano algorithm \cite{ekert2008complex} which provides the optimal price to each desktop user. Since the number of mobile users connecting to a specific desktop user may vary in each time period due to price changes by other desktop users, the desktop user needs to continuously adjust the bandwidth price to maximize its utility. The desktop users may change their bandwidth prices until their total non-decreasing utility is unchanged. However, the price adjustments did not consider the constraints on the available bandwidth of the desktop users.

\begin{table*}
\caption{Applications of economic and pricing models for resource management in cloud-based VoD systems.}
\label{table_VoD_system}
\scriptsize
\begin{centering}
\begin{tabular}{|>{\centering\arraybackslash}m{0.2cm}|>{\centering\arraybackslash}m{0.4cm}|>{\centering\arraybackslash}m{1.6cm}|>{\centering\arraybackslash}m{1cm}|>{\centering\arraybackslash}m{0.8cm}|>{\centering\arraybackslash}m{1.2cm}|>{\centering\arraybackslash}m{5.4cm}|>{\centering\arraybackslash}m{2.4cm}|>{\centering\arraybackslash}m{1.2cm}|}
\hline
\multirow{2}{*} {\textbf{}} & \multirow{2}{*} {\textbf{Ref.}} & \multirow{2}{*} {\textbf{Pricing model}} & \multicolumn{3}{c|} {\textbf{Market structure}} & \multirow{3}{*} {\textbf{Mechanism}} & \multirow{2}{*} {\textbf{Objective}} & \multirow{2}{*} {\textbf{Solution}} \tabularnewline
\cline{4-6}
 & & & \textbf{Seller} & \textbf{Buyer} & \textbf{Item} & & &\tabularnewline
\hline
\hline
\parbox[t]{2mm}{\multirow{9}{*}{\rotatebox[origin=c]{90}{ \hspace{-3cm} Cloud-based VoD system}}}
&\cite{cong2014lbas}& Combinatorial auction& Cloud providers&VoD provider&Video instances &Given sellers' asks, the buyer selects the winner based on the difference between the price evaluated by the buyer minus sellers' asking prices. The payment is based on the Vickrey auction&Cost minimization, and truthfulness&Optimal
solution \tabularnewline \cline{2-9}

&\cite{niu2012theory} & Generic pricing& Public cloud providers&VoD providers&Probabilistic bandwidth guarantees &A broker determines the bounds of price offered to each buyer based on the Gaussian distribution of the buyer's demand and the pricing policy of the sellers &Bandwidth reservation cost minimization, and cloud resource efficiency optimization&Nash equilibrium \tabularnewline \cline{2-9}

&\cite{niu2012pricing} & Non-cooperative game& Broker&VoD providers&Cloud bandwidth &Buyers submit their pricing strategies, and optimal prices are achieved using the Cauchy-Schwarz inequality and the proof by contradiction &Utility maximization for buyers&Nash equilibrium \tabularnewline \cline{2-9}

&\cite{niuasynchronous} & Utility maximization& Cloud provider&VoD providers&Egress network bandwidth &Seller updates the price using the first-order partial derivative of its total cost. Then, the buyer determines its optimal resource allocation based on the updated price &Social welfare maximization, and profit maximization&Optimal solution \tabularnewline \cline{2-9}

&\cite{niu2013efficient} & Utility maximization& Cloud provider&VoD providers&Egress network bandwidth &Buyers update the price using the first-order derivative of their utility functions. The seller determines the optimal resource allocation for buyers &Social welfare maximization, and profit maximization&Optimal solution \tabularnewline \cline{2-9}
\hline
\parbox[t]{2mm}{\multirow{9}{*}{\rotatebox[origin=c]{90}{ \hspace{-4cm} P2P-assisted cloud-based VoD system}}}
& \cite{lin2015autotune} & Stackelberg game& Cloud provider&Users&Bandwidth &The VoD provider sets the price using the usage-based pricing. Buyers select bit rates to maximize their utilities, and then the VoD provider sets the price so as to maximize its revenue. Each buyer recalculates the optimal bit rate& Optimal bit rate for buyers, and revenue maximization for the seller&Stackelberg
equilibrium \tabularnewline \cline{2-9}

& \cite{feng2010peer} & Double auction& Peers&Peers&Video segment &If an ask of a seller is less than a bid of a buyer, there is a transaction at which the price is the average of the ask and the bid & Ex-post individual rationality, budget balance, and cloud bandwidth reduction&Market equilibrium \tabularnewline \cline{2-9}

& \cite{wu2012incentivizing}& Generic pricing& Peers&VoD provider&Video caching service&Buyer sets the price for caching video depending on the refreshing probability, the desired number of replicas of the video in the system, and the storage probability of the video & Ex-post individual rationality, budget balance, and cloud bandwidth reduction&Market equilibrium \tabularnewline \cline{2-9}

&\cite{wu2014distributed}& Cost minimization& Peers&VoD provider&Video caching service&Buyer calculates its operation cost involving its upload cost to the cloud and the reward price paid to peers. The reward price is the solution of the operation cost minimization problem& Operation cost minimization&Optimal solution \tabularnewline \cline{2-9}

& \cite{chakareski2015cost} & Stackelberg game& Peers&Cloud&Network bandwidth, storage space, and CPU power&Buyer offers the price for contributing its video contents, and then sellers decide their optimal resource contributions. Given the total resource contribution, the buyer determines the offered price by using the first derivative& Payoff maximization for sellers, and utility maximization for buyer &Stackelberg equilibrium \tabularnewline \cline{2-9}

& \cite{mostafavi2015game} & Stackelberg game& Peers&Streaming server&Upload bandwidth&Sellers decide the number of bandwidth units based on the buyer's budget. Then, the buyer determines the optimal budget using the first-order derivative& Utility maximization for sellers and buyer &Stackelberg equilibrium \tabularnewline \cline{2-9}

& \cite{mostafavi2015game} & Reverse auction& Peers&Streaming server&Upload bandwidth&The buyer computes its utility based on sellers' asks. The winner determination and the payment rules are implemented based on the budget feasible mechanism design for submodular functions &Truthfulness, and utility maximization for buyer&Nash equilibrium\tabularnewline \cline{2-9}
\hline
\parbox[t]{2mm}{\multirow{9}{*}{\rotatebox[origin=c]{90}{ \hspace{-1cm}CWMSN}}}
& \cite{nan2014stackelberg} & Stackelberg game& Desktop users&Mobile users&Bandwidth&Each buyer in a group will change the current connection to the seller until all buyers in the group achieve the equal utility. Then, sellers compete with each other by deciding the amounts of bandwidth and the corresponding prices& Equal utilities for buyers, and payoff maximization for sellers &Stackelberg equilibrium \tabularnewline \cline{2-9}

& \cite{nan2014distributed}& Stackelberg game& Desktop users&Mobile users&Bandwidth&Same as \cite{nan2014stackelberg}, but a punishment coefficient is introduced to avoid the cheating behavior of buyers. The coefficient is the ratio of a buyer's bid and its minimum bandwidth requirement & Equal utilities for buyers, payoff maximization and fairness for sellers &Stackelberg equilibrium \tabularnewline \cline{2-9}

&\cite{nan2015pricing}& Generic pricing& Desktop users&Mobile users&Bandwidth&Each seller determines the pseudo-demand function of the bandwidth demand of buyers and then formulates its benefit maximization problem & Utility maximization for sellers, and efficient allocation &Utility equilibrium \tabularnewline \cline{2-9}

\hline
\end{tabular}
\par\end{centering}
\end{table*}

\begin{table*}[h]
\caption{A summary of advantages and disadvantages of major approaches for the resource management in cloud-based VoD systems.}
\label{table_sum_advantage_VoD_system}
\scriptsize

\begin{centering}
\begin{tabular}{|>{\centering\arraybackslash}m{2cm}|>{\centering\arraybackslash}m{5.5cm}|>{\centering\arraybackslash}m{5.5cm}|}
\hline
\cellcolor{myblue} &\cellcolor{myblue} &\cellcolor{myblue} \tabularnewline
\cellcolor{myblue} \multirow{-2}{*} {\textbf{Major approaches}} &\cellcolor{myblue} \multirow{-2}{*} {\textbf{Advantages}} &\cellcolor{myblue} \multirow{-2}{*}{\textbf{Disadvantages}} \tabularnewline
\hline
\hline
\cite{cong2014lbas} &\begin{itemize} \item  Achieve economic efficiency \item Avoid collusion \end{itemize} & \begin{itemize}  \item Have high computational complexity \item Support only one VoD provider \end{itemize}\tabularnewline \cline{2-3}
\hline
\cite{niu2013efficient} &\begin{itemize} \item Support multiple VoD providers \item Minimize communication overhead \end{itemize} & \begin{itemize}  \item Have slow convergence \item Support only one cloud provider  \end{itemize}\tabularnewline \cline{2-3}
\hline
  \cite{lin2015autotune}&\begin{itemize} \item Have stable equilibrium \item Achieve win-win solution\end{itemize} & \begin{itemize}  \item Have slow convergence \end{itemize}\tabularnewline \cline{2-3}
 \hline
\cite{wu2014distributed} &\begin{itemize} \item Adaptive to nonasymptotic system and highly dynamic video popularity \end{itemize} & \begin{itemize}  \item  Be challenging to learn the feature of system dynamics and adjust the pricing scheme \end{itemize}\tabularnewline \cline{2-3}
 \hline
 \cite{mostafavi2015game} &\begin{itemize} \item Achieve computational efficiency  \end{itemize} & \begin{itemize}  \item Do not support online situations, i.e., the available bandwidth changes
dynamically \end{itemize}\tabularnewline \cline{2-3}
  \hline
 \cite{nan2014distributed} &\begin{itemize} \item Prevent cheating behaviors  \end{itemize} & \begin{itemize}  \item  Have unstable equilibrium \item Require having information about each other's strategies  \end{itemize}\tabularnewline \cline{2-3}
 \hline
\end{tabular}
\par\end{centering}
\end{table*}

\textbf{Summary:} In this section, we have discussed two major issues in cloud-based VoD systems. We have reviewed economic and pricing approaches for these issues. A summary of these approaches is given in Table~\ref{table_VoD_system}. As shown in the table, most approaches address the bandwidth allocation issue in the cloud-based VoD systems since the bandwidth consumption for the video distribution in these systems is crucial. Moreover, the Stackelberg game is considered to be an efficient pricing model in guaranteeing utility maximization for both sellers and buyers. In the next section, we review economic and pricing approaches for the resource management in cloud-based Software Defined Wireless Network (SDWN) model. SDWN is a combination of the cloud (i.e., the cloud data center networking), Software-Defined Networking (SDN), and wireless networks.

\section{Applications of economic and pricing models for resource management in cloud-based SDWN}
\label{sec:SDN_NFV}
\begin{figure}[ht]
 \centering
\includegraphics[width=\linewidth]{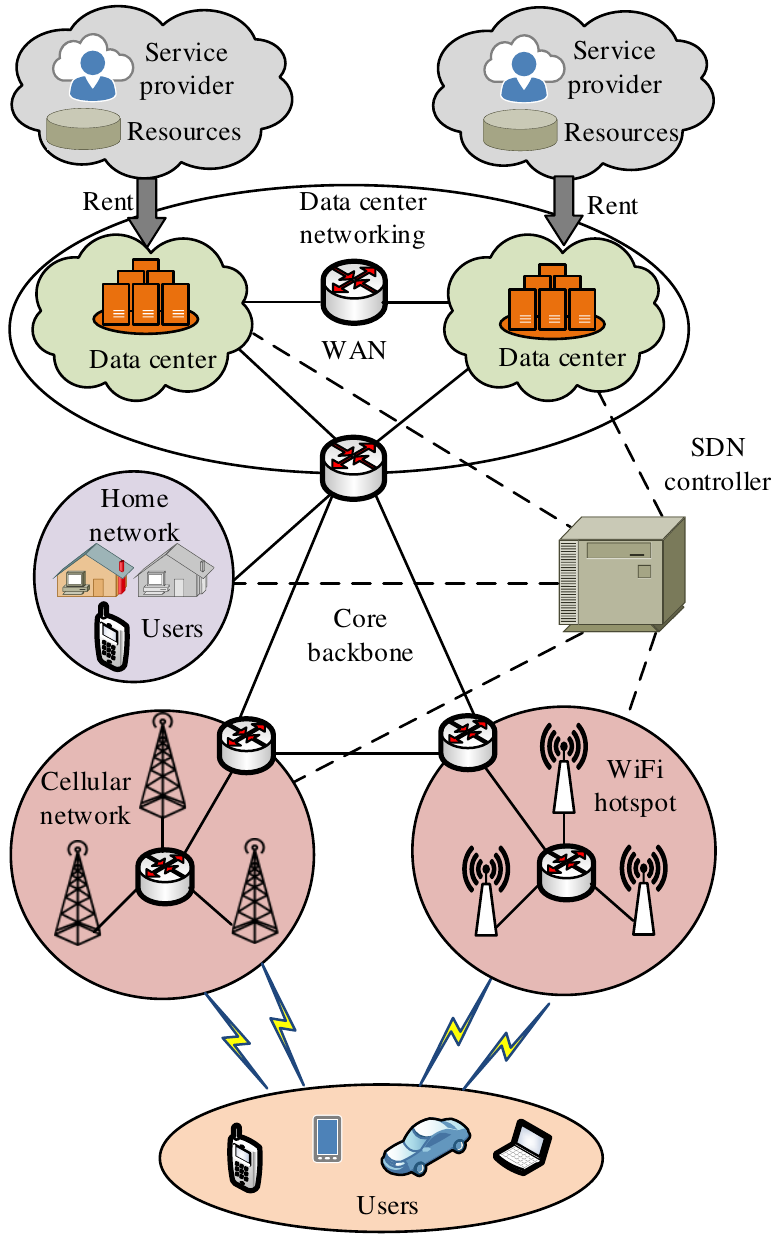}
 \caption{An illustration of cloud-based software-defined wireless network.}
 \label{SDWN_model}
\end{figure}

Distributed data centers in cloud data center networking as discussed in Section \ref{sec:cloud_data_center} can reduce data transfer cost and delay for users. However, the geo-distributed networks increase the difficulty of global resource management unless there is a centralized control. SDN (see Section \ref{sec:cloud_SDN_definition}) provides a real-time centralized control based on both instantaneous network status and user defined policies. In practice, SDN has been adopted in wireless networks to form the Software Defined Wireless Network (SDWN) \cite{li2012toward}, \cite{chaudet2013wireless}, \cite{jagadeesan2015software} for the centralized control and global optimization \cite{bernardos2014architecture}. Thus, data center networking can be combined with SDWN, namely cloud-based SDWN, for complex network management as illustrated in Fig.~\ref{SDWN_model}. In the network model, the SDN controller acts as a ``brain'' of the network. It monitors and allocates resources from data centers to users via wireless networks, i.e., cellular networks and WiFi hotspots. Such centralized resource management usually aims at optimizing the total benefit of all stakeholders. Thus, economic and pricing models such as bargaining game, network utility maximization, and stackelberg game, are appropriate solutions since their outcomes can guarantee maximizing the overall utilities of all stakeholders. In particular, the economic and pricing models have been used to address the following issues.


\begin{itemize}
\item{\textit{Bandwidth allocation}: Bandwidth allocation in the cloud-based SDWN is to allocate bandwidth from data centers owned by cloud providers to Service Providers (SPs) and mobile users in a centralized manner at the SDN controller. Economic and pricing models have been used to maximize the payoffs for cloud providers, service providers, and mobile users simultaneously.}
\item{\textit{Mobile data offloading}: Mobile data offloading, also known as WiFi offloading, is the use of complementary network technologies to reduce the amount of data being transferred through cellular networks. In the cloud-based SDWN, the mobile data offloading is enabled by SDN at an edge of the network to dynamically position or reposition the traffic. However, since the complementary networks and cellular networks may belong to different parties, the traffic offloading is implementable only if the benefits of the parties are satisfied. Economic and pricing models such as the contract theory have been adopted to guarantee the condition.}
\end{itemize}
Before discussing these approaches, the cost analysis is introduced to evaluate the cost of wireless networks when SDN is enabled.

\subsection{Cost analysis}
\label{sec:SDN_NFV_SDWN_cost}
The authors in \cite{zhang2015cost} analyzed the costs of the LTE network owned by a Mobile Network Operator (MNO) with and without SDN, denoted as SDN-LTE and non-SDN-LTE, respectively. The Finnish mobile network topology was adopted as a reference model. The costs generally are divided into the CAPital EXpenditure (CAPEX) and OPerational EXpenditure (OPEX). CAPEX involves costs of network equipments and their deployment cost. The SDN-LTE configuration is implemented in a more centralized manner, and thus the deployment cost is lower than that of non-SDN-LTE. OPEX includes energy consumption, site visits, and network management expenses. Since SDN increases the automation of fault detection, the network management expenses in SDN-LTE are expected to decrease compared with those in non-SDN-LTE. Moreover, the site visit costs in the non-SDN-LTE model arise since base stations, i.e., eNBs, and switches are distributed across the country which may incur more travel expenses and time on site. The quantitative results showed that SDN-LTE can reduce the annual CAPEX by around 7.72\% and the annual OPEX by 0.31\% compared with non-SDN LTE. However, these savings are relatively small compared with the total annual cost of MNO.

Similarly, the authors in \cite{knoll2015life} evaluated the total cost of the LTE network through CAPEX and OPEX when the Network Function Virtualization (NFV)/SDN approaches are implemented. Three possible scenarios were considered. The first scenario is that MNO owns the Virtualized Network Functions (VNFs)/SDN and hardware. CAPEX involves initial investments of the VNF/SDN software licences and the hardware. OPEX includes the maintenance cost, variable costs such as energy consumption and cooling cost, software update, certificate update and bug fixing costs. In the second scenario, MNO owns VNF/SDN and rents hardware. CAPEX is still the VNF/SDN software licenses, but OPEX is the annual rental fee for the required hardware. In the third scenario, MNO rents both VNF/SDN and hardware. In this case, there is no CAPEX while OPEX involves the cost incurred for VNF/SDN, annual service charge, e.g., the hardware and software maintenance, and energy consumption. The simulation showed that the total cost when VNF/SDN are used is always lower than that without VNF/SDN. Moreover, the total cost in the second scenario is the lowest over the period from 2014 to 2019.

The evaluations using the above cost model show the benefit in terms of cost minimization when SDN is enabled. In the following, we provide some other economic and pricing models for the SDN-based resource management in the cloud-based SDWN. Note that although most approaches in this section addressed the issues in the cloud-based SDWN, a few wireline-enabled SDN models, e.g., SDN-enabled home networks, will be also discussed as an extension.
\subsection{Bargaining Game for Bandwidth Allocation}
\label{sec:SDN_NFV_SDWN_bargaining}
Considering the cloud-based SDWN, the authors in \cite{ding2015service} investigated the issue of sharing resources, i.e., CPU, memory and bandwidth, among service providers to extend their capabilities in serving mobile users. The service providers can be divided into sellers and buyers. The main objective is to guarantee QoS for users while maximizing the total increased utility of all sellers and buyers. Thus, the Nash bargaining game with cooperation policy was adopted. The sellers cooperate with each other to form a resource pool, and the buyers bid for the resources in the resource pool with a unit price. The increased utility of the seller is obtained from leasing resources and the price while the increased utility of the buyer is defined by the difference of its utility after and before renting resources from the sellers. The objective function is the total increased utility of all buyers and sellers. The deal price is then determined based on the Cardano's formula \cite{dunham1990cardano}, and the basic optimization is used to obtain the total allocated resource. The buyers share the total resource corresponding to their demands. The simulation results showed that the proposed approach increases both allocated resources for buyers and revenue for sellers compared with the bargaining game in which buyers compete with each other. However, the bargaining game with the competition policy is more appropriate in the realistic scenarios due to the resource scarcity.

\subsection{Pricing Models for Mobile Data Offloading}
\label{sec:SDN_NFV_SDWN_Pricing_Models}
As shown in Fig.~\ref{SDWN_model}, Access Points (APs) can be integrated in the cloud-based SDWN to minimize the cloud bandwidth consumption for the base stations \cite{3GPP_IPFLOW}, and thus reducing the service cost and latency for mobile users. In this context, two economic and pricing models were adopted corresponding to two specific scenarios. In the first scenario, the access points and base stations are owned by one service provider, and the aim of the mobile data offloading is to maximize utility functions of all mobile users. Thus, the NUM framework was used. In the second scenario, the access points and base stations are owned by different providers, and the access points will only admit the traffic from base stations under some conditions. The negotiation on these conditions between the access points and the base stations can be modeled using a contract theory. Further details of the two pricing models are as follows.

\subsubsection{Utility maximization}
\label{sec:SDN_NFV_SDWN_utility}
In the first scenario, the authors in \cite{kang2016sdn} considered allocating the heterogeneous bandwidth including the cloud and WiFi bandwidth to the mobile users by using the NUM framework (see Section \ref{subsec:Utility_maximization}). The utility of each mobile user is a strictly concave function of the heterogeneous bandwidth allocated to the mobile user. The method of Lagrange multipliers is used to solve the optimization problem with their interpretations as cloud bandwidth and WiFi bandwidth prices that the users are willing to pay. At each iteration, the bandwidth prices are updated by using the gradient projection method. Then, the allocated bandwidth is updated by the SDN controller. It was proved that there always exists a unique optimal solution for the allocated bandwidth and the prices. The simulation showed that the proposed approach outperforms the baseline approach in terms of the total network utility. The baseline approach did not use SDN as well as heterogeneous bandwidth. However, as stated in \cite{niuasynchronous} and \cite{niu2013efficient}, multiple SDN controllers need to be considered to enhance the scalability of the system.


\subsubsection{Contract theory}
\label{sec:SDN_NFV_SDWN_contract}

 In the second scenario, the contract model was adopted which allows to construct several traffic-payment bundles, e.g., the amount of traffic that the access point needs to offload and the corresponding payment that the base station needs to offer. The authors in \cite{zhang2015offloading} used three contract theoretic models for the service trading between the base station and the access points in SDN, namely \textit{perfect discrimination}, \textit{linear pricing}, and \textit{anti adverse selection}. In the \textit{perfect discrimination}, there does not exist the information asymmetry, meaning that the idle capacities of the access points are known by the base station. In this scenario, the base station can solve its payoff optimization problems separately for each access point. The base station's payoff associated with an access point is defined as the monetary gain through the offloaded traffic minus its payment to the access point, given the constraint that the access point's payoff is equal to or greater than zero. By taking the derivative of the objective function, an optimal payment to the access point and the amount of offloaded traffic can be obtained. The optimal payment is that the marginal valuation equals the marginal cost. On the contrary, in the \textit{linear pricing}, the base station does not know access points' idle capacities, but it has knowledge of the probability that the access point has a certain idle capacity. Therefore, the base station only specifies a unit of traffic per payment for the offloading process. In particular, the base station formulates its payoff optimization problem by calculating the expected payment requested by the access points through the probability functions. Then, the optimal amount of traffic per payment to the access points is obtained by taking the first derivative of the objective function.

Obviously, the payoff of the base station in the \textit{perfect discrimination} is higher than that in the \textit{linear pricing}. However, in the \textit{perfect discrimination}, there is no compatible incentive for the access points since their payoffs are zero. To obtain an incentive compatible contract, the \textit{anti adverse selection} is used. The \textit{anti adverse selection} is similar to the \textit{perfect discrimination}. However, the constraint of the incentive compatibility for access points is added into the optimization problem of the \textit{perfect discrimination} (i.e., the base station's payoff maximization). This problem was then solved by using the Lagrange multiplier method to determine the optimal contract, i.e., the traffic-payment, for each access point. As shown in the simulation results, the \textit{linear pricing} gives the access points the highest payoff, followed by the \textit{anti adverse selection}, and then the \textit{perfect discrimination}. On the contrary, the base station gets the maximum payoff in the \textit{perfect discrimination} since it has full knowledge of access points' idle capacities.
\subsection{Stackelberg Game for Bandwidth Allocation in SDN-Enabled Home Networks}
\label{sec:SDN_NFV_SDWN_stackelberg}
SDN can be combined with the existing home networks as shown in Fig.~\ref{SDWN_model} to maintain users' Quality of Experience (QoE) and guarantee the QoE \cite{krishnan2013video}. In this setting, home networks are connected to the service providers via a digital subscribe line or broadband cable link, and users request content services from the service provider through the subscribe lines.

Typically, service providers rent cloud bandwidth from cloud providers to deliver their contents to users. The authors in \cite{eghbali2015bandwidth} investigated maximizing payoffs for the service provider and users by adopting the Stackelberg game. The service provider, i.e., the leader, charges the users, i.e., followers, for their requesting services via SDN according to the time-dependent usage-based pricing strategy \cite{zhang2014time}. The service provider also gives a reimbursement to the user, which is proportional to the amount of traffic that the user shares with other neighboring users. Thus, the payoff of the service provider is the difference between the total service price from the users and the total reimbursement. The user's payoff is the utility function of the allocated resources minus the charge that it pays the service provider. Given the service provider's pricing and reimbursement strategy, the users aim to maximize their payoffs by choosing traffic consumption and the amount of sharing bandwidth. The optimization problems of users are solved based on the Lagrange multiplier with the subgradient projection method. Based on these optimal solutions, the service provider determines the pricing and reimbursement strategy so as to maximize its payoff function. The numerical results showed that the payoffs of both the user and the service provider from the proposed approach improve 400\% compared with that in the best-effort home network with usage-based pricing \cite{zhang2014time}. However, the impact of limited backhaul capacity on the payoff functions needs to be considered in the future work.

We close this section with an approach which addresses the price-based bandwidth allocation for control applications in SDN as proposed in \cite{feng2014joint} and \cite{tao2015allocation}. The aim is to maximize the rate of control applications while guaranteeing the fairness of allocation among them. The fairness criterion means that the rate of a control application is proportional to the price of bandwidth paid by the control application. The optimization problem is to maximize the sum of the rate of each control application multiplied with the corresponding price given the limited capacity. The optimization problem is a strictly convex function of the allocated rate. Therefore, there exists a unique optimal solution for the rate for each control application. However, the flow table, an essential network resource in SDN, needs to be considered in the future work.

\begin{table*}
\caption{Applications of economic and pricing models for resource management in cloud-based SDWN}
\label{table_SDN}
\scriptsize
\begin{centering}
\begin{tabular}{|>{\centering\arraybackslash}m{0.1cm}|>{\centering\arraybackslash}m{0.4cm}|>{\centering\arraybackslash}m{1.6cm}|>{\centering\arraybackslash}m{1cm}|>{\centering\arraybackslash}m{0.9cm}|>{\centering\arraybackslash}m{1.2cm}|>{\centering\arraybackslash}m{5.4cm}|>{\centering\arraybackslash}m{2.4cm}|>{\centering\arraybackslash}m{1.3cm}|}
\hline
\multirow{2}{*} {\textbf{}} & \multirow{2}{*} {\textbf{Ref.}} & \multirow{2}{*} {\textbf{Pricing model}} & \multicolumn{3}{c|} {\textbf{Market structure}} & \multirow{3}{*} {\textbf{Mechanism}} & \multirow{2}{*} {\textbf{Objective}} & \multirow{2}{*} {\textbf{Solution}} \tabularnewline
\cline{4-6}
 & & & \textbf{Seller} & \textbf{Buyer} & \textbf{Item} & & &\tabularnewline
\hline
\hline
\hline
\parbox[t]{2mm}{\multirow{9}{*}{\rotatebox[origin=c]{90}{ \hspace{-5cm}Cloud-based SDWN}}}
&\cite{zhang2015cost} & Cost model& Mobile network operator& Users&Networking resources &Buyer evaluates the costs of the LTE network when SDN is enabled. Two types of cost are introduced, i.e, CAPEX (fixed costs) and OPEX (variable costs)&Seller's profit maximization&Cost optimization \tabularnewline \cline{2-9}

& \cite{knoll2015life} & Cost model& Mobile network operator& Users&Networking resources &Same as \cite{zhang2015cost}, but three scenarios are considered when evaluating the operation cost of the LTE network&Seller's profit maximization&Cost optimization \tabularnewline \cline{2-9}

&\cite{ding2015service} & Bargaining game& Selling SPs& Buying SPs&CPU, memory, and bandwidth &The objective is to maximize the total increased utility of all buyers and sellers. The deal price is determined based on the Cardano's formula, and the total allocated resource is obtained using the first derivative&Increased utility maximization for both buyers and sellers&Nash bargaining solution \tabularnewline \cline{2-9}

&\cite{kang2016sdn} & Utility maximization& SP&Users&Bandwidth &The method of Lagrange multipliers is adopted. At each iteration, the bandwidth prices are updated by using the gradient projection method, and then the allocated bandwidth for buyers is updated by the seller&Total network utility maximization&Optimal solution \tabularnewline \cline{2-9}

&\cite{zhang2015offloading} &Contract theory& Access points&Base station&Mobile data offloading service&Three contract theoretic models for the service trading are used, that are \textit{perfect discrimination}, \textit{linear pricing}, and \textit{anti adverse selection}. In particular for the \textit{anti adverse selection}, the seller determines the optimal prices and the amounts of bandwidth using the Lagrange multiplier method & Payoff maximization, incentive compatibility, and individual rationality&Optimal solution \tabularnewline \cline{2-9}

&\cite{eghbali2015bandwidth} & Stackelberg game& SP&Users&Bandwidth&Seller sets the bandwidth price based on the time-dependent usage-based pricing strategy. Then, buyers choose the amount of bandwidth based on the Lagrange multiplier method. The seller finally determines the bandwidth price & Payoff maximization for sellers and buyers&Stackelberg equilibrium \tabularnewline \cline{2-9}
\hline
\end{tabular}
\par\end{centering}
\end{table*}

\begin{table*}[h]
\caption{A summary of advantages and disadvantages of major approaches for the resource management in cloud-based SDWN.}
\label{table_sum_advantage_SDWN_system}
\scriptsize

\begin{centering}
\begin{tabular}{|>{\centering\arraybackslash}m{2cm}|>{\centering\arraybackslash}m{5.5cm}|>{\centering\arraybackslash}m{5.5cm}|}
\hline
\cellcolor{myblue} &\cellcolor{myblue} &\cellcolor{myblue} \tabularnewline
\cellcolor{myblue} \multirow{-2}{*} {\textbf{Major approaches}} &\cellcolor{myblue} \multirow{-2}{*} {\textbf{Advantages}} &\cellcolor{myblue} \multirow{-2}{*}{\textbf{Disadvantages}} \tabularnewline
\hline
\hline
\cite{ding2015service}&\begin{itemize} \item Achieve flexible resource management
and demand-driven resource distribution  \end{itemize} & \begin{itemize}  \item Be not appropriate in the realistic scenarios \end{itemize}\tabularnewline \cline{2-3}
\hline
\cite{kang2016sdn} &\begin{itemize} \item Manage resource in a centralized and holistic manner \item Enable reliable functional verification \ \end{itemize} & \begin{itemize} \item Be unscalable \item Do not support heterogeneous resource allocation \item Support only one SDN controller  \end{itemize}\tabularnewline \cline{2-3}
\hline
\cite{zhang2015offloading} &\begin{itemize} \item Overcome the
information asymmetry \item Support multiple traffic-payment bundles  \end{itemize} & \begin{itemize} \item Support only one SDN controller  \end{itemize}\tabularnewline \cline{2-3}
\hline
\cite{eghbali2015bandwidth} &\begin{itemize} \item Adapt to both real-time and non-real-time service requests \item Be resilient to demand fluctuations \end{itemize} & \begin{itemize} \item Support only one service provider  \end{itemize}\tabularnewline \cline{2-3}
 \hline
\end{tabular}
\par\end{centering}
\end{table*}

\textbf{Summary:} This section discusses the applications of economic and pricing models in cloud-based SDWN. The existing approaches which address the resource management in the network are summarized in Table~\ref{table_SDN}. In general, the number of existing approaches is relatively small, and most of them investigated the bandwidth allocation. More studies need to be done, for example, for the mobile data offloading.
\section{Review summary, open issues and future research directions}
\label{sec:Open_issues}

A number of approaches reviewed in this survey show the importance of economic aspects not only for business development, but also for system design and optimization. Evidently, economic and pricing models have addressed various issues in cloud networking models in which traditional algorithms becomes less effective or cannot be applied. Apart from the existing approaches, there are still challenges, open issues, and new research directions as discussed in the following.


\subsubsection{False-name bidding in double auction}
\label{sec:Open_issue_false_bidding}
 In the reviewed approaches based on double auction \cite{zheng2015star}, \cite{zhang2014community}, \cite{Jin}, \cite{jinauction}, \cite{wu2016scalable}, multiple cloud resources are sold simultaneously. As such, one bidder can submit multiple bids using different identities so as to gain an additional profit. The bidding process which is implemented under fictitious identities is called false-name bidding \cite{yokoo2004effect}, and the double auction with the false-name bidding is no longer dominant-strategy incentive compatible. Therefore, robust mechanisms against the false-name bidding, e.g., the Threshold Price Double (TPD) auction protocol \cite{yokoo2005robust}, for cloud bandwidth reservation need to be investigated.

\subsubsection{Collusion in auction}
\label{sec:Open_issue_collusion}
 Apart from the false-name bidding cheating, bidders in the reviewed approaches based on auction, i.e., the VCG auction \cite{gui2014soar}, \cite{khantowards}, the combinatorial auction \cite{forde2011exclusive}, and the double auction \cite{zheng2015star}, \cite{zhang2014community}, \cite{Jin}, \cite{jinauction}, \cite{wu2016scalable}, may collude with each other through coordinating their bids. This suppresses the competition for cloud resource, thus reducing the price that the bidders must pay for the cloud resource. However, the collusion behavior will degrade the efficiency of the resource allocation as well as the cloud provider's revenue. Collusion-resistant mechanisms thus need to be applied. Pricing strategies can be used to provide the bidders incentives not to perform the collusion behavior.

\subsubsection{Resource demand forecasting methods}
\label{sec:Open_issue_Forecasting}
 As discussed in \cite{wanis2015modeling} (see Section \ref{sec:cloud_data_center_bandwidth_other}), resource demand fluctuation of cloud tenants impacts the availability of resources, the pricing policy as well as the profit of cloud providers. Therefore, it is important for the cloud provider to predict the workload fluctuation. However, there was only one technique introduced for the demand forecast, i.e., the Markov chain model. Therefore, more advanced techniques need to be adopted to improve the performance of resource demand prediction in cloud networking environment. Some candidate methods are fuzzy logic \cite{hagras2004creating}, neural networks \cite{zhang2010artificial}, machine learning \cite{li2011predicting}, and a joint statistical learning and optimal decision-making \cite{bogacz2007optimal}. 

\subsubsection{Pricing models for multipath routing in cloud data center networking}
\label{sec:Open_issue_data_center_networking}
\begin{figure}[ht]
 \centering
\includegraphics[width=7.6cm, height=5.8cm]{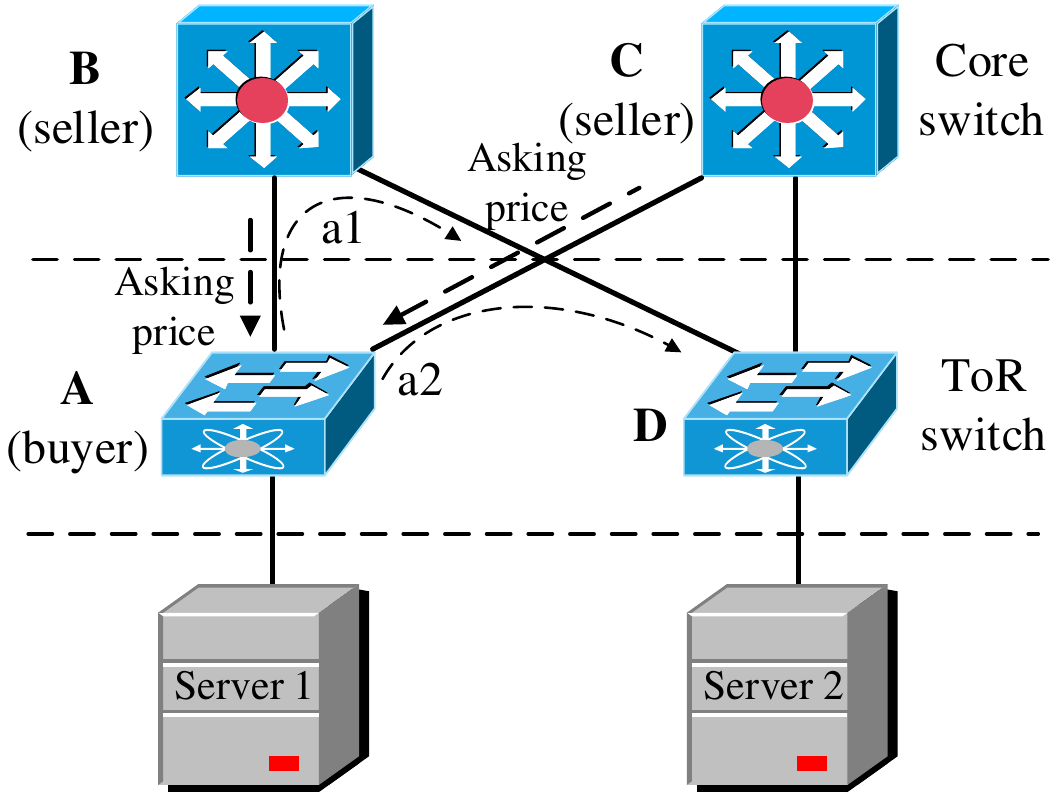}
 \caption{Multipath routing in data center networks based on reserve auction.}
 \label{multipath_routing_data_center}
\end{figure}

Data centers consist of several hundreds and thousands of servers \cite{benson2010network}. The servers are connected with each other through multi-rooted hierarchical topologies, e.g., the fat tree topology \cite{al2008scalable}, which provide multipath between each pair of servers. A simple example is shown in Fig.~\ref{multipath_routing_data_center} in which there are two different paths, i.e., a1 and a2, between servers 1 and 2. Efficient routing strategies are thus necessary. Traditional routing algorithms such as distance vector \cite{he2002destination} and link state \cite{clausen2003optimized} do not support the multipath routing efficiently because they make routing decisions only based on packet destinations. This means that all packets to the same destination are routed through the same path which may result in congestion. By taking into account the multipath feature, pricing approaches such as auction are efficient solutions for the path selection to achieve a highly load-balanced network and minimize network latency. For example, the reverse auction can be used at the switch A which acts as a buyer to select path a1 or a2 based on asking prices submitted from switches B and C, i.e., sellers. The asking price of switch B or C is proportional to the total congestion level of links connected to the switch and the number of hops from the switch towards the destination. The path containing the lowest asking price is selected for forwarding the packet. 
\subsubsection{Pricing models in cloud robotics}
\label{sec:Open_issue_Robotics}
\begin{figure}[ht]
 \centering
\includegraphics[width=6.6cm, height=8cm]{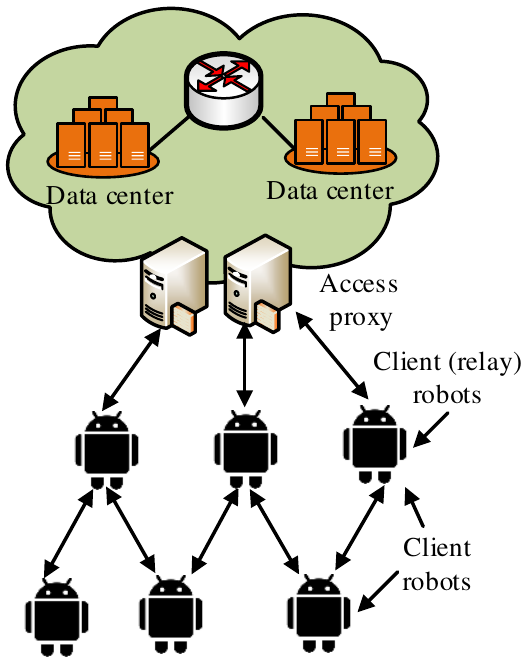}
 \caption{An architecture of typical cloud robotic systems.}
 \label{robotic_system}
\end{figure}

Cloud robotics is a combination of robotic technology with network and cloud computing infrastructure \cite{li2016usage}, \cite{hu2012cloud}. Robots benefit from powerful computing, storage, and communication resources of data centers in the cloud. This integration enables robotic systems to be smarter, faster, and less expensive. A common cloud robotic system is shown in Fig.~\ref{robotic_system}. Multiple robots are connected with data centers through wired or wireless networks to perform tasks. The proxy allocates resources in the cloud to the client robots. Since robotic applications require real-time execution, the key challenge is the low-latency response given network bandwidth constraint. Therefore, efficient bandwidth allocation among robots is of great importance. Resource management mechanisms such as auction or Stackelberg game are promising solutions since they model an interaction among client robots. For example, auctions can be adopted among client robots, i.e., buyers, which compete for the connectivity provided by a relay robot, i.e., a seller \cite{wang2014hierarchical}. To maximize the utility of relay or client robots and cloud provider's profits, the Stackelberg game can be used \cite{wang2012game}, \cite{wang2016pricing} in which the cloud provider, i.e., the leader, optimizes bandwidth price, and client robots, i.e., followers, choose their transmission rates. 

\subsubsection{Economic and pricing models in Network Function Virtualization (NFV)}
\label{sec:Open_issue_NFV}
Network Function Virtualization (NFV) is a concept that leverages virtualization technologies to offer a new way for designing, deploying, and managing networking services. NFV and SDN are closely related technologies which are implemented through the software running on physical equipments. While SDN decouples the network control and forwarding functions on a physical network equipment, NFV decomposes network functions, e.g., a firewall, from the physical network equipment \cite{mijumbi2015network}. NFV brings several benefits such as reduction of OPEX and CAPEX due to consolidating networking appliances \cite{bhaumik2012cloudiq}, facilitating the deployment of new services with increased agility and faster time-to-value, and achieving better system scalability according to users' demand. However, implementing NFV introduces several challenges. For example, physical resource sharing among multiple users can lead to congestion on the physical infrastructure as well as an unfair use of the resources \cite{elias2014optimization}, \cite{elias2016efficient}. Pricing approaches such as auction or smart data pricing can be adopted to offer incentives to users to use resources efficiently. 

\section{Conclusions}
\label{sec:Conclusion}
This paper has presented a comprehensive survey of the applications of economic and pricing theories to resource management in cloud networking. Firstly, we have described a general architecture of the cloud networking including its components and corresponding services. Then, we have introduced and analyzed various pricing models with the objectives to understand the motivations of using the economic and pricing theory in cloud networking. Afterwards, we have provided detailed reviews, analyses, and comparisons of the approaches using economic and pricing theories to solve a variety of issues in specific systems of cloud networking, i.e., the cloud data center networking, mobile cloud networking, edge computing, cloud-based VoD system, and cloud-based SDWN. Finally, we have outlined open issues as well as future research directions.


\bibliographystyle{IEEEtran}
\bibliography{CloudNetworkingDatabase}{}
\vspace{-1cm}

\end{document}